\documentclass[a4paper,11pt]{article}
\pdfoutput=1 

\usepackage{jcappub} 

\usepackage[T1]{fontenc} 

\usepackage[switch]{lineno} 
\usepackage{graphicx}	
\usepackage{color,xcolor}
\usepackage{upgreek}
\usepackage{xspace}
\usepackage{aas_macros}

\newcommand\Msun{M_\odot}

\newcommand{\degr}{\ensuremath{^\circ}\xspace}

\renewcommand{\citep}{\cite}
\renewcommand{\citet}{\cite}
\renewcommand{\citealt}{\cite}

\newcommand\changereftwo{}  
\newcommand\changeref{}  
\newcommand\change{}
\newcommand\changetwo{}
\newcommand\changethree{}
\newcommand\bkg{{\change CR background}\xspace}

\newcommand\BO[1]{#1}
\newcommand{\BOout}[1]{}
\newcommand\MC[1]{#1}
\newcommand{\MCout}[1]{}
\newcommand\LT[1]{#1}
\newcommand\FA[1]{#1}
\newcommand\QR[1]{#1}

\title{\boldmath Prospects for a survey of the Galactic plane with the Cherenkov Telescope Array}

\author[1]{S.~Abe,}
\author[2]{J.~Abhir,}
\author[3]{A.~Abhishek,}
\author[4,5,*]{F.~Acero,%
\note[*]{Corresponding authors\\
F.~Acero (\url{fabio.acero@cea.fr})\\
M.~Chernyakova (\url{masha.chernyakova@dcu.ie})\\
B.~Olmi (\url{barbara.olmi@inaf.it})\\
Q.~Remy (\url{quentin.remy@mpi-hd.mpg.de})\\
L.~Tibaldo (\url{luigi.tibaldo@irap.omp.eu})\\}
}
\author[6]{A.~Acharyya,}
\author[7,8]{R.~Adam,}
\author[9]{A.~Aguasca-Cabot,}
\author[10]{I.~Agudo,}
\author[11,12]{A.~Aguirre-Santaella,}
\author[13]{J.~Alfaro,}
% ...
\affiliation[1]{Institute for Cosmic Ray Research, University of Tokyo, 5-1-5, Kashiwa-no-ha, Kashiwa, Chiba 277-8582, Japan}
\affiliation[2]{ETH Z\"urich, Institute for Particle Physics and Astrophysics, Otto-Stern-Weg 5, 8093 Z\"urich, Switzerland}
\affiliation[3]{INFN and Universit\`a degli Studi di Siena, Dipartimento di Scienze Fisiche, della Terra e dell'Ambiente (DSFTA), Sezione di Fisica, Via Roma 56, 53100 Siena, Italy}
\affiliation[4]{Universit\'e Paris-Saclay, Universit\'e Paris Cit\'e, CEA, CNRS, AIM, F-91191 Gif-sur-Yvette Cedex, France}
\affiliation[5]{FSLAC IRL 2009, CNRS/IAC, La Laguna, Tenerife, Spain}
\affiliation[6]{University of Alabama, Tuscaloosa, Department of Physics and Astronomy, Gallalee Hall, Box 870324 Tuscaloosa, AL 35487-0324, USA}
\affiliation[7]{Universit\'e C\^ote d'Azur, Observatoire de la C\^ote d'Azur, CNRS, Laboratoire Lagrange, France}
\affiliation[8]{Laboratoire Leprince-Ringuet, CNRS/IN2P3, \'Ecole polytechnique, Institut Polytechnique de Paris, 91120 Palaiseau, France}
\affiliation[9]{Departament de F{\'\i}sica Qu\`antica i Astrof{\'\i}sica, Institut de Ci\`encies del Cosmos, Universitat de Barcelona, IEEC-UB, Mart{\'\i} i Franqu\`es, 1, 08028, Barcelona, Spain}
\affiliation[10]{Instituto de Astrof{\'\i}sica de Andaluc{\'\i}a-CSIC, Glorieta de la Astronom{\'\i}a s/n, 18008, Granada, Spain}
\affiliation[11]{Institute for Computational Cosmology and Department of Physics, Durham University, South Road, Durham DH1 3LE, United Kingdom}
\affiliation[12]{Instituto de F{\'\i}sica Te\'orica UAM/CSIC and Departamento de F{\'\i}sica Te\'orica, Universidad Aut\'onoma de Madrid, c/ Nicol\'as Cabrera 13-15, Campus de Cantoblanco UAM, 28049 Madrid, Spain}
\affiliation[13]{Pontificia Universidad Cat\'olica de Chile, Av. Libertador Bernardo O'Higgins 340, Santiago, Chile}
% ...

\abstract{
Approximately one hundred sources of very-high-energy (VHE) gamma rays are known in the Milky Way,  detected with a combination of targeted observations and 
\change{surveys}. A survey of the entire Galactic Plane in the energy range from a few tens of GeV to a few hundred TeV has been proposed as a Key Science Project for the upcoming Cherenkov Telescope Array Observatory (CTAO). This article presents the status of the studies towards the Galactic Plane Survey (GPS). We build  and make publicly available a sky model that combines data from recent observations of known gamma-ray emitters with state-of-the-art physically-driven models of synthetic populations of the three main classes of established Galactic VHE sources (pulsar wind nebulae, young and interacting supernova remnants, and compact binary systems), as well as of interstellar emission from cosmic-ray interactions in the Milky Way. We also perform an optimisation of the observation strategy (pointing pattern and scheduling) based on recent estimations of the instrument performance. We use the improved sky model and observation strategy to simulate GPS data corresponding to a total observation time of 1620 hours spread over ten years. Data are then analysed using the methods and software tools under development for 
real data. Under our model assumptions and for the realisation considered, we show that the GPS has the potential to increase the number of known Galactic VHE emitters by almost a factor of five. This corresponds to the detection of more than two hundred pulsar wind nebulae and a few tens of supernova remnants at average {\changetwo integral} fluxes one order of magnitude lower than in the existing sample above 1 TeV, therefore opening the possibility to perform unprecedented population studies. The GPS  also has the potential to provide new VHE detections of binary systems and {\changetwo pulsars}, to confirm the existence of a hypothetical population of gamma-ray pulsars with an additional TeV emission component, and to \changereftwo{detect bright} {\changetwo sources capable of accelerating particles to PeV energies (PeVatrons)}. Furthermore, the GPS will constitute a pathfinder for deeper follow-up observations of these source classes. Finally, we show that we can extract from GPS data an estimate of the contribution to diffuse emission from unresolved sources, and that there are good prospects of detecting interstellar emission and statistically distinguishing different scenarios.
Thus, a  survey of the entire Galactic plane carried out from both hemispheres with CTAO will ensure a transformational advance in our knowledge of Galactic VHE source populations and interstellar emission.
}

\begin{document}
\maketitle
\flushbottom

\section{Introduction}
Current gamma-ray observations in the very-high-energy (VHE) domain (above a few tens of GeV) have revealed a wide range of particle acceleration and interaction phenomena in Galactic astrophysical objects including pulsar wind nebulae (PWNe, e.g., \citealt{2018A&A...612A...2H}), supernova remnants (SNRs, e.g., \citealt{2018A&A...612A...3H}),  and {\change gamma-ray binaries \citep[e.g.][]{grlbCTA19}}, as well as many sources {\change whose nature is} still unknown.
{\change Important} results include the prevalence of extended sources powered by pulsars in the VHE sky {\change \cite[e.g.][]{2018A&A...612A...1H}}, the {\change measurement of soft gamma-ray spectra for most SNRs which challenges} the standard paradigm for the origin of Galactic cosmic rays (CRs) {\change up to the knee \cite[e.g.][]{funk2015}}, and the existence of particle confinement or slow diffusion around many of these sources {\change \cite[e.g.][]{2021Univ....7..141T}}.
{\change Despite these results, }the mechanisms of particle acceleration and their efficiencies, the maximum {\change achievable} energies, the nature of the accelerated particles and the radiation processes, the role played by the source's environment and the evolution with the source's age, and the contributions to the population of Galactic CRs remain uncertain in most cases.

Surveys, which provide large samples of sources with a small selection bias, have enabled major advances in astronomy in general, and in the field of VHE astronomy in particular. VHE surveys were carried out with imaging atmospheric Cherenkov telescopes (IACTs) which were the first instruments with sufficient sensitivity to detect the brightest TeV sources \cite{CrabWhipple}, and with ground-based air shower arrays that only recently reached sensitivity to detect larger numbers of sources. Air shower arrays have a worse angular resolution and worse sensitivity for short observations, but they reach higher energies and have a larger field of view so that they naturally survey the entire observable sky and are better suited to detect very extended \BO{sources} and \BO{Galactic} diffuse emission.

Historically, the first survey of a quarter of the Milky Way (with \BO{Galactic} longitude range $-2\degr < l < 85\degr$) at {\changetwo very high energies} was performed by the HEGRA system of imaging atmospheric Cherenkov telescopes (IACTs), with no significant detection of gamma-ray sources \citep{2002A&A...395..803A}. The largest IACT survey to date was performed by H.E.S.S. \citet{2018A&A...612A...1H}, and covered the longitude range $-110\degr < l < 65\degr$  and latitudes $|b| \lesssim 3\degr$, with a total of 78 sources detected. 
{\changetwo Additionally, VERITAS performed a survey of the nearby Cygnus star-forming region, leading to the detection of a new source and the detailed study of three other VHE sources \citep{2018ApJ...861..134A}.}

Among air-shower arrays, Milagro observed the Galactic plane at longitudes $30\degr < l < 216\degr$ and detected TeV emission towards 14 previously known Galactic gamma-ray sources 
\citep{2009ApJ...700L.127A}. 
ARGO-YBJ surveyed the Milky Way at longitudes $-10\degr < l < 70\degr$ detecting four gamma-ray excesses with {\changetwo statistical significance greater than five standard deviations ($> 5\sigma$)} toward the Galactic plane \citep{2013ApJ...779...27B}. 
The third catalog of sources detected by the HAWC observatory \citep{2020arXiv200708582A}, which covers \BO{Galactic} longitudes $0\degr < l < 110\degr$ and $150\degr < l < 240\degr$, includes 65 sources located mostly towards the Galactic plane.
{\changetwo The LHAASO collaboration have recently released their first catalog of gamma-ray sources \citep{LHAASO1cat:2023} with 90 sources detected with high significance ($>5\sigma$) above 1 TeV, including 32 new TeV sources, as well as 43 sources with lower significance ($> 4\sigma$) above 100 TeV.} LHAASO \BO{has }detected photons from Galactic sources up to a maximum of 1.4 PeV  \citep{2021Natur.594...33C}.
 
Furthermore, due to its wide field of view and decade-long exposure, the Large Area Telescope (LAT) aboard the \textit{Fermi Gamma-ray Space Telescope} has reached into the VHE domain. The LAT has surveyed the entire sky, detecting 189 high-significance sources at energies above 30 GeV at $|b|<10\degr$ \citep{2022arXiv220111184F}. Of those, 61 are VHE emitters known from previous surveys.

Diffuse TeV gamma-ray emission was also observed towards the Galactic plane \cite{2008ApJ...688.1078A,2014PhRvD..90l2007A,HAWCdiffuse,TibetASgamma:2021tpz,Zhao:2021dqj}. While at GeV energies Galactic diffuse emission is interpreted as the result of CR interactions with interstellar gas and radiation fields, at TeV it is likely a superposition of interstellar emission and collective emission from populations of sources that cannot be detected individually {\changetwo given the sensitivities of the current instruments \cite[e.g.][and references therein]{2021Univ....7..141T}.

The Cherenkov Telescope Array (CTA) is the next-generation IACT system \citep{2011ExA....32..193A,2023arXiv230512888H}. {\changetwo CTA will consist of  a southern array located at the Paranal Observatory (Chile) and a northern array located at} Roque de los Muchachos Observatory (Spain), {\change allowing observations of} \change{almost} the entire sky. Its energy range will extend from 20~GeV to beyond 200~TeV, with a sensitivity {\change improvement of} an order of magnitude with respect to current IACT systems. With a large field of view, reaching {\change $>8\degr$ diameter} at the highest energies\footnote{For an IACT array the effective field of view is determined by the camera's geometrical field of view, and also by a decreasing detection efficiency at increasing angular distances from the centre of the cameras. The webpage
\url{https://www.cta-observatory.org/science/ctao-performance/\#1472563544190-020879e1-468f}
shows how the CTA sensitivity to a point-like source varies as a function of angular distance from the centre of the field of view and of photon energy.}, and  an angular resolution of a few arcminutes, {\changetwo CTA is  well-suited to perform large surveys.}

A survey of the entire Galactic plane was proposed as a Key Science Project for the CTA Observatory (CTAO) \citep{2013APh....43..317D,2019CTAscience} with the goals of:
\begin{enumerate}
\item providing a census of Galactic VHE emitters, such as SNRs and PWNe, through the detection of hundreds of new sources and, therefore, enabling population studies;
\item identifying a list of promising candidates for follow-up observations, such as new gamma-ray binaries,  PeVatron candidates (objects capable of accelerating particles to PeV, i.e., up to the knee of the CR spectrum);
\item determining the properties of TeV diffuse emission from the Galactic plane;
\item searching for new and unexpected high-energy phenomena in the Milky Way;
\item providing a legacy dataset to the astronomical and astroparticle physics communities.
\end{enumerate}
{\changetwo The results from the Galactic Plane Survey (GPS) during the first years of operations will also be instrumental in seeding other observational programmes, including several Key Science Projects and the general observer programme. Therefore, periodic releases of data and scientific results to the public during the execution of the GPS are proposed \cite{2019CTAscience}.} We note that other dedicated Key Science Projects are proposed to study transient and flaring sources, including Galactic sources and extragalactic sources observed towards the Galactic plane \cite{2019CTAscience}, for which GPS observations can also be exploited. 
\changereftwo{Galactic transients are discussed in a companion publication \cite{CTAGalTrans}.}

In this article we provide an updated and improved assessment of {\change prospects for} the GPS compared to \cite{2019CTAscience}. We describe the proposed implementation and expected scientific returns, as well as the development of analysis pipelines.
In section~\ref{sec:skymodel} we describe the sky model used in our simulations, based on recent gamma-ray source catalogues and on observations of individual sources, as well as state-of-the-art models of the main populations of VHE gamma-ray sources (SNRs, PWNe, binaries) and interstellar gamma-ray emission.  In section~\ref{sec:observations} we describe the implementation of the 
survey in terms of pointing pattern and observation scheduling and the simulations of the observations used for this paper.  In section~\ref{sec:catalogs} we describe how we created catalogues of sources in the entire Galactic plane based on the results of the analyses of the simulated data.  In section \ref{sec:population}  we discuss the properties of the two dominant source classes detected in the survey: PWNe and SNRs.  In section~\ref{sec:dedicatedsources} we present analyses of {\change the simulated data aimed at studying} gamma-ray binaries, pulsars, and PeVatrons. 
In section \ref{sec:diffuse} we discuss Galactic diffuse emission, including {\changetwo the emission} from unresolved sources and {\changetwo from} interstellar processes.  Finally, in section \ref{sec:summary}, we present a summary of the results and discuss perspectives for the future.

\section{Sky Model}\label{sec:skymodel}

Gamma-ray emission from the Milky Way in the VHE domain is dominated by individual emitters, which include SNRs, pulsars, PWNe and pulsar halos, {gamma-ray binaries}, star-forming regions (SFRs), and sources of still unknown nature \citep[e.g.,][]{2018A&A...612A...1H}. A large-scale diffuse component is also detected and interpreted as a mix of interstellar emission from the interactions of CRs with interstellar gas and radiation fields and of collective emission from populations of sources that cannot be detected individually {\changetwo given the sensitivities of the current instruments} \citep[e.g.,][]{2014PhRvD..90l2007A}.

The sky model used in this paper consists 
of three main components that are described in this section:
\begin{enumerate}
\item a set of {\change known} sources modelled on the basis of observations from current and past instruments;
\item synthetic population models tuned to existing observations for the three main classes of Galactic VHE emitters (SNRs, PWNe, and {\change gamma-ray} binaries) to model \BO{the sources that will be detected  by CTA as well as those that will remain undetected {\changetwo and will then constitute an unresolved diffuse emission component};}
\item models for interstellar emission.
\end{enumerate}

\subsection{Known sources}\label{sec:known-sources}

\FA{We model the population of known sources with a compilation of existing catalogues from IACTs, the {\it Fermi}-LAT space telescope, and air-shower arrays. For the known VHE sources detected by IACTs we used the compilation provided by gamma-cat\footnote{\url{https://gamma-cat.readthedocs.io}}} that we enhanced based on recent results not yet included in the official gamma-cat release.
 The completeness of this source list was checked against TeVCat\footnote{\url{http://tevcat2.uchicago.edu/}} \citep{2008ICRC....3.1341W}.
 
 \FA{Galactic variable sources detected at {\changetwo very high energies} {\change (i.e. binaries and pulsars)} are also included in our sky model.}
For pulsars, we include dedicated spectral and temporal models for 38 objects preselected among those detected by \textit{Fermi}~LAT  \citep{2013ApJS..208...17A, 2017ApJS..232...18A}. We arbitrarily implement two {spectral classes}, either extending from GeV to TeV energies as simple steep power laws extrapolated from  \textit{Fermi}-LAT measurements (Crab-like pulsars), or with an additional TeV component modelled as  a power-law with an exponential \BO{cutoff} whose brightness is randomly generated assuming a TeV to GeV flux ratio ranging from $5\times 10^{-5}$ to $10^{-2}$, while complying at the same time with upper-limits available in the literature (Vela-like pulsars). More details and the full list of pulsars simulated are provided in appendix~\ref{app:skymodelpsr}. Gamma-ray  binaries  are  known to be variable on orbital and sometimes superorbital timescales \citep[see, e.g.,][for a review]{grlbCTA19}. Some of them also display spectral variability as a function of the orbital phase. Therefore, for six known gamma-ray binaries and one candidate gamma-ray binary within the sky area covered by the GPS, we introduce dedicated temporal and spectral models as detailed in table~\ref{tab:grlb}.

We then add sources detected by \textit{Fermi}~LAT  \citep[3FHL catalogue,][]{2017ApJS..232...18A} and by HAWC \citep[2HWC catalogue\footnote{The 3HWC catalogue, based on a 30\% longer integration time, { \changetwo and the first catalogue of LHAASO sources} became available only when this work was already in an advanced state.},][]{2017ApJ...843...40A}. {\change Sources are added only if {\changetwo they are} not associated with already included TeV sources. 
Furthermore, we exclude 3FHL and 2HWC sources for which a source among those discussed above lies within their $3\sigma$ position uncertainty.
Finally,} to limit the number of very soft sources of {\change limited interest for CTA}, we include only 3FHL sources detected at energies $> 50$~GeV with a significance $>3\sigma$, {\changetwo and associated with }at least one photon of measured energy $> 100$~GeV.

In gamma-cat, 2HWC, and 3FHL\BO{,} the morphology of extended gamma-ray sources is modelled using simple geometrical templates (disc or Gaussian {\changetwo shapes}). In order to test source detection and characterisation algorithms on more realistic source models\BO{,} we include 14 sources modelled by dedicated templates\BO{,} that either replace the corresponding catalogue model or are added to the sky model for a few recent detections. Templates are also used to model some diffuse features not captured by the large-scale interstellar emission models described below. The list and a description of the templates \BO{are} provided in table~\ref{tab:templates}.

Furthermore, the gamma-ray catalogues describe the source spectra using simple analytical functions. Often the limited statistics above a few TeV prevent current instruments from detecting spectral curvature, therefore the catalogues report the spectra as simple power laws. However, based on the maximum energy {\change achievable} by the accelerators, radiative cooling of the particles, and the physics of the gamma radiation mechanisms, \BO{a power law is not expected}
to be a good approximation of the gamma-ray spectrum up to several hundreds of TeV for many sources {\change(see e.g. \citealt{2022Univ....8..505C})}. 
In order to assess the capabilities of CTA to measure spectral curvature in a more realistic {\change scenario},  we add a physically-motivated exponential cutoff to the power-law spectra with {\changetwo  a hard spectral index} ($< 2.4$) {\change as described in appendix~\ref{sec:cutoffs}. 
The cutoff is applied to the spectrum} of known SNRs, PWNe, active galactic nuclei\footnote{AGNs are background sources for Galactic studies. Those detected by current instruments within the sky area covered by the GPS are included in the sky model.} (AGNs), and unidentified sources, which are treated as if they belong to the dominant source class according to the energy band of the observations (PWNe for gamma-cat and 2HWC, AGNs for 3FHL). 
We do not modify the spectrum of the two PeVatron candidates Westerlund~1 \citep{2019NatAs...3..561A} and HESS~J1641$-$463 \citep{2015ICRC...34..834O}, which are therefore included with the currently measured power-law spectrum without any curvature.

\subsection{Synthetic source populations}

CTA is expected to extend our horizon of detectability and {\changetwo to observe} a number of presently undetected faint sources.
Therefore, our sky model needs to include a population of synthetic sources representing {\change 
fainter, and yet undetected objects}. The following subsections describe the methodology we use to create synthetic populations of the three main classes of VHE sources, namely SNRs, PWNe, and binaries.

\subsubsection{Supernova Remnants}\label{sec:snrpop}
%
SNRs are simulated using the approach described in \citet{2013MNRAS.434.2748C,2017MNRAS.471..201C}:
a \BO{supernova (SN)} event is randomly placed in the Galaxy in time and position, assuming a constant rate of  3 explosions per 100 years \citep{2011MNRAS.412.1473L} \BO{over the last $10^5$ years}.

Four types of SNe are considered: \BO{thermonuclear (TN), high-ejecta-mass core-collapse (CC-HEM),  low-ejecta-mass core-collapse (CC-LEM) and high-explosion-energy core-collapse (CC-HEE).}  \changeref{The following relative rates for the various SN types are assumed \citep{2010ApJ...718...31P}: 0.32, 0.44, 0.22, and 0.02 respectively.}
Each type is characterised by a typical range of physical parameters: the mass of the ejecta (from $1.4\Msun$ to $20\Msun$), the mass loss-rate and the velocity of the wind (as in  \citealt{2010ApJ...718...31P}).
{\changethree  The explosion energy is fixed to $10^{51}$ erg.}
For a better description of core-collapse systems associated with PWNe, we consider the mass of the ejecta to have a Gaussian distribution with peak at $13\,\Msun$ and $\sigma=3\,\Msun$, truncated at a minimum value of $5\,\Msun$ and  a maximum of $20\,\Msun$ \citep{2009ARA&A..47...63S}. 

{\change The Galaxy is modelled} with four logarithmic spiral arms as in \citet{2006ApJ...643..332F}; {\change core-collapse SNe follow the distribution of pulsars  \citep{2004IAUS..218..105L} {\changethree characterised by the Galactocentric radial distribution described in \citet{2006ApJ...643..332F}}, while thermonuclear SNe are placed considering a flat  distribution for the inner region \citep{2006MNRAS.370..773M}, plus a radial {\changetwo distribution} in the outer zone \citep{2005ApJ...629L..85S, 2011RvMP...83.1001P}. The height above the Galactic plane follows the distribution of the molecular gas}.

{\change The evolution of SNRs requires {\change us} to follow the time variation of the position of the forward shock ($R_{\mathrm{sh}}$), marking the shell boundary in the ambient medium, modelled following the three-dimensional distribution of the {\changetwo atomic and molecular hydrogen} presented in \citet{2003PASJ...55..191N, 2006PASJ...58..847N}, with density varying {\changetwo between} $10^{-5}$ and 10 cm$^{-3}$.}

Gamma-ray emission is produced by particles accelerated at the SNR shock, both via neutral-pion decay (in case of nuclei) and inverse-Compton (IC) scattering (in case of electrons). 
{\change Their spectra are modelled as} a power-law in the particle momentum $p$, following \citet{2013MNRAS.434.2748C,2017MNRAS.471..201C}:  $f(R_{\mathrm{sh}},\,p,\,t) = A(t) p^{-\alpha}$ , {\change where} $\alpha$ is considered as a free parameter, ranging {\change over} $4.1\leq \alpha \leq 4.4$ (e.g. \citealt{2005A&A...429..755P, 2007ARNPS..57..285S, 2008ApJ...678..939Z, 2010A&A...513A..17O, 2010APh....33..160C,2011MmSAI..82..760G}).
{\change The normalisation $A(t)$ is determined by requiring that the CR pressure at the shock is a fraction of the shock ram pressure (fixed to 0.1, \citealt{2008ApJ...678..939Z}). {\changetwo The maximum energy is determined} by the assumption that particles escape the acceleration region when their diffusion length equals 1/10 of the shock radius.}

Electrons are accelerated at the same rate {\changetwo as} protons, and the two types of particles have identical spectral shapes at low enough energies, where radiation losses affecting electrons can be neglected. {\change The ratio between the electron and proton spectra is assumed to be  $10^{-4}$} \citep{2013MNRAS.434.2748C,2017MNRAS.471..201C, 2010ApJ...712..287E, 2012A&A...538A..81M}. 
The electron spectrum is shaped by radiation losses above a threshold energy (where the {\change characteristic energy-loss time becomes} shorter than the SNR age), and it steepens by {\changetwo 1.0} in spectral index with respect to the injection spectrum \citep{2012A&A...538A..81M}.

{\changethree The spatial distribution of the accelerated particles inside the SNR is determined by solving a transport equation, assuming  a 20~$\upmu$G magnetic field inside the SNR shell, a momentum-dependent Bohm-like diffusion coefficient, advection and radiative losses (details can be found in  \citealt{2013MNRAS.434.2748C}).}

The gamma-ray emission from each SNR of the population is finally obtained by calculating: (i) the hadronic component from proton-nuclei interactions (see \citealt{2006PhRvD..74c4018K} and \citealt{2009APh....31..341M} for nuclei heavier than hydrogen); (ii) the {\change (dominant)} leptonic component from IC scattering of the accelerated electrons on CMB photons \citep{1970RvMP...42..237B}.


\subsubsection{Interacting Supernova Remnants (iSNRs)}
%
{\changetwo 
In recent years, SNRs interacting with molecular clouds (iSNRs) have emerged as a new class of gamma-ray emitters with observational characteristics which distinguish them from other members of the SNR class.}
In these systems, the gamma-ray emission originates from inelastic interactions of CRs accelerated by the SNR shock with the matter in the cloud (through neutral-pion decay).
These systems are characterised by emission {\changetwo that is} spatially coincident with the molecular clouds \citep[e.g.][and references therein]{Funk2016}.

In order to produce a synthetic population of such systems, three ingredients are needed: a synthetic population of molecular clouds, the properties of CRs inside {\changetwo these} clouds, and the probability that a molecular cloud has an SNR close enough to interact. A synthetic population of Galactic molecular clouds is created using the {\changethree spiral arm} distribution from \citet{2006ApJ...643..332F}.
A mass is assigned randomly to each cloud, assuming a power-law mass function of {\changetwo spectral} index $-1.6$ for clouds with masses $> 10^3 \, M_\odot$ \citep{Rice2016}. 
A broken power-law model is assumed for the CRs inside the cloud. The spectral parameters ({\changetwo i.e. the slopes and the} energy breaks) as well as the CR densities inside the clouds are extracted randomly from the distributions obtained for the observed SNR-molecular cloud systems and assigned to each molecular cloud {\changetwo \citep{silvia2023interacting}}. 
The gamma-ray spectrum is then calculated assuming hadronic emission via the proton-proton interaction.
The probability of interaction with an SNR is assumed to be 1.5$\%$ in order to reproduce the observed flux distribution for known systems. A possible leptonic component associated with these systems is not considered.

\subsubsection{Pulsar Wind Nebulae}
\FA{Synthetic PWNe are produced by associating a pulsar to every core-collapse SN from the population described in section~\ref{sec:snrpop}. The properties of each pulsar are drawn following the prescription by \citet{2011ApJ...727..123W}, based on gamma-ray emitting pulsars.}
Each source is evolved in time {\changetwo assuming} a pure dipole braking for the pulsar spin-down (i.e. a braking index $n=3$). {\changetwo We note that no important variations are expected in the population if the index is varied over the range $2.3\leq n \leq 3$ \citet{2023MNRAS.520.2451B}.}
A three-dimensional kick velocity is associated with each pulsar, following a double-sided exponential distribution with mean value $\sim 380$ km s$^{-1}$  \citep{2006ApJ...643..332F}.

Previous attempts to provide a population of Galactic PWNe, and {\changetwo to estimate} its gamma-ray contribution, were based on the assumption of an ``average'' PWN, {\change with properties tuned} to match observations (see e.g. \citealt{2018A&A...612A...2H}).
{\change Here instead, we use the same approach as {\changetwo in} \citet{2022MNRAS.511.1439F}, where the (approximated) dynamical and radiative evolution of each source in the population is reproduced using a one-zone model. {\changetwo  This provides} a physically-informed description of the intermediate phase of the PWN evolution, when the nebula starts to interact with the shocked ejecta of the SNR (usually known as the \textit{reverberation} phase). This phase may cause important variations in the PWN spectral energy distribution at all energies, due to the system compression, and {\changetwo so} must be properly treated in order not to introduce biases in the population study \citep{2020MNRAS.499.2051B, 2023MNRAS.520.2451B}.}

{\change Leptons} responsible for the 
emission are injected in each PWN following a broken power-law in energy: $Q(E,t)= Q_0(t) (E/E_b)^{\alpha_i}$, with the break energy {\change randomly drawn} from a lognormal distribution centred at $E_b=0.28$ TeV  and with a 0.12 TeV spread.
{\changetwo  The} injection indices 
range {\change over}  $1.0\leq \alpha_1\leq 1.7$ for $E\leq E_b$, and $2.0\leq \alpha_2 \leq 2.7$ otherwise. 
The normalisation $Q_0(t)$ is determined by requiring that the power injected in particles equals a fraction $(1-\eta)$ of the  pulsar luminosity $L(t)$, with  the magnetic fraction $\eta$ ranging {\change over} $0.02-0.2$.
The radiation properties of each PWN are  then computed {\change by} evolving the injection function in time, considering both adiabatic and radiation losses (synchrotron and IC) plus the possible escape through diffusion. 
The magnetic field is modelled as in \citet{2009ApJ...703.2051G}, considering the variation of the magnetic energy due to both the adiabatic expansion/contraction of the nebula, and the energy injection from the pulsar.

Gamma-ray emission through IC is finally evaluated using the Klein-Nishina cross-section for the photon-electron interaction \citep{1970RvMP...42..237B}, considering different local photon fields: photons from the CMB, synchrotron photons, and near and far IR photons from the stellar background, described by a normal distribution centred on the value {\change estimated} from the GALPROP model at the specific position of each source \BO{in the Galaxy} \citep{2017ApJ...846...67P}.

A \BO{considerable} {\change fraction} of the pulsars escape their SNR on a time scale smaller than the final age of the simulation. 
The escaping  pulsars leave a relic nebula at their original position, which remains an active source of gamma-ray emission through IC radiation as long as the pulsar-injected particles survive against losses and adiabatic expansion. Escaping pulsars form then a bow shock nebula {\change that inject high-energy particles in the ambient medium \citep{2019MNRAS.490.3608O} and are possibly associated with the formation of TeV halos. 
The number of expected halos in the Galaxy is still debated -- {\changetwo ranging} from a few to hundreds \citep{2019PhRvD.100d3016S, 2020A&A...636A.113G,2022arXiv220711178M} -- and a conclusive physical  {\changetwo model is still lacking}. For this reason, here we have only included two \textit{ad hoc} models  for the halos surrounding Monogem and Geminga \citep{2017Sci...358..911A}, based on their observational properties (table~\ref{tab:templates}).}

\subsubsection{Gamma-Ray Binaries}

The synthetic population of binaries consists of a list of simulated systems giving the location in the Galaxy, the orbital period, and the orbital lightcurve above 1 TeV. The population is modelled following \citet{2017A&A...608A..59D}, to which we refer for details and references. The binary is assumed to be a $1.4\,\mathrm{M}_{\odot}$ neutron star in orbit with a $30\,\mathrm{M}_{\odot}$ massive star, i.e. a generic pulsar-driven gamma-ray binary. The location in the Galaxy is randomly drawn from the radial distribution of OB stars and following a model of the spiral arms \citep{2004A&A...422..545Y, 2006ApJ...643..332F}. 
The gamma-ray emission assumes that high-energy electrons injected close to the neutron star inverse Compton scatter UV photons from the massive star to VHE energies. The model also takes into account the absorption of the VHE emission due to pair production with the UV photons from the star following \cite{Dubus2006}, as well as physical eclipses of the VHE emission region (assumed pointlike) by the massive star. The orbital period is drawn from a flat distribution in log~P$_{\textrm{orb}}$ from 1 to $10^{4}$ days while the eccentricity is drawn from a thermal eccentricity distribution \citep{Ambartsumian37} corrected for circularisation at short orbital periods. To normalise the lightcurves, the injected power in VHE-emitting electrons is assumed to be $\approx 1\%$ of the total pulsar spin-down power. This was estimated from the VHE observations of the rotation-powered pulsars PSR J2032$+$4127 and PSR B1259$-$63, which both orbit a massive star. The radiative efficiency is taken into account by scaling the power to the ratio of escape timescale to radiative timescale.  The total available power is randomly drawn from the observed distribution of spin-down power of young pulsars.  

The resulting lightcurves display a variety of complex shapes as the relative position of the star, the particles and {\change the} observer changes as a function of orbital phase. The shapes range from sinusoidal behaviour when the orbit is circular to flare-like behaviour for highly-eccentric orbits (see \citealt{2017A&A...608A..59D} for examples). Hence, even though this model was built to reproduce a binary population consisting of young rotation-powered pulsars in orbit around a massive star (gamma-ray binaries), the resulting lightcurves also provide reasonable test cases for the outburst-like behaviour expected of other classes of binary sources, such as microquasars.

\subsubsection{Comparison of the synthetic source populations with the known gamma-ray sky and assembly of the source model}

\FA{We have described in the previous subsections the construction of a sky model including known sources and \LT{populations} of synthetic sources.} In order to reduce computing time and resources we eliminate the low-flux tails of the populations and only retain for the following synthetic sources above an integral photon flux threshold: $6 \times 10^{-16}$~cm$^{-2}$~s$^{-1}$ in the 100 GeV-1 TeV energy interval for steady sources, $2 \times 10^{-17}$~cm$^{-2}$~s$^{-1}$ $>1$~TeV at the lightcurve peak for binaries. \LT{Figure~\ref{fig:lognlosin} compares source counts as a function of {\changetwo integral} flux $> 1$~TeV for the different populations.} {\change To generate the synthetic populations and, in particular, to tune the properties of PWNe, we make the simplifying hypothesis that all currently unidentified sources are PWNe. Therefore, as expected, our source model is largely dominated by PWNe.}

The first LHAASO catalogue \citep{LHAASO1cat:2023} became available when this work was already advanced and was not included in our sky model. However, our synthetic PWN model naturally accounts for a comparable number of sources extending beyond 100~TeV. Indeed, the PWN population contains 72 sources distributed over the entire sky with flux $>100$~TeV similar to the 43 sources detected by LHASSO in the Northern hemisphere only.

In order to build the sky model used as input for realistic simulations, we need to remove sources from the synthetic populations that have equivalents in the known source part of our model to avoid double-counting some sources.
Indeed, as the synthetic source models account for the entire populations, the brightest sources have {\changetwo most likely} already been detected with existing telescopes}. For each source already detected (SNR, iSNR, PWN, composite SNR-PWN system, binary){\changetwo, therefore,} we exclude the most similar synthetic source belonging to the same class. The similarity is established based on source position, {\changetwo angular} extension, and {\changetwo integral} flux $> 1$~TeV. The exact procedure used to identify the most similar source is detailed in appendix~\ref{sec:syn_source_del}. 
{\change After removing those sources, we obtain the global population model of Galactic sources that will be used for the rest of this paper. 
{\changetwo This model} is also made publicly available to the community}\footnote{The model is available at \url{https://doi.org/10.5281/zenodo.8402519}.}. We stress that, for sources below the detection threshold for current instruments, the model is based on a single realisation of the synthetic populations and not on the mean expectation.

\begin{figure}
\centering
  \includegraphics[width=0.8\columnwidth]{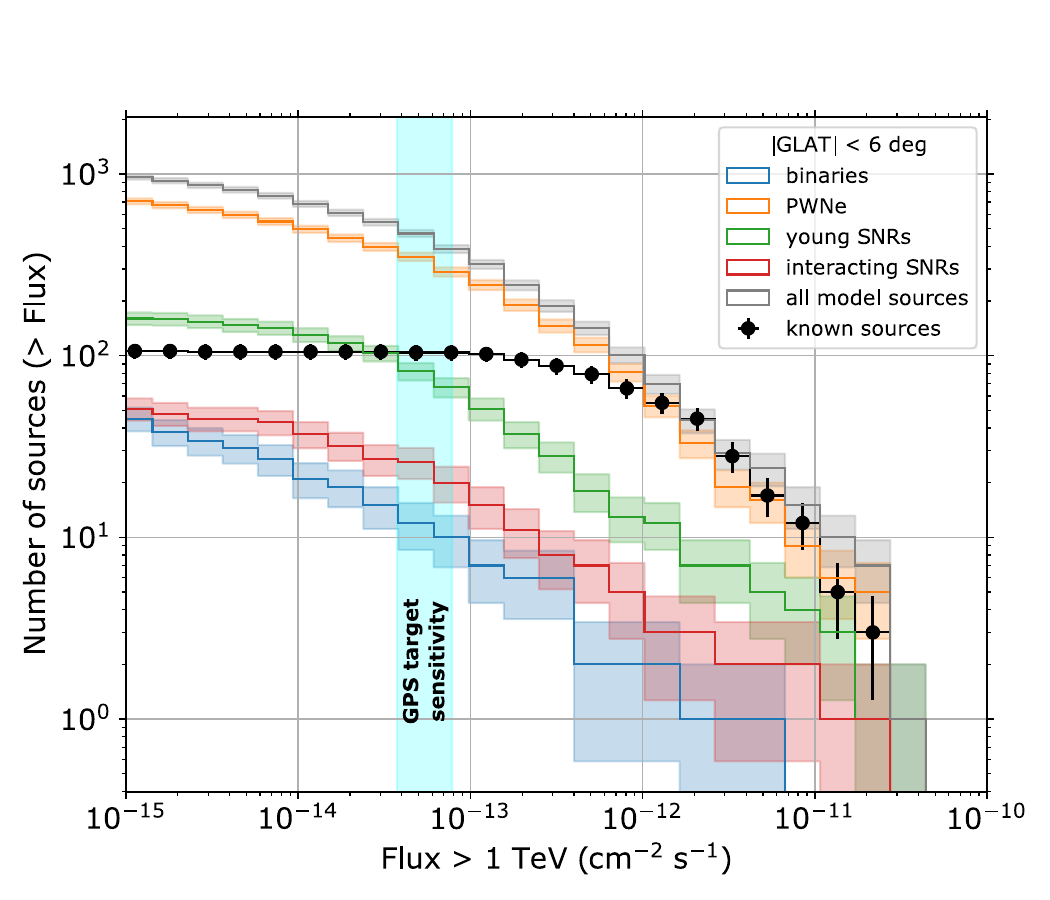}
  \caption{Cumulative number of sources {\changetwo for Galactic latitude $< 6\degr$} as a function of integral source flux above 1 TeV, showing known emitters and the synthetic source populations from our sky model for different source classes. For known emitters the vertical bars show the estimated Poisson fluctuations. The {\changetwo shaded blue vertical band} shows the target sensitivity of the GPS defined in \citet{2019CTAscience}.} \label{fig:lognlosin}
\end{figure}

\subsection{Interstellar emission}\label{sec:iem-simu}

{\change Uncertainties in the predictions for interstellar emission across the CTA energy range are remarkably large \cite[e.g.][]{Luque:2022buq,marinos2023}. Therefore we use a small set of models to asses the uncertainties in our results. The reference interstellar emission model (IEM) used to produce the simulated CTA dataset} (hereafter ``base model'', or IEM-base) is based on the {\tt DRAGON} code {\changetwo that evaluates} the propagation of charged CRs in the Galaxy \citep{Evoli2017jcap,Evoli2018jcap}.
The key assumption behind the base model is that CR transport is homogeneous and isotropic, {\changethree while the Galaxy is axisymmetic}.
The \LT{interstellar} emission, originating from neutral-pion decay,{ \changetwo IC scattering} on the diffuse low-energy photon background, and bremsstrahlung, is computed with the {\tt HERMES} code \citep{Dundovic:2021ryb}. 
The hadronic component \LT{is based on} a gas model composed of a set of column density maps in $(l,b)$ Galactic coordinates for atomic and molecular gas, associated {\change with} Galactocentric rings\footnote{\change The details of the \LT{gas model} are provided in \citet{Galview1} and \url{https://fermi.gsfc.nasa.gov/ssc/data/analysis/software/aux/4fgl/Galactic_Diffuse_Emission_Model_for_the_4FGL_Catalog_Analysis.pdf}.}. \LT{For IEM-base we model the CO-to-H$_2$ conversion factor, $X_{\rm CO}$, using the functional form from \citet{arimoto1996} with parameters tuned to be consistent with recent observations of the Milky Way \cite[e.g.][]{2015ARA&A..53..199G}:
 \begin{equation}
 X_{\rm CO} = 0.6 \times 10^{0.4\left(\frac{R}{5\,\rm{kpc}} - 1\right)} \times 10^{20} \; \rm{cm}^{-2} \, (\rm{K}\, \rm{km}\, \rm{s}^{-1})^{-1},
 \end{equation}
 where $R$ is the Galactocentric radius.
 }

\LT{{\change The base model provides a minimal expectation for interstellar emission in the TeV energy range.
A first alternative model} (IEM-varmin) features an inhomogeneous CR transport including a linear gradient in the index of the diffusion coefficient. \QR{This model is discussed in a recent paper}  \citet{Luque:2022buq}, and predicts fluxes in the multi-TeV domain larger \QR{than the base model} by a factor of a few to ten. 
The normalisation of this model is adjusted mainly by tuning the masses of target molecular gas via $X_{\rm CO}$ \LT{over large regions of the sky}.  The resulting  $X_{\rm CO}$ factors are 2-5 times higher than the values usually measured in \LT{the Milky Way}. Therefore, we introduce a second {\change alternative} model (IEM-varmin rescaled) with the same CR transport setup, but with a rescaling of the $X_{\rm CO}$ values and gamma-ray emissivity per gas nucleon at 8 GeV that match \textit{Fermi}-LAT measurements \citep{Acero2016apjs}.}
\QR{This model predicts an intermediate flux between the base and varmin models at TeV energies.}

\section{Observation plans}\label{sec:observations}

The GPS will consist of observations of the entire Galactic plane using both the southern and northern arrays \cite{2019CTAscience}. The 
{\change target sensitivity for isolated point-like sources} is, at {\changetwo integral} photon fluxes above 1 TeV, $\approx 5 \times 10^{-14}$~cm$^{-2}$~s$^{-1}$, a factor deeper than existing surveys, and coherently covering the full Milky Way. The survey will be graded so that regions with a higher expected number of sources, especially the inner part of the Galaxy, will receive significantly more observation time \BO{from a total allocated budget of 1620~hours over ten years, partitioned between the two arrays and different regions, as described in table~6.3 of  \cite{2019CTAscience} {\change and summarised in table~\ref{tab:hours}. }
\begin{table}
\centering
\begin{tabular}{lccc}
\hline
Region & STP (h) & LTP (h) & Total (h) \\
\hline
{\bfseries SOUTH} \\
300\degr-60\degr, Inner region & 300 & 480 & 780 \\
240\degr-300\degr, Vela, Carina &  & 180 & 180 \\
210\degr-240\degr &  & 60 & 60 \\
{\bfseries NORTH} \\
60\degr-150\degr, Cygnus, Perseus &  180 & 270 & 450 \\
150\degr-210\degr, anticentre &  & 150 & 150 \\
\hline
\end{tabular}
\caption{Observing times in different regions of the Galactic plane. The total times are split in a short-time programme (STP) and a long-time programme (LTP), {\changetwo see main text for more details}. Abridged from table~6.3 of \cite{2019CTAscience}. }
\label{tab:hours} 
\end{table}
These observing times will be used as a working hypothesis for the rest of this paper.}

\subsection{Instrument description, simulation strategy and software}\label{sec:tools}

{\change The results} presented in the paper are based on the full-scope array described in \citet{CTAMClayout}, also known as the CTA {\changetwo baseline} or Omega configuration. {\change This includes 4 Large-Sized Telescopes (LSTs), 25 Medium-Sized Telescopes (MSTs), and 70 Small-Sized Telescopes (SSTs) in the Southern hemisphere, plus 4 LSTs and 15 MSTs in the Northern hemisphere. We refer the reader to \citet{CTAMClayout} for a detailed description of the telescopes and their characteristics.}

{\changethree The data discussed and analysed in the rest of the paper are produced through simulations of the sky model described in section \ref{sec:skymodel}.}
{\change Simulated data in the context of this work consist of science-ready event lists, that is, lists of candidate photons with their estimated arrival directions and energies. Such lists will be produced from an analysis of the Cherenkov images recorded by the telescopes. They are one of the main products expected from CTAO (dubbed Data Level 3, or DL3) and are planned to be openly distributed to the community for the scientific exploitation of the data \citep[see, e.g.][]{zanin2022}.}

{\change We simulate the event lists based on instrument response functions (IRFs) extracted from detailed Monte~Carlo simulations of the interactions of gamma~rays (and background CRs) with the atmosphere and the subsequent collection and recording of their Cherenkov light signal by the telescopes, followed by image analysis, shower reconstruction, and event selection. The IRFs include a description of the gamma-ray effective area, point spread function (PSF), and energy dispersion, as well as a description of the residual CR background as a function of position with respect to the camera centre and measured energy. We used the IRFs} \textit{prod3v2} \citep{prod3b} optimised for observation durations of
50 hours.\footnote{ Specifically, the {\tt South\_z20\_50h}, {\tt South\_z40\_50h} and {\tt South\_z60\_50h} IRFs were
used for the Southern array, while the {\tt North\_z20\_50h} and {\tt North\_z40\_50h} IRFs were
used for the Northern array. {\change The IRF optimisation takes into account a trade-off between background rejection and gamma-ray efficiency that depends on the duration of the observations. In practise, for the observations described in this article, the IRFs optimised for 50 hours are always the most appropriate among those currently available.}} For the rest of the paper we will always assume that the IRFs are perfectly known.

{\changetwo During the preparation of this work, an initial array configuration with a reduced number of telescopes (the Alpha configuration) was approved for construction \citep{zanin2022,2023arXiv230512888H}. While the telescope reduction will impact the prospects of the survey outlined in this paper to some degree, this is expected to be mitigated by ongoing improvements in the event-level analysis tools. In particular, we expect to maintain the off-axis sensitivity which is critical for survey observations. A comparison between current performance curves for the Alpha configuration and those corresponding to the IRFs used in our paper is presented in \cite{irfcomparison}.}

\LT{For easier reading we also list here the simulation and analysis software tools used in the rest of the paper with the corresponding versions and references.}
\begin{itemize}
	\item \textit{ctools}, version 1.6.3 or higher \citep{ctools};
	\item \textit{Gammapy} version 0.17 or higher \citep{gpy}.
\end{itemize}
Simulations were performed with \textit{ctools}. Catalogues were produced with both \textit{ctools} and \textit{Gammapy} independently.
For other parts of the analysis one of the two software packages was conveniently chosen.

\subsection{Pointing pattern}
\begin{figure}
\centering
  \includegraphics[width=.7\textwidth]{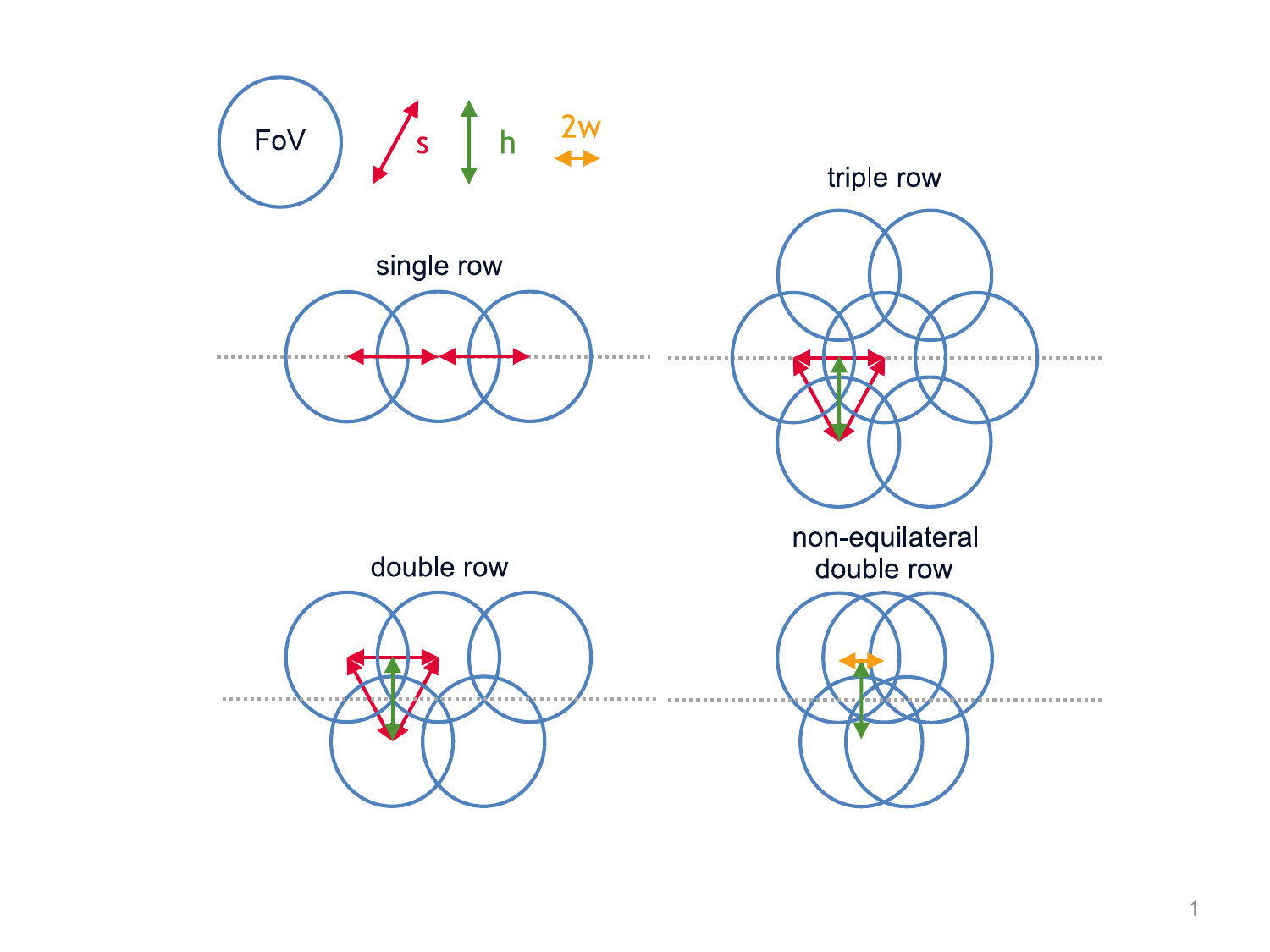}
  \caption{Schematic view of the pointing patterns considered. Dashed lines represent the Galactic equator. Circles represent the CTA field of view (FoV) for an individual pointing. Red arrows show the pattern step ($s$), i.e., the distance between pointing directions of adjacent pointings for single row and equilateral patterns. Green arrows show the latitude spacing ($h$) for patterns with multiple rows. The orange arrow shows twice the longitude spacing ($w$) for the non-equilateral double row pattern. {\changetwo The figure is not to scale. Notably the actual CTA field of view, in relation to the chosen step size, is larger than shown in this schematic view.}\label{fig:pointpat}}
\end{figure}

%
The GPS pointing pattern has been discussed in \citet{2013APh....43..317D} and  in \BO{\cite{2019CTAscience}.}
Two pointing patterns were considered, either with pointing directions aligned along the Galactic equator (single row), or 
alternating above and below the equator following an equilateral triangle pattern (double row). 
Here we revisit the GPS pointing strategy by considering more pattern candidates \BO{(shown in figure~\ref{fig:pointpat})}, the IRFs {\it prod3b-v2}, and a lower energy threshold of 70~GeV (above which CTA observations {\change of a typical duration of 50~h become competitive with 10~years of \textit{Fermi}-LAT observations} for steady point-like sources).
Besides the aforementioned single-row and double-row patterns we consider two more possibilities:
\begin{itemize}
\item a triple-row pattern with pointing directions alternating on and above/below the Galactic equator, aimed at obtaining more exposure at higher latitudes;
\item a non-equilateral double-row pattern, with pointing directions alternating above and below the equator but with different spacing in latitude and longitude {\change (see figure~\ref{fig:pointpat} for an illustration)}; {\change this is motivated by the need to split the total observing times into short individual observations (with a typical duration of 30 minutes)  in order to sample the lightcurves of periodic sources (see below) and, at the same time, to allow to use a smaller spacing in longitude (which yields a more uniform coverage) while still preserving good coverage at higher latitudes.}
\end{itemize}

For consistency with  {\change \cite{2013APh....43..317D,2019CTAscience}}  we refer to the distance between adjacent pointing directions of a single-row or equilateral pointing pattern as the pattern step $s$ (see figure~\ref{fig:pointpat}). As we will discuss later, the latitude spacing $h$ is one of the most critical parameters. For easier comparison with the other patterns we define the step of the non-equilateral double row pattern as $s = \sqrt{4/3} \, h$, i.e., the step of an equilateral double row pattern with the same latitude spacing $h$. To characterise the non-equilateral double row pattern we also introduce a parameter $w$ that corresponds to the longitude distance between adjacent pointings in alternate rows.

We calculate exposure and effective (PSF) maps for all the patterns under consideration and for energy thresholds of 70~GeV, 125~GeV, and 1~TeV, using the IRFs described above\footnote{Results are unchanged in terms of optimal pattern if we use IRFs optimised for 5~hours of observations, while the latter yield slightly worse sensitivities overall.} and for observation zenith angles of $20\degr$ and $40\degr$. 
\BO{The calculations described here have been performed using the {\it ctexpcube} and {\it ctpsfcube} tools from {\it ctools}.}
For this step and the rest of the pattern optimisation procedure the time per pointing is always adjusted to result in a fixed time spent per unit longitude.

{\change Steps $\lesssim 3\degr$ yield exposure fluctuations as a function of longitude $< 7\%$ for latitudes $|b| < 2\degr$ and, for multiple-row patterns, a spill-over of exposure $<10\%$ at high latitudes $|b| > 5\degr$ (where the number density of Galactic sources and the intensity of interstellar Galactic emission are very small).
The instrument PSF varies as a function of photon energy and inclination with respect to the pointing direction. The effective PSF, calculated for each position in the sky by taking into account the pointing pattern, is improved for smaller steps, but the effect in terms of variations of the 68\% containment angle is $< 5\%$ for the patterns considered here with $s \lesssim 5\degr$.}

Then we investigate the sensitivity to an isolated point-like source for all the configurations described above. We calculated the sensitivity using {\it cssens} {\changethree (from  {\it ctools})} assuming 6.5~hours of time spent per degree in longitude (foreseen for the inner Galaxy) and $w = 0.1\degr$ as a representative value of longitude spacing for the non-equilateral double-row pattern\footnote{This value corresponds to the approximate spacing to be used for the inner Galaxy region based on the observing time allocated for 30-minute observations.}.

\begin{figure}
\centering
	\includegraphics[width=.67\textwidth]{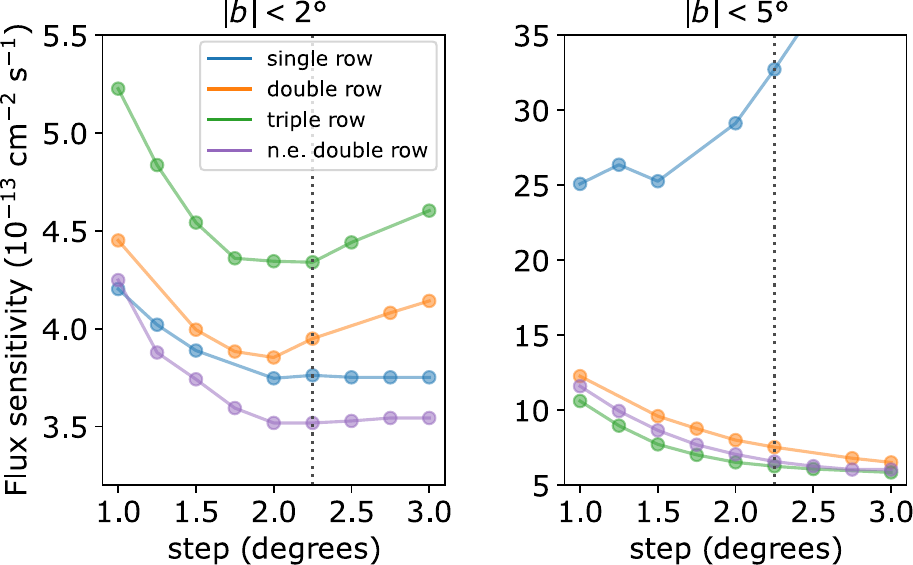}
  \caption{{\changetwo Integral} flux sensitivity to an isolated point-like source at energies \mbox{$E> 125$~GeV} averaged over two latitude bands \mbox{ $|b| < 2 \degr$} and \mbox{$|b| < 5\degr$} {\change for observations with the southern array at 40\degr zenith angle}. We assume an observing time of 6.5~hours per degree in longitude. The dashed line highlights the step chosen for the non-equilateral double row pattern used in the rest of the paper.}
  \label{fig:pointsens}
\end{figure}

In figure~\ref{fig:pointsens} we show the sensitivity at energies $>  125$~GeV {\change for observations with the southern array at 40\degr zenith angle} for all the pointing patterns considered in two latitude bands: $|b| < 2 \degr$ and $|b| < 5\degr$.
The non-equilateral double-row pattern yields the best sensitivity in the Galactic plane, with a broad minimum for steps $s$ between $2\degr$ and $3\degr$ and the least dependence upon the exact choice of step. At higher latitudes, the single-row pattern yields much worse sensitivity than multiple-row schemes. Multiple-row patterns yield a sensitivity that improves as a function of increasing step size and tends to a plateau for $s > 2\degr$. The non-equilateral double-row pattern is only marginally worse than the triple-row scheme for the same step. Results are similar for the other energy ranges considered, {\change for other zenith angles, for the northern array}, and for sources with moderate {\changetwo angular} extensions ($<0.2\degr$). 

\begin{figure}
\centering
	\includegraphics[width=.8\textwidth]{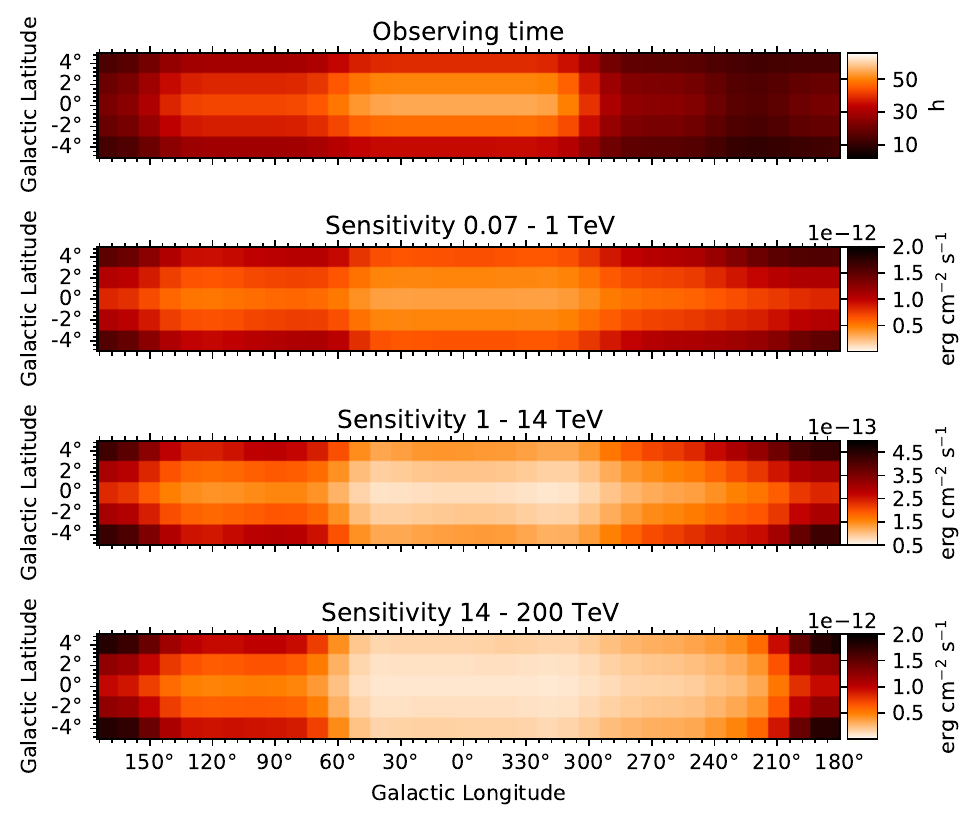}
  \caption{Top panel: observing time as a function of position in the sky, including all observations at an offset $<5\degr$ for a given direction. Other panels: sensitivity to a point-like source with power-law spectrum of index 2.34 in three energy bands. The sensitivity is defined as the minimum energy flux integrated over the energy range of interest to achieve a $5\sigma$ detection. The sensitivity estimates assume that the longitude range 210\degr-60\degr is observed using the southern array, the rest of the plane using the northern array. \label{fig:sensmap}}
\end{figure}

\BO{Based on these considerations, for the rest } of the paper we adopt the non-equilateral double-row pattern with a step $s = 2.25\degr$ ($h = 1.95\degr$). We split the total observing times per longitude range given in table~\ref{tab:hours} in individual pointings of 30 minutes each. The longitude spacing $w$ in each longitude range is set to the ratio of the longitude span over the total number of pointings in that range within each year of the programme (see the next section on scheduling and appendix~\ref{app:scheduling}).
\LT{Figure~\ref{fig:sensmap} shows the observing time and resulting sensitivity for a point-like source as a function of position in the Galactic plane and in three broad energy ranges. Including observations at an offset $<5\degr$ for a given direction, the total observing time {\changetwo at any given position} on the Galactic equator varies from 15~h to 60~h as a function of longitude, with an average offset with respect to the field of view centre of 2.5\degr.} We note that the differences between the two arrays in the number of telescopes and their optimisation for different energy ranges are clearly visible when comparing the observing time to the achieved sensitivity as a function of longitude. Regions observed from the southern hemisphere {\change (210\degr-60\degr)} benefit from better performance, especially at the highest energies.

\subsection{Observation scheduling}

The scheduling of the observations affects the detectability of variability in source fluxes. We implement a realistic scheduling strategy that follows the overall plan outlined in \BO{\cite{2019CTAscience}. }
Specifically, the simulated GPS observations were spread over ten years, with a short-term programme (STP) for
which 480 h of observing time were allocated over the first two years, and a long-term programme (LTP) for
which 1140 h of observing time were allocated over the following eight years (table~\ref{tab:hours}).
The simulated observations were scheduled from 1 January 2021 in dark conditions only, requiring that the Sun is at least
$15^\circ$ below the horizon and that the Moon is also below the horizon.
Observations were scheduled so that a given pointing is observed as close as possible to its
minimum zenith angle. {\change Other factors affecting the observability such as weather and hardware failures were not taken into account for simplicity.}

In addition, observations were distributed over the year so that a given location of the sky
is revisited at different time intervals, enabling the detection of periodic flux variations that
cover periodicities between a few days and up to a few years.
In this way, the periods of all known, and most expected, gamma-ray binaries should be
accessible using the GPS data. Practically this was achieved by tentatively distributing the observations as a function of time, and then altering the time distribution algorithm until we reached an exposure as a function of time difference between pointings at all longitudes as uniform as possible and without any gaps in the interval between 5 days  and about 1 year. A detailed description of the algorithm is provided in appendix~\ref{app:scheduling}.

\subsection{Observation simulation}

For each pointing, {\it ctobssim} {\changethree(from {\it ctools})}
 was used to generate mock event lists, based on the sky model described in section~\ref{sec:skymodel} and the IRFs (taking into account the instrument energy dispersion).
To cope with the limited number of zenith angles for which IRFs were available, pointings with a zenith angle smaller than $30\degr$ were then assigned the $20\degr$ IRF, pointings with a zenith angle between $30\degr$ and $50\degr$ were assigned the $40\degr$ IRF, and other pointings were assigned the $60\degr$ IRF. Only a few pointings were finally scheduled which required the $60\degr$ IRF. {\change Most of the results presented below, unless stated otherwise, are based on a single realisation of the simulated event lists.
\section{Source catalogues} \label{sec:catalogs}

{\changethree The source catalogues are planned to be one of the major products delivered by CTAO to the community. In the following we present the developments done to prepare a catalogue production workflow and discuss its potential outcome in term of detectable sources from the CTA GPS survey.}

\subsection{Analysis outline}

{\changethree In figure \ref{fig:survey} we show the excess counts above the true \bkg for the full GPS survey at latitudes $|b|<6^\circ$ and for energies between 0.07 and 200~TeV. Using the simulated data and the true IRFs we build catalogues of sources in the entire Galactic plane in this energy range.}

The first step of the catalogue production is to build, in a short amount of computational time, a list of candidate objects from the structures found in the excess or significance maps. The candidate objects are then fitted with different models in order to determine the best-fit model and its optimal parameters. The fitted candidates are filtered such that, for each object, the test statistic $\mathrm{TS} = 2 \, \Delta \ln(L) > 25$ with $\Delta \ln(L)$ \MC{being} the difference in log-Likelihood between the best-fit model including the source and the model for the null hypothesis (no source). {\changethree An additional step would be} needed to determine the optimal threshold in order to minimise both false positive and false negatives rates. Furthermore, the threshold should be set in terms of significance, since the statistical meaning of the TS value depends on the number of degrees of freedom associated with the source model. However, these aspects are not critical for the results we present in the following {\changethree as we can compare them to the true sky model}. 

The catalogue production was performed independently using two different pipelines. In the following we will refer to the output catalogues as A and B, respectively. The main differences between the output catalogues result from the analysis strategies used for the initial object detection (different algorithms), and the model fitting refinement (order and type of models tested). Catalogue A is based on the work presented in \cite{2018sf2a.conf..205C} \QR{and summarised in appendix \ref{app:catalog_A}}.  Catalogue B partly relies on the techniques and ideas discussed in \cite{2020APh...12202462R}, but the complete work-flow is detailed in appendix \ref{app:catalog_B}. In the next section we will comment on how the differences between the two strategies affect the results.

\begin{figure*}
    \centering
    \includegraphics[width=\textwidth]{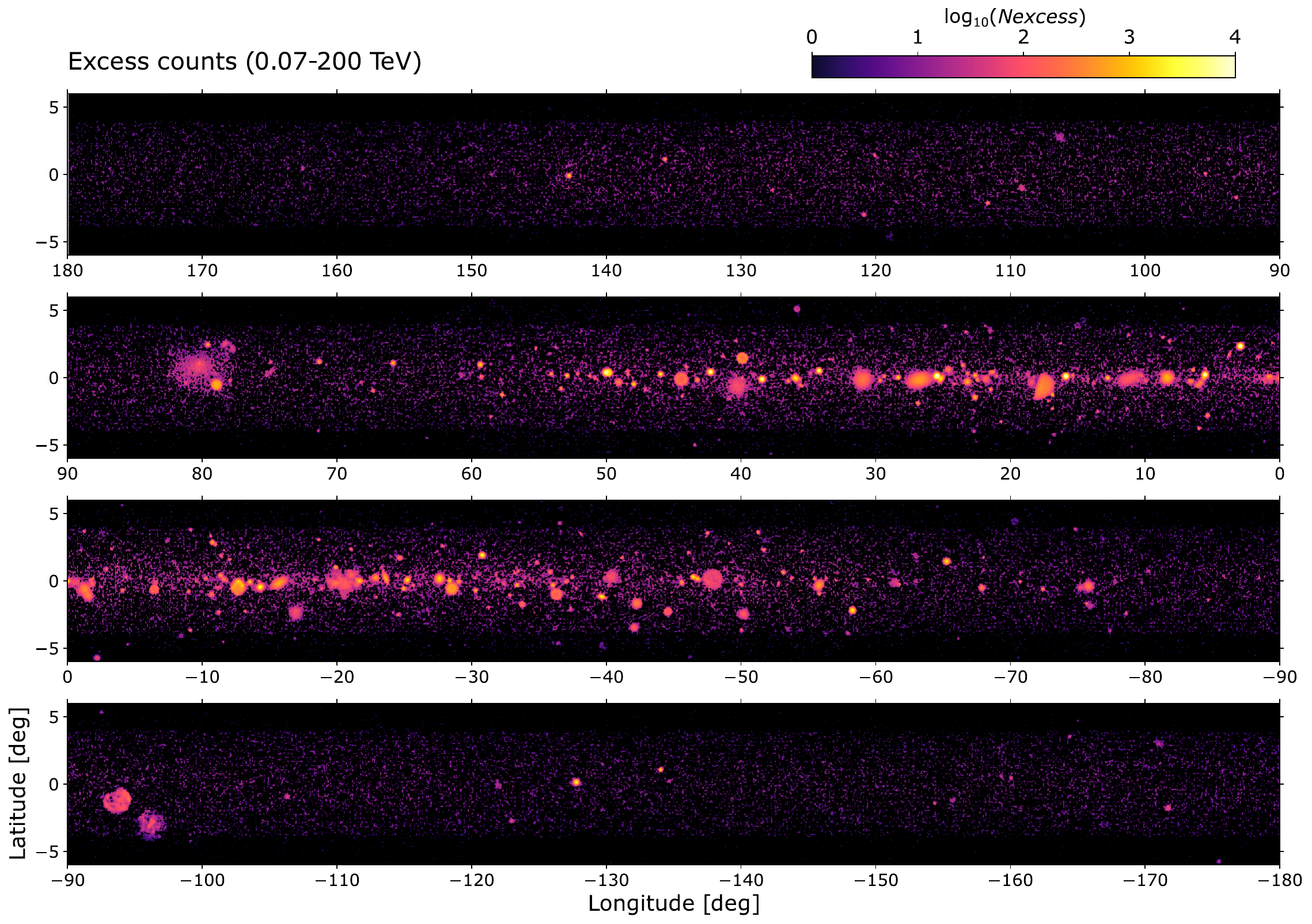}
    \caption{Excess counts above the \bkg in the 0.07-200 TeV energy range from the entire CTA GPS. The spatial bin\MCout{s} width is $0.06^\circ$ and a Gaussian smoothing with  $\sigma=0.03^\circ$  is applied.}
    \label{fig:survey}
\end{figure*}

\begin{figure*}
    \includegraphics[width=\textwidth]{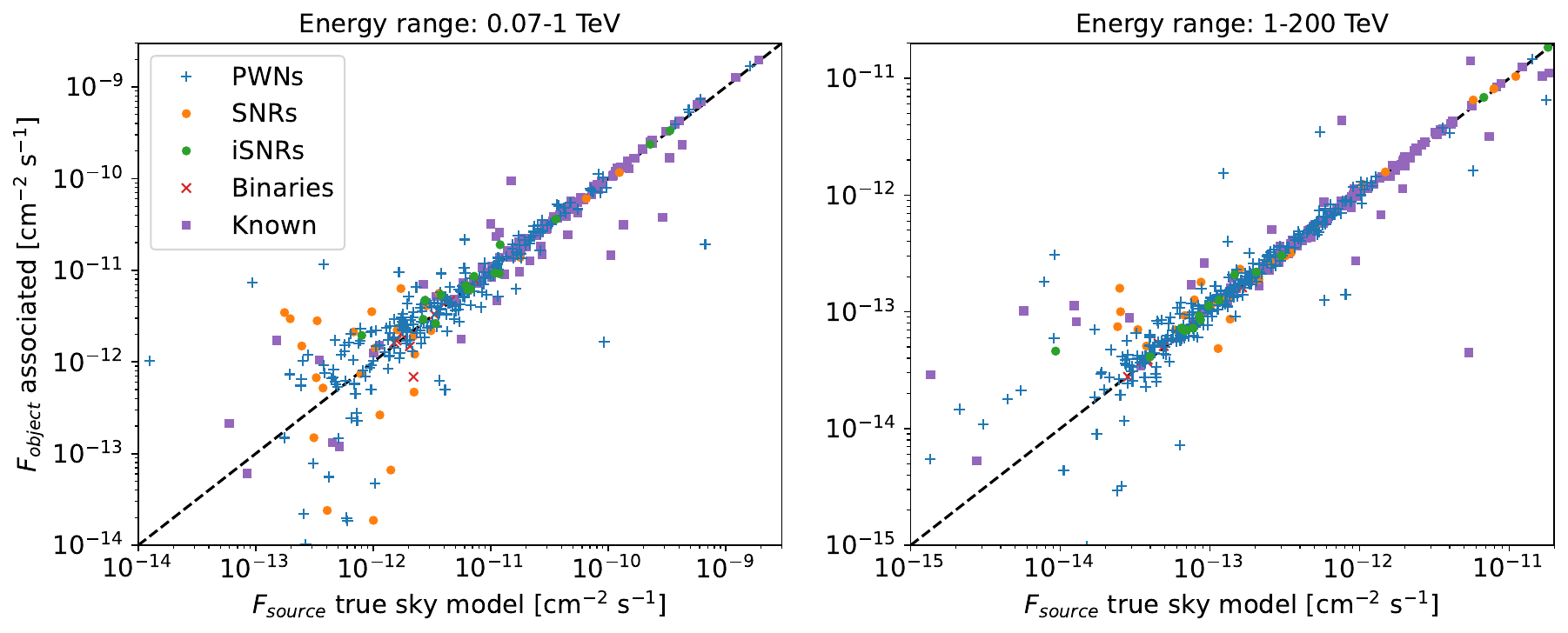}
    \caption{Integral flux of the sources {\change in the true sky model} versus integral flux of the associated objects from catalogue B in two energy bands 0.07-1 TeV (left) and 1-200 TeV (right). The dashed black line correspond\MC{s} to a one-to-one match. The ``Known'' label corresponds to a compilation of sources detected by the current generation telescopes at GeV or TeV energies (see section \ref{sec:known-sources})}
    \label{fig:flux_sources}
\end{figure*}    

\renewcommand{\arraystretch}{1.0}
\setlength{\tabcolsep}{0.1cm}
\begin{table*}
\centering
\begin{tabular}{c| c c c c c c c | c c c}
\hline
Name &  \small{PWN }& \small{SNR} & \small{iSNR} & \small{Binaries} & \small{Known} & {\change \small{Unmatched}} & \small{Total} & $\Delta F/F$ &$f_{\rm match}$ & $f_{\rm reco}$\\
\hline
True detectable  & 294 & 37 & 24 & 10 & 134 & - & 499 & - & - & -\\                                                         
Catalogue A  & 241 & 16 & 20 & 10 & 111 & 169 & 567  & -12.5\% &0.70 & 0.80\\
Catalogue B  & 257 & 31 & 14 & 10 & 122 & 36 & 470  & 3.8\%  &0.92 & 0.87\\                      
\hline
\end{tabular}
\caption{{\change Number of detectable sources (TS > 25 for the true sky model{\changethree, labelled True detectable}) and detected objects (TS > 25 in the catalogue) in the 0.07-200 TeV energy range. Unmatched} corresponds to detected objects without direct match with any simulated sources, $\Delta F/F$ is the relative error on the total flux of the detectable sources, $f_{\rm match}$ is the fraction of the detected objects matching a true source, and $f_{\rm reco}$ is the fraction of the true detectable sources matching a detected object.{\changetwo The first columns correspond to the synthetic populations and the ``Known'' column to sources already detected by the current generation telescopes (at GeV or TeV energies). Further details on the content of the true sky model are given in section \ref{sec:skymodel} and appendix \ref{app:skymodel}.} }
\label{tab:detections}
\end{table*}

\begin{figure}
\centering
    \includegraphics[width=0.505\columnwidth]{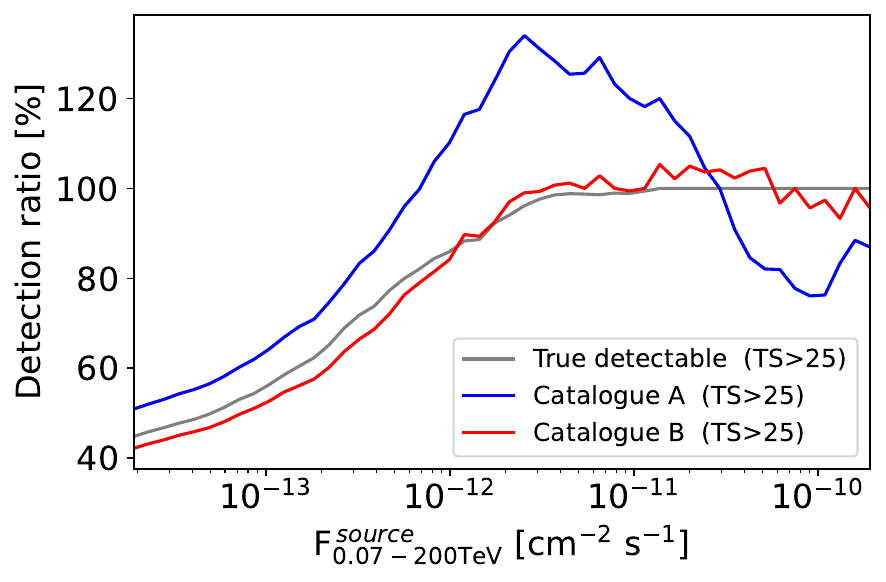}
        \includegraphics[width=0.48\columnwidth]{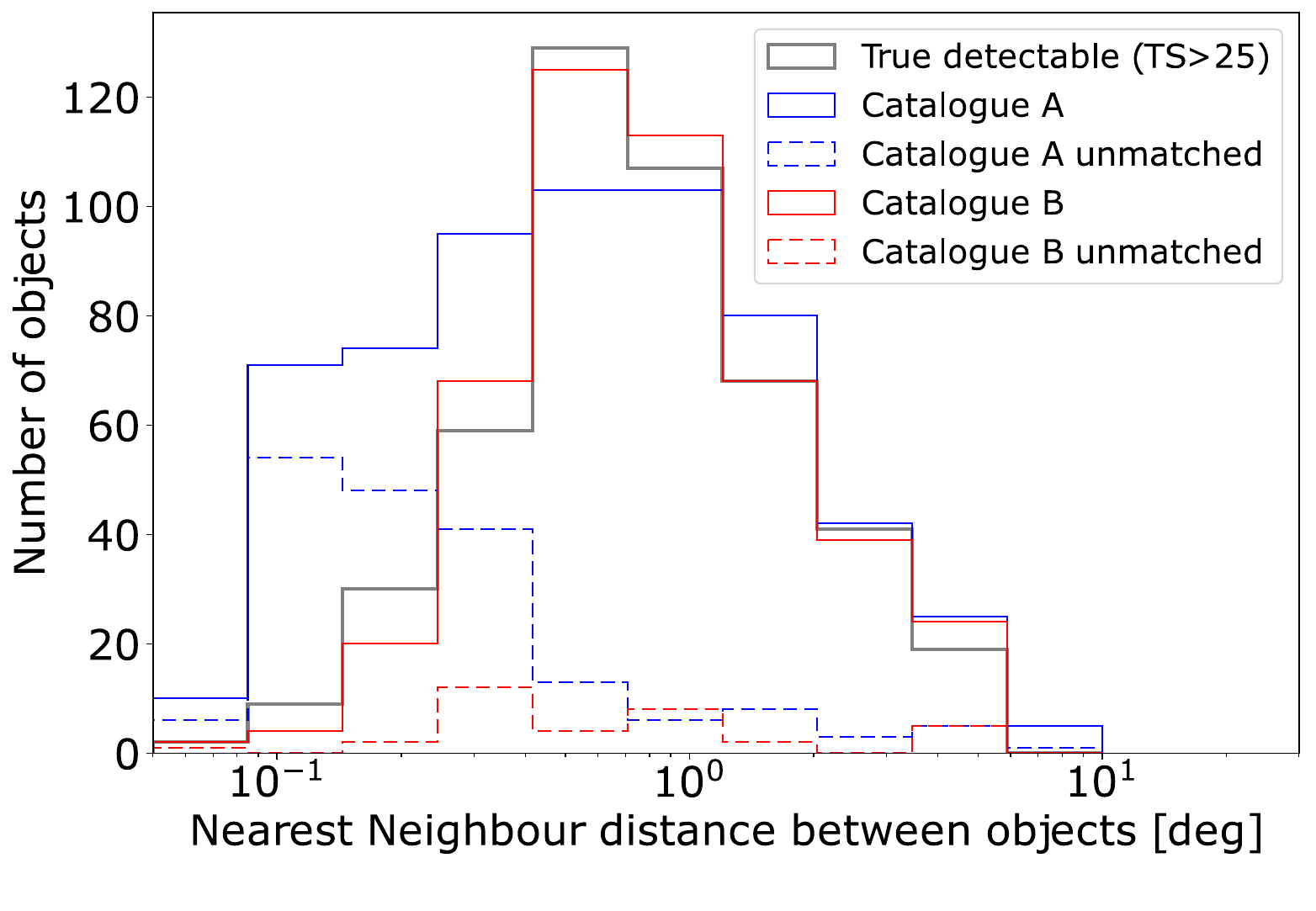}
\caption{Left panel: Detection ratio {\changethree of sources }above a given integrated flux (for \mbox{E = 0.07-200 TeV}). The grey curve corresponds to the expected detections with \mbox{TS > 25} for the true sky model, the coloured curves correspond to the detections reported in the catalogues at the same threshold. Note that the detection ratio of a catalogue can exceed 100\% because of the confusion bias or modelling bias (see details in the text). Right panel : Histogram of the nearest neighbour distance between detected objects.}
    \label{fig:det_frac}
\end{figure}

\subsection{Diagnostics and results}\label{sec:diag}

In order to match the detected objects in the catalogue with the sources in the true sky model we test for spatial coincidence using two criteria.
For each object we first select true sources within a given angular separation to the object centre defined as \mbox{$d_{\rm inter-center}<d_{min}+f_R \times R_{\rm object}$}, where  $R_{\rm object}$ is the 68\% containment radius of the fitted model convolved by the PSF.
We set $d_{min}=0.1^{\circ}$ which corresponds to about twice the 68\% containment radius of the PSF at 1 TeV and $f_R = 30\%$ which is close to the dispersion in relative error on the radius guess from the first step of the catalogue production.
After this first selection we report the association maximising the surface overlap fraction, defined as the ratio of the intersection to the union of the candidate object and true source surfaces:
\mbox{$\rm{SF_{\rm overlap}}= \left( S_{\rm object\,  \cap  \, source} \right) / \left( S_{\rm object \, \cup \, \rm source} \right)$},
where the surfaces are delimited by the iso-contours at 68\% containment in flux of the model convolved by the PSF.
We choose to report only the association {that maximises the surface overlap fraction} in order to limit the possible associations for extended objects. We also enforce that each detected object can be associated with only one true source and vice-versa. Moreover, only the associations with \mbox{$\rm SF_{ overlap}>0.25$} are reported in order to limit spurious associations. We checked on a sample of mock catalogues that this threshold maximizes the balanced accuracy defined as the average of true positive and true negative rates for associations.

In order to test the overall quality of the catalogue produced we introduce the matching fraction,  \mbox{$f_{\rm match}=N_{\rm match}/N_{\rm objects \, detected}$}, defined as the fraction of the detected objects matching a true source;  and the reconstruction fraction, \mbox{$f_{\rm reco}=N_{\rm match}/N_{\rm sources \, detectable}$}, defined as the fraction of the true detectable  sources that match a detected object. {A source} is defined as detectable if it has a $\mathrm{TS} > 25$ {\change for the true sky model} ({the same threshold as for the detected objects in the catalogue)}. $f_{\rm match}$ and $f_{\rm reco}$ are indicators of purity and completeness, respectively.

Table \ref{tab:detections} gives the number of sources simulated that are detectable for the different synthetic source populations and for the known sources. Based on the matching criterion previously defined\BO{,} we also report the potential detections from the catalogues associated to the same populations. These results show that we may detect up to {\change 500  
sources} in the 0.07-200 TeV energy range from the CTA GPS which is more than 6 times the number of objects in the HESS-GPS \citep{2018A&A...612A...1H} or the 3HWC \citep{2020ApJ...905...76A} catalogues.

Overall, the relative error on the {\changetwo integral}  flux of the detectable sources is about 4$\%$ (for catalogue B); 
{\changethree the good agreement in flux of those objects spatially associated with true sources can be seen in figure \ref {fig:flux_sources}.}
The larger deviation for the fainter sources can be explained by source confusion\MC{,} in particular in the composite systems where the PWN and/or the SNR are not significant enough to be individually detected. Outliers are expected for the known sources as some of these are simulated with complex templates or multiple models, rather than with a single parametric model, which make\MC{s} the one-to-one associations more ambiguous. In a few cases, sources simulated with multiple components of similar spatial extent but with different spectra can be associated with the wrong fitted component, as the association criterion is purely spatial.
In the most common case, one complex source is fitted with multiple simpler models. This can be seen as a modelling bias (further discussed below). {\change Finally, some of these deviations may be peculiar to the individual realisation of the simulated dataset considered.}

We also define the detection ratio as the number of objects detected with $\mathrm{TS} > 25$
{\changethree above a given integrated flux, divided by the total number of simulated sources above the same flux (figure \ref{fig:det_frac}, left).}
The detection ratio of a catalogue can exceed 100\% because of confusion or modelling biases. In the case of the confusion bias, the emission from the unresolved sources biases  the flux of the sources near the threshold upwards and so enhances their detection probability. In the case of the modelling bias, we see that sources simulated with more complex models (shell, elliptical Gaussian, or template) than those considered in the catalogue construction (Gaussian, disc, and point-like) can be fragmented into multiple smaller objects of lower flux. 
Another indicator of this effect is the increase of unmatched object detections with a small distance to their nearest neighbour, as shown in the right panel of figure \ref{fig:det_frac} {\changethree (with reference to catalogue A)}.

The fragmentation of complex sources into multiple sub-structures explains most of the discrepancy observed for catalogue A in figure \ref{fig:det_frac}. It also explains the large number of objects detected but the low matching fraction, as shown in \mbox{table \ref{tab:detections}}. At this stage, the filtering and merging of the sub-structures is more a classification than a statistical problem. Thus, the solution introduced in the production of catalogue B to solve this issue is to use pattern recognition techniques: (i) to determine \textit{a priori} the best-suited morphological model between a shell or a generalized Gaussian (see appendix \ref{app:catalog_B5}); (ii) to identify  \textit{a posteriori} the groups of objects that could be merged together or replaced by a different model (see appendix \ref{app:catalog_B6}).
The use of a larger variety of spatial models and the identification of multiple objects as a single entity allows a better description of the complex sources in the true sky model. This leads to a {\changethree better} estimate of the number of source detections and their individual fluxes which explains the closer agreement of the detection ratio in catalogue B seen in figure~\ref{fig:det_frac}.
The results of catalogue B in terms of detection ratio and matching fraction show that we can produce a catalogue close to the limit of detections expected from the true sky model.

 For the following sections one of the two catalogues, A or B, was conveniently chosen. Differences between the quality of the catalogues are such that they do not affect the main conclusions discussed.

\section{Population studies: PWNe and SNRs}
\label{sec:population}

\begin{figure*}
    \centering
    \includegraphics[width=0.55\textwidth, bb= 0 0 420 320, clip]{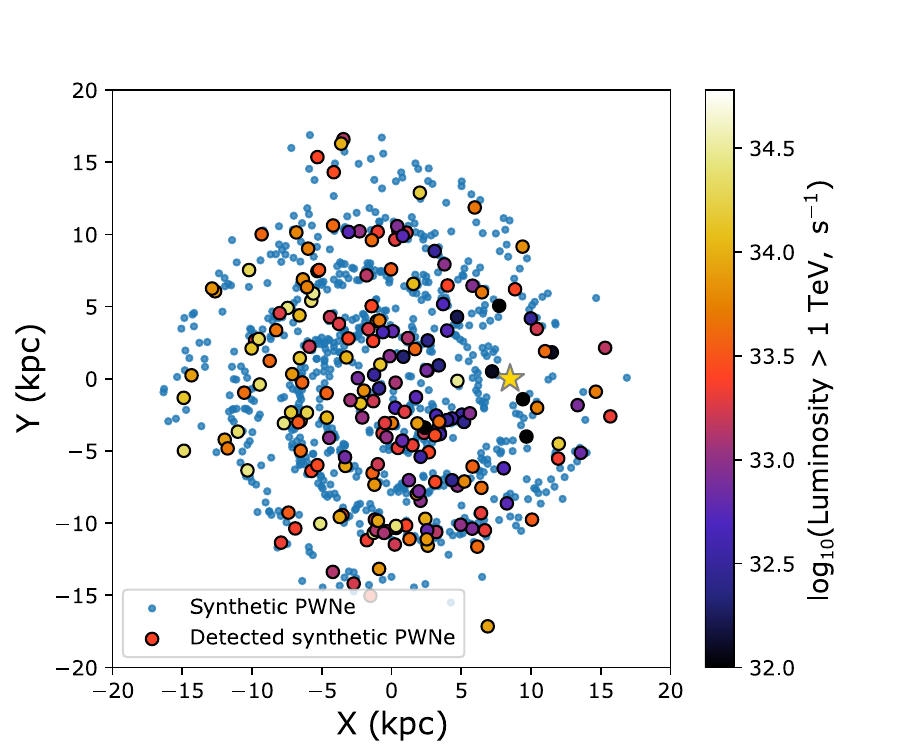}
    \includegraphics[width=0.44\textwidth, bb= 5 0 410 398,clip]{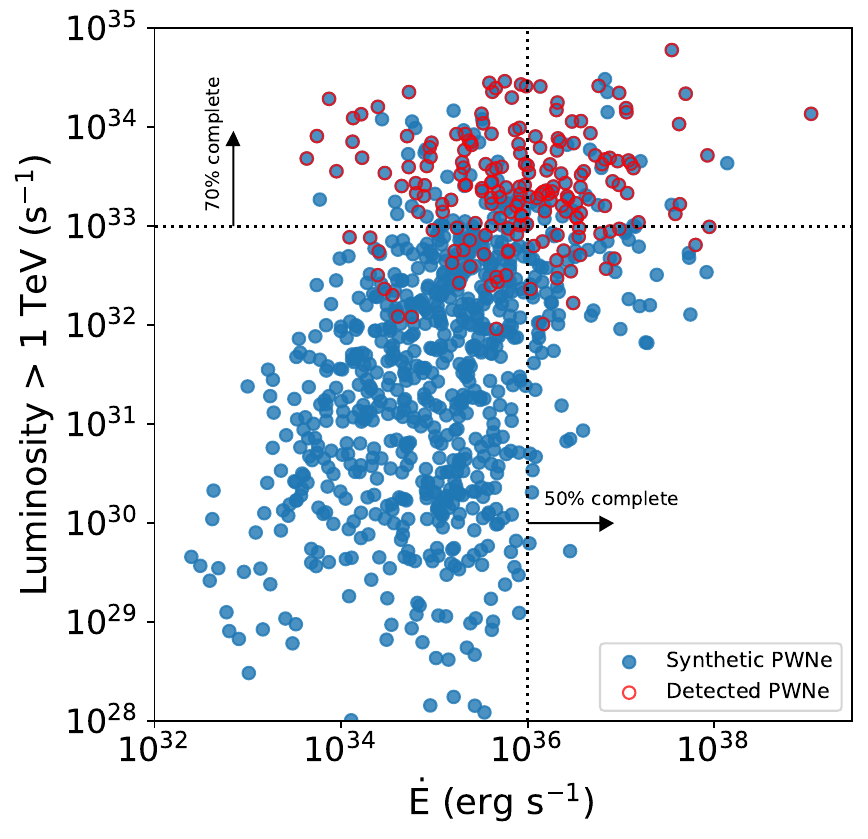}  \\
      \includegraphics[width=0.42\textwidth, bb= 0 0 335 328,clip]{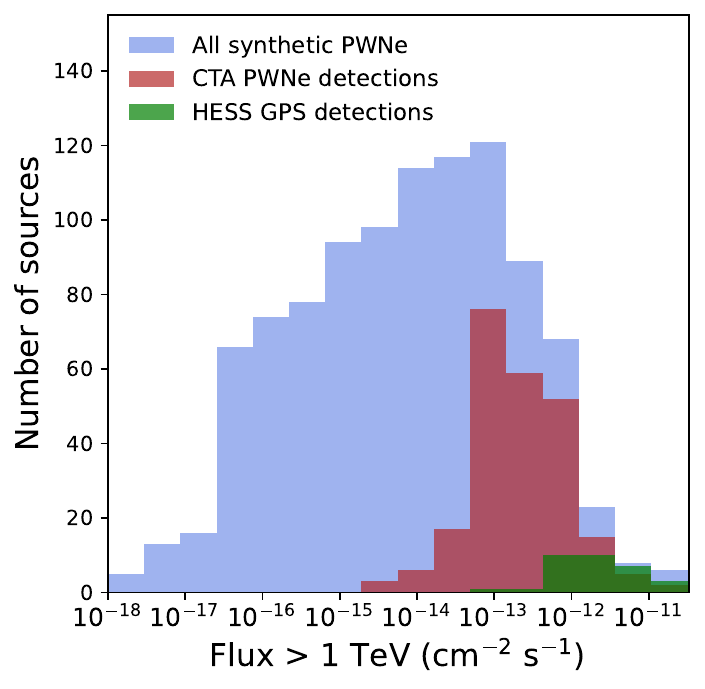}  

    \caption{Properties of the synthetic PWN population and corresponding detections. 
     Left panel, upper row: Galactic distribution of the synthetic (shown with blue circles) and detected objects (circles with colour coding in terms of luminosity). The Sun is represented with a yellow star at a distance of 8.5 kpc from the Galactic centre.
       Note that the lack of detections in the solar neighbourhood is due to nearby synthetic sources having similar properties to known sources being removed from the population (see section~\ref{sec:skymodel}).
        Right panel, upper row: Distribution of the synthetic and detected PWNe in the luminosity-$\dot{E}$ space. The dotted lines indicate that the survey could detect 50\% of PWN with $\dot{E} > 10^{36}$ erg s$^{-1}$ and 70\% of sources with luminosity $> 10^{33}$ s$^{-1}$.
        Bottom row:  integral flux histogram of the CTA GPS detections compared to the entire synthetic PWN population and the HESS GPS PWN catalogue \citep{2018A&A...612A...2H}. }
    \label{fig:pop}
\end{figure*}

In this section we will discuss the properties of the two dominant source classes detected in the simulated survey: PWNe and SNRs.  We caution the reader that the absolute detection numbers depend on the individual realisations of the population models and the CTA configuration considered. However, this will not affect the main conclusions.

PWNe are the dominant source class at TeV energies in the Galactic plane. About 250 synthetic PWN detections in addition to the known ones are included in the catalogues described above and a brief overview of their properties is discussed below. We will focus on the population study of the synthetic sources, as opposed to real sources observed with current instruments, in order to have a uniform sample for which all physical parameters are known ($\dot{E}$, age, Galactic coordinates, distance, etc.).

Figure \ref{fig:pop} (left panel, upper row) shows the spatial distribution of  the entire synthetic PWN population generated, along with the  objects detected in catalogue B and associated with a synthetic PWN. The CTA GPS sensitivity makes it possible to detect a large number of objects even at the far edge of our Galaxy. As expected, a selection bias is observed in the Galactic distribution of detected sources as a function of heliocentric distance. Only the most luminous objects are detected at the farthest edges of the Galaxy.

About one third of the detected synthetic PWNe are found to be spatially extended at a statistical significance greater than 3$\sigma$ which is an important feature to help identify the source category (i.e. likely PWN or SNR origin).
The completeness of this survey is illustrated in figure \ref{fig:pop} (right panel, upper row) in terms of the intrinsic properties of the sources.
We can see that this survey is able to detect half of all the PWNe in the Galaxy currently powered by an energetic pulsar with spin-down luminosity $\dot{E}> 10^{36}$ erg s$^{-1}$.
Concerning detection limits, 95\% of the sources we have detected have an  $\dot{E} > 2.2 \times10^{34}$ erg s$^{-1}$ and a luminosity >  $2.7 \times10^{32}$ s$^{-1}$.
Note that the $\dot{E}$ value used here is the present $\dot{E}$ value (not the value at birth). This, combined with environmental factors, explains the scatter in the Luminosity-$\dot{E}$ correlation as PWNe at TeV energies act as calorimeters and reflect the integral of power injected into the system since birth.

The comparison in figure \ref{fig:pop}  (bottom row) of our current population of known PWNe \citep[HESS PWN catalogue:][]{2018A&A...612A...2H} with the members of our synthetic population detected in the simulation and the entire synthetic population emphasises the transformational jump in population size that CTA will bring to the field of PWN population studies (about 7 times the current sample, or 2.5 times if we consider that most of the unidentified sources are PWNe  as was done in the construction of the population model). In terms of sensitivity, 95\% of the HESS PWNe have an integral flux above 1 TeV $> 2.2\times10^{-13}$ cm$^{-2}$ s$^{-1}$ compared to  $> 2.4\times10^{-14}$ cm$^{-2}$ s$^{-1}$ for the CTA sample, an order of magnitude improvement. \changereftwo{For a flux $> 2.4 \times 10^{-14}$~cm$^{-2}$~s$^{-1}$ ,  the completeness factor is $\sim$55\%, meaning that CTA is expected to detect more than half of the PWN population in our Galaxy above that flux.}
 \changeref{We note a small discrepancy at the high-flux end in figure \ref{fig:pop} (bottom row) where the population of the brightest simulated PWNe is not entirely detected. This is mainly due to a bias in the catalogue analysis. Some extended PWNe can end up being split into multiple sources with smaller extension by the catalog pipeline, leading to a lower reconstructed flux. In addition, in composite SNRs cross talk between the SNR and the PWN can lead to an inaccurate reconstruction of the PWN flux.} 

The second-most numerous class detected in this survey is SNRs. Focusing only on the synthetic shells and interacting SNRs, 45 (31 and 14 respectively) sources were detected in catalogue B. This suggests that the CTA GPS may be able to increase, by a factor larger than two, the number of SNRs observed at TeV energies. About half of the synthetic SNRs detected are significantly spatially extended, which is a valuable feature for the identification of sources with multiwavelength counterparts.  Among the sources simulated as shells (young SNRs), 19 are effectively detected as shells.
The distribution of SNRs in integral true simulated flux and distance in figure \ref{fig:snr} shows that new sources can be detected out to the other side of our Galaxy (up to 20 kpc in this realisation). The flux sensitivity is 5-10 times better than the current TeV SNR sample, with sources being detected down to an integral flux of a few $ 10^{-14}$ cm$^{-2}$ s$^{-1}$ .
 The newly detected shell-type SNRs have an age range from 0-10 kyrs with a detection efficiency of 15-30\% in each age bin and an average of 3 SNRs per 1 kyr age bin. This wider distribution of ages than current samples will pave the way to a more detailed population study.

This is a major step forward which allows to explore the population of Galactic SNRs by discovering and measuring the spatial extension of new SNRs. This is possible for CTA even with the relatively limited observing times (< 50~h at an average offset of $2.5\degr$) provided by the GPS. Conversely, the current population of SNRs was established through dedicated deep ($>100$~h) observations for the faintest objects.

\begin{figure}
    \centering
    \includegraphics[width=0.7\columnwidth]{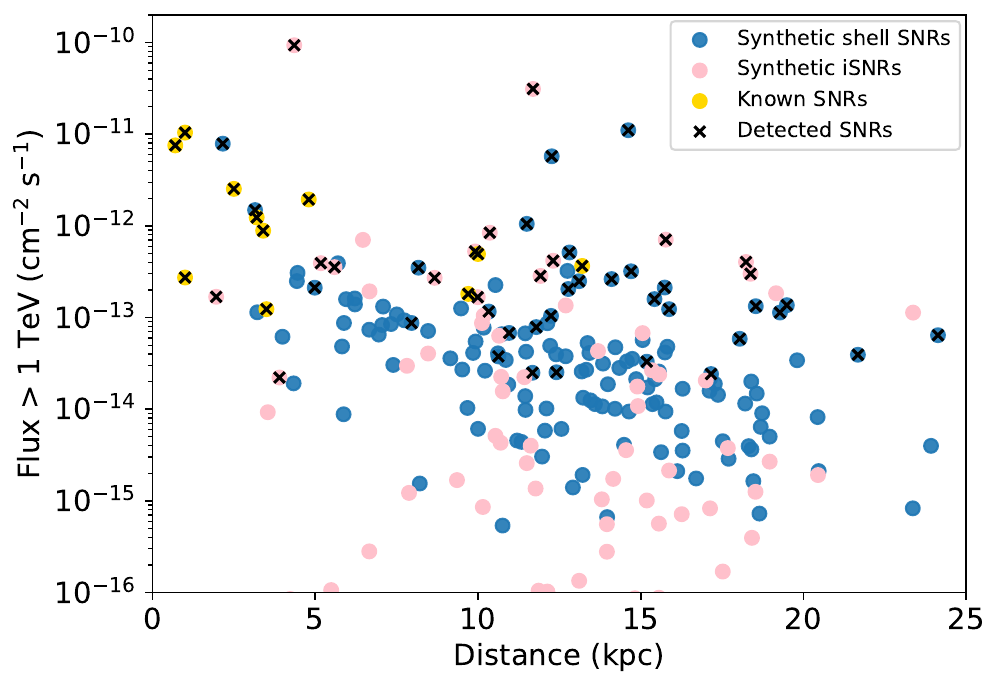}
    \caption{Comparison of the known SNRs and the population of synthetic shell and interacting SNRs in the integral true simulated flux-distance parameter space. New objects can be detected up to a distance of 20 kpc and down to an integral flux of a few 10$^{-14}$ cm$^{-2}$ s$^{-1}$.  Note that for a given flux, the detectability of a SNR depends on its extension.}
    \label{fig:snr}
\end{figure}

\section{Dedicated analyses of other source classes}
\label{sec:dedicatedsources}

\subsection{Gamma-ray binaries}

\subsubsection{\BO{Known sources}}
\label{sect:knownbin}
Six known gamma-ray binaries and one candidate gamma-ray binary (see table~\ref{tab:grlb}) were included in the simulation. 
{\change For all of these we derived a phase-folded lightcurve. }
In the phase-resolved analysis we included all sources detected in catalogue A within 3.5\degr from the source of interest. The flux and spectral parameters were left free to vary only for the source of interest and for nearby sources (within 1$^\circ$), while they {\changetwo were} frozen to the catalogue A values for more distant ones.

\begin{figure}
\centering
\includegraphics[width=0.7\columnwidth]{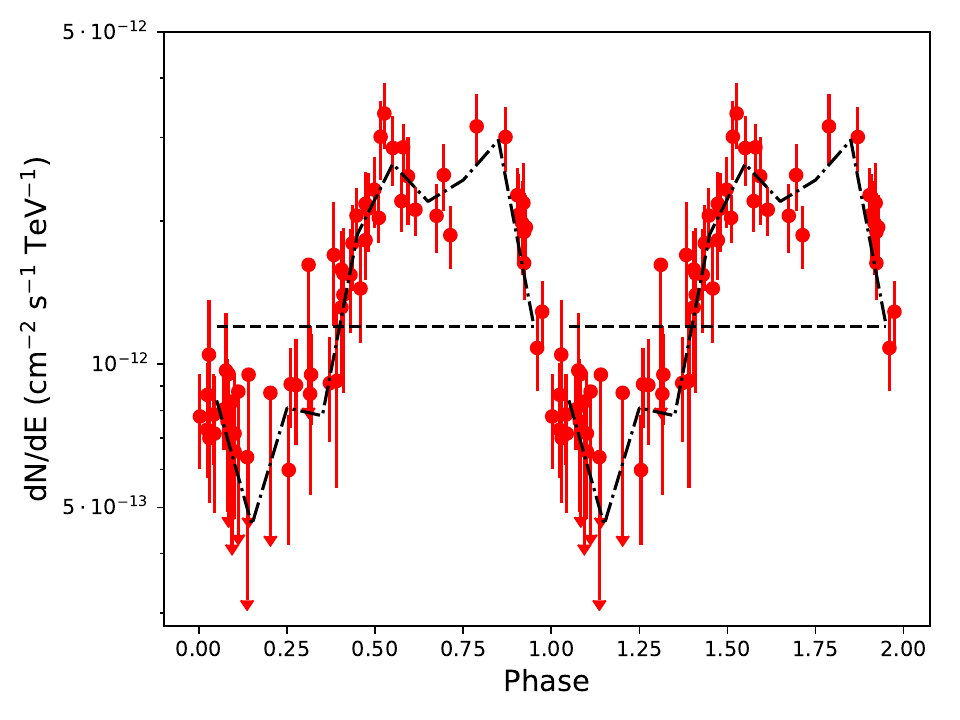}
\caption{Phase-folded lightcurve above 100 GeV of  LS 5039 on a 30-minute time scale (one point per individual observation) as reconstructed from simulated data. The flux normalisation dN/dE is given at 1 TeV. The dash-dotted black line shows the simulated profile and the black dashed line shows the mean {\changetwo differential}  flux. See~\ref{sec:app-othermodels} for a description of the input to the simulation.}
\label{fig:ls3039x}
\end{figure}

All gamma-ray binaries included in the simulations are bright and are clearly detected by CTA in a 30-minute observation during the orbital phases with {\changetwo differential}  fluxes at 100 GeV higher than $\sim$2.5$\times 10^{-13}$ cm$^{-2}$ s$^{-1}$ TeV$^{-1}$ {\change (the exact flux threshold depends on the spectral index of the source)}. The simulated orbital and spectral variability is clearly detected using the GPS observations (see e.g. the phase-folded lightcurve of LS 5039 in figure \ref{fig:ls3039x}).
{\change In phase bins corresponding to 5 -- 10 hours of observations}, spectral parameters of the binaries were reconstructed to better than 10\% level.

\subsubsection{Blind search for variable sources}
{\change Catalogue A contains 73 sources, that either were classified by the automatic pipeline as point sources, or associated during the subsequent cross-correlation with point-like sources in the true sky model.}
For each of these we created a lightcurve using events with energies between 0.1 and 100 TeV from the observations pointed within $3^\circ$ of the source. 
For this purpose we fixed the spectral shape of each source to the catalogue results. The flux was left free for the source of interest and nearby sources (located within $1^\circ$), and was fixed to the catalogue value for more distant sources. 
A simple $\chi^2$ test was used to analyse the lightcurves and, if the probability to have a constant flux was less than 5\% (without accounting for trials), then the source was classified as being variable. The lightcurves were generated for temporal intervals corresponding to one individual observation (30 minutes), one week, and, for sources simulated with periods longer than a week, one month and one year (see  figure \ref{fig:bin095} for an example).

\begin{figure*}
\centering
\includegraphics[width=0.495\textwidth, bb= 0 0 420 330, clip]{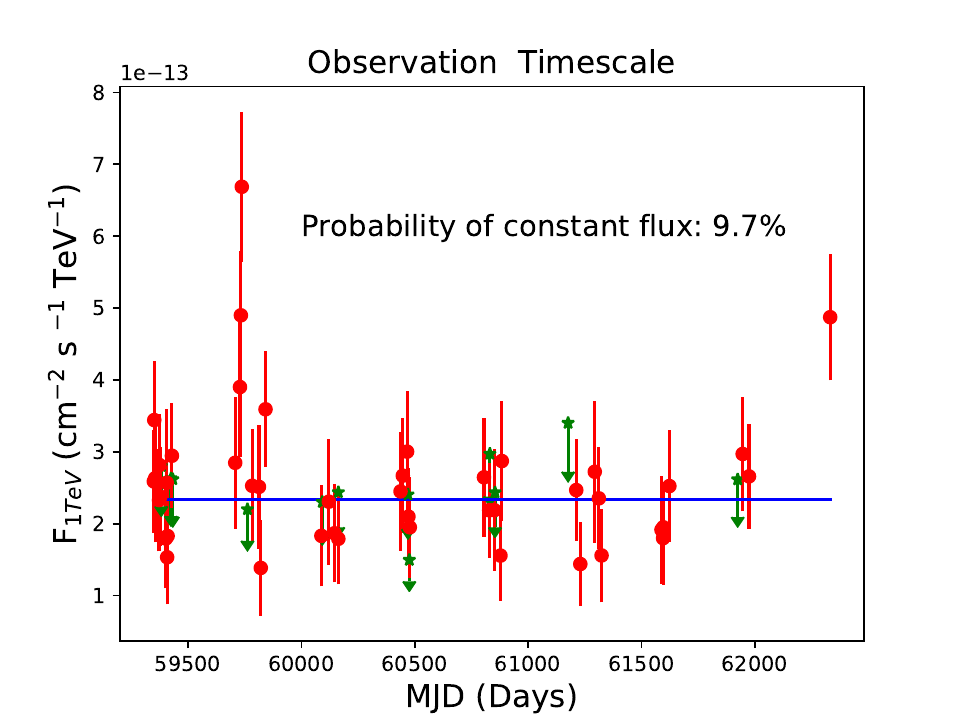}
\includegraphics[width=0.495\textwidth, bb= 0 0 420 330, clip]{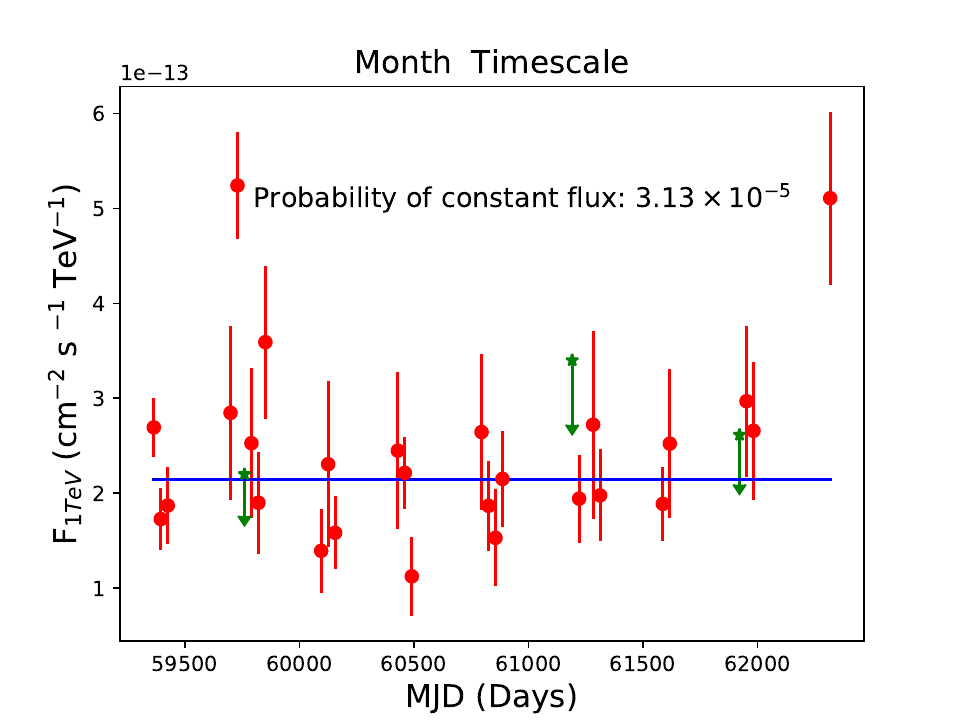}
\caption{Variability of the {\changetwo differential} flux at 1 TeV of source bin095 on timescales of a single observation (30 minutes, left) and a month (right).
Red dots show $>5\sigma$ detections and green stars {\changetwo show} 95\% c.l. upper limits when a significant detection was not achieved. The blue line shows the average flux.}
\label{fig:bin095}
\end{figure*}

Table \ref{tab:varsrc} shows the results for { \change sixteen} sources simulated as variable on a timescale longer than 30 minutes. {\change Twelve} of them were classified as variable in the analysis, according to the criterion described above. 
On the other hand, four of them were not detected as variable, either because their mean flux is too low or because they have a flat orbital flux profile. 

Table \ref{tab:varsrc} also includes two sources that showed a probability of constant flux below 5\%  in spite of being simulated as constant. In the case of 3FHL J1855.3$+$0751 we verified that the false variability detection is related to the imperfect modelling of a bright extended shell SNR nearby that affects differently the flux determination depending on epoch/pointing direction. By selecting events only within 3\degr of the source for the analysis we obtain probabilities to have a constant flux $> 5\%$. In the case of the synthetic source pwn2059, the variability detection is determined by a single {\change 30-minute observation for the individual realisation of the simulated dataset considered}. A detailed investigation based on 10000 realisations of the observation simulations showed that the low-flux point that drives the variability detection stems from a combination of downward statistical fluctuations in the number of events from the source and the background at the source position. The said fluctuations consist of variations $< 10$~events, therefore the simple $\chi^2$ test based on Gaussian statistics employed here may not be fully appropriate in the context of blind searches for variable sources with CTA and further developments are needed.

Nevertheless, this study shows that the GPS can be used to find variable sources on different time scales, so that follow-up observations of individual sources to study in detail their properties can be performed.

\begin{table*}
\centering
 \setlength{\tabcolsep}{2pt}
\begin{scriptsize}
\begin{tabular}{ccccccccc}
 \hline
 Source Name& $\sigma_{obs}$ & F$_{mean}$ &P$_{\mathrm{Obs}}$ &P$_{\mathrm{Day}}$ &P$_{\mathrm{Week}}$ &P$_{\mathrm{Month}}$ &P$_{\mathrm{Year}}$ &Period \\
 &  &  & & & & & & d \\ 
 \hline
\textbf{ LS I +61$^\circ$ 303}  &15.6   & \ 0.97$\pm$0.03&    \textbf{2$\times$10$^{-295}$}&
\textbf{8$\times$10$^{-302}$}&\textbf{5$\times$10$^{-273}$}&-&-&26.5\\

\textbf{ PSR B1259-63}&  11.1 &  \ 0.33$\pm$0.01 &\textbf{3$\times$10$^{-251}$}&\textbf{4x10$^{-121}$}&\textbf{9$\times$10$^{-128}$}&\textbf{10$^{-288}$}&\textbf{10$^{-165}$}&1241\\

\textbf{bin040}& 5.3 &   0.06$\pm$0.01 &6.5$\times$10$^{-1}$&6.3$\times$10$^{-1}$&1.6$\times$10$^{-1}$&
\textbf{5$\times$10$^{-3}$}&\textbf{5.7$\times$10$^{-6}$}& 3358\\

\textbf{ 
1FGL J1018.6-5856} &4.3 &  \ 0.41$\pm$0.02 &  \textbf{9$\times$10$^{-22}$}&  \textbf{5$\times$10$^{-25}$}&  \textbf{7$\times$10$^{-29}$}& -& -& 16.6\\

 \textbf{LS5039}& 4.0 & \ 4.15$\pm$0.03&\textbf{  10$^{-37}$}& \textbf{10$^{-37}$}&-&-&-&3.9\\

\textbf{bin095}&   3.9 &  \ 0.23$\pm$0.01&9.6$\times$10$^{-2}$&\textbf{4.9$\times$10$^{-2}$}&\textbf{1.8$\times$10$^{-3}$}&\textbf{3.13$\times$10$^{-5}$}&-&200\\ 

\textbf{HESS J1832-093}&  3.5 & \ 2.28$\pm$0.02&\textbf{ 6.6 $\times$10$^{-7}$}& \textbf{1.8$\times$10$^{-6}$}& \textbf{1.5$\times$10$^{-5}$}& \textbf{1.2$\times$10$^{-4}$}&-&365\\

\textbf{PSR J2032+4127}&  3.4 & \ 3.95$\pm$0.05&\textbf{ 4.1 $\times$10$^{-26}$}& \textbf{1.2$\times$10$^{-26}$}& \textbf{5.1$\times$10$^{-27}$}& \textbf{8.23$\times$10$^{-29}$}&
\textbf{5.36$\times$10$^{-32}$}&
1.8$\times 10^4$\\

\textbf{HESS J0632+057} &3.0 & \ 0.30$\pm$0.06&\textbf{ 10$^{-8}$}& \textbf{10$^{-8}$}& \textbf{7$\times$10$^{-9}$}& \textbf{6$\times$10$^{-10}$}&-&315\\

\textbf{bin074}& 2.8 &  \ 0.05$\pm$0.01 &9.9$\times$10$^{-1}$&9.9$\times$10$^{-1}$&9.3$\times$10$^{-1}$&\textbf{6$\times$10$^{-3}$}&4.9$\times$10$^{-1}$&840\\
 
\textbf{bin159}& 2.6 & \ 0.11$\pm$0.01 &1.6$\times$10$^{-1}$&\textbf{1.6$\times$10$^{-2}$}&\textbf{4.7$\times$10$^{-2}$}&-&-&5.2\\

\textbf{bin123}& 2.3 &  \ 0.09$\pm$0.01 &3.9$\times$10$^{-1}$&5$\times$10$^{-1}$&3.7$\times$10$^{-1}$&\textbf{1.8$\times$10$^{-2}$}&\textbf{2$\times$10$^{-4}$}&522\\

bin162& 1.9 & \  0.10$\pm$0.01 &9.9$\times$10$^{-1}$&9.9x10$^{-1}$&9.3$\times$10$^{-1}$&6$\times$10$^{-2}$&4.9$\times$10$^{-1}$&1387\\

bin154& 1.9 &  \ 0.05$\pm$0.01 &9.9$\times$10$^{-1}$&9.7$\times$10$^{-1}$&8.2$\times$10$^{-1}$&6.1$\times$10$^{-1}$&-& 35.8\\

bin093& 1.4 &  \ 0.06$\pm$0.01&9.9$\times$10$^{-1}$&9.9$\times$10$^{-1}$&9.9$\times$10$^{-1}$&-&-&7\\

bin146& 1.0 &  \ 0.07$\pm$0.01 &9.9$\times$10$^{-1}$ &9.9$\times$10$^{-1}$&-&-&-&1.5\\

\hline
\textbf{pwn2059}& 2.5& \  2.30$\pm$0.17&\textbf{2.3$\times$10$^{-5}$}&\textbf{4.8$\times$10$^{-3}$}&\textbf{2.8$\times$10$^{-5}$}&-&-&N/A\\
\textbf{3FHL J1855.3+0751}& 1.1&  \ 1.55$\pm$0.37  &9.9$\times$10$^{-1}$&\textbf{10$^{-5}$}&\textbf{2.7$\times$10$^{-4}$}&\textbf{1.1$\times$10$^{-3}$}&-&N/A\\
 \hline
\end{tabular}
\end{scriptsize}
 \caption{
 List of variable sources. Sources simulated as variable are given in the upper section, while sources simulated as constant but detected as variable are given in the bottom section. Sources detected as variable and the probabilities $P$ for the corresponding time {\change binning} are marked in bold.
 {\change F$_{mean}$ is the mean {\changetwo differential} flux at 1 TeV  in units of 10$^{-12}$ cm$^{-2}$ s$^{-1}$ TeV$^{-1}$.
  $\sigma_{obs}=(F_{max}-F_{mean})/\sqrt{(\Delta F_{max}^2 + \Delta F_{mean}^2 )}$ is the significance of the maximum flux deviation on  the 30 minutes time scale.}
  }
\label{tab:varsrc}
\end{table*}

\subsection{Pulsars}

We analysed data for all known \textit{Fermi}-LAT pulsars included in the simulations. To this end we selected observations in which the pulsar has a maximum offset of 5$^\circ$ from the centre of the field of view. No relativistic effects or delay in the generation of event time stamps were considered in the simulations. Therefore, no barycentric corrections are included in the analysis chain.

{\changetwo As a first step, we searched for pulsed emission at the known positions of the pulsars. Events were extracted from a region of 5$^\circ$ around the pulsar and weighted by the probability to originate from the pulsar position. This probability was estimated by using the PSF model of the instrument as a function of angular distance to the source and energy of each event.
The significance when testing for a periodic signal based on the known ephemerides was estimated using the weighted H-test \cite{1989A&A...221..180D, 2011ApJ...732...38K, 2013ApJS..208...17A}.

Among the 25 Crab-like pulsars, pulsations from PSR~J1833$-$1034 and PSR J1838$-$0537 were detected at significance levels (H-test values) of 5.1$\sigma$ (37) and 7.2$\sigma$ (70.3), respectively.}
We note, however, that the quoted significance levels are subject to caution due to a possible contribution of the PWN in the tens of GeV range, owing to the use of phase-averaged spectra to model the pulsed emission in the true sky model (see appendix~\ref{app:skymodelpsr}).
Some of the brightest \textit{Fermi}-LAT pulsars are not detected due to the limitation in latitude coverage of the CTA GPS  (see figure~\ref{fig:sensmap}). Indeed, the latter provides scarce or no observing time/sensitivity  at the positions of the Crab and Geminga pulsars, PSR~J0007$+$7303, PSR~J1057$-$5226, and PSR~J1514$-$4946
(for instance, the closest pointing in the simulated GPS lies at 4.8\degr from the Crab pulsar{\changetwo , located at a Galactic latitude of $-5.78\degr$). These objects are however expected to be well-studied through dedicated observations independent of the GPS.}

{\changetwo The Vela pulsar, as well as several Vela-like pulsars covered with at least few hours of observing time by the GPS were significantly detected.}
\BO{Figure \ref{fig:VelaPSR} shows an example of the reconstructed phase profile for the Vela pulsar for both GeV and TeV components, as compared to the input \textit{Fermi}-LAT phase profile.}  

\begin{figure}
        \centering
        \includegraphics[width=0.75\textwidth]{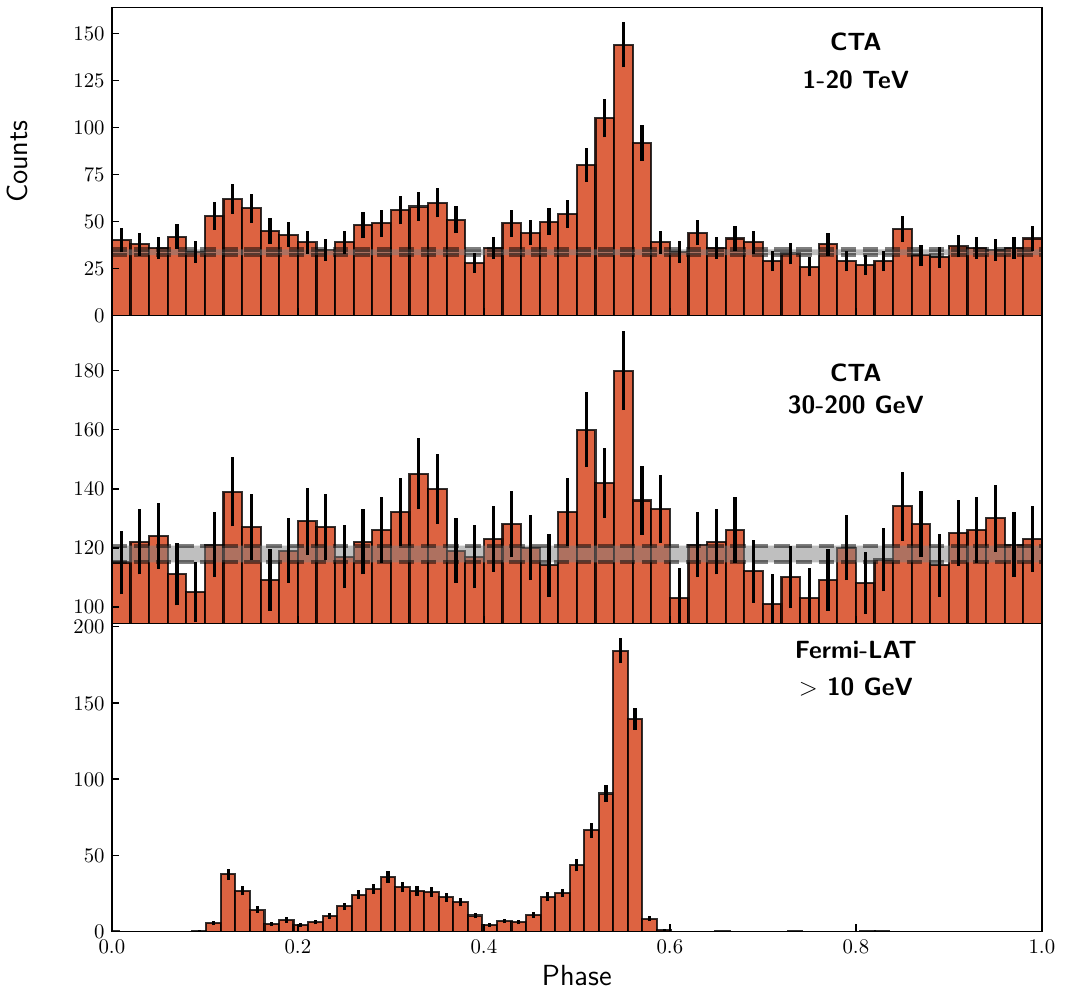} 
        \caption{Phasogram of the Vela pulsar from simulated GPS data for two energy ranges: 1-20 TeV (top panel, $0.1\degr$ around the pulsar position) and 30-200 GeV (middle panel, $0.25\degr$ around the pulsar position). 
The \textit{Fermi}-LAT profile above 10 GeV used as template in the simulations is shown for comparison in the bottom panel.
The grey shaded areas on the two upper panels show the estimated level of the background evaluated in the off-phase interval [0.7$-$1.0]}
        \label{fig:VelaPSR}
\end{figure}

{\changetwo In a second step, we performed a maximum likelihood fit to the data without using the timing information in order to
quantify the CTA sensitivity to phase-averaged emission of pulsars.}     
Events with energies from 30 GeV to 100 TeV were selected from a square region of $4\degr$ size
around each pulsar. The model spectrum was assumed to be either a simple power-law function (PWL) or
a power law with exponential cutoff (PLEC). The spectral parameters of nearby sources known from
other observations and/or present in Catalogue A were left free. 

{\changetwo For all Vela-like pulsars with significant periodic emission we also detected the phase-averaged TeV component (TS$\ge$25),
whereas the GeV component was bright enough to be detected only for the Vela pulsar.
Although the observing time for the latter pulsar is small (Galactic latitude of $-2.79\degr$), we were able to
reconstruct with fair precision its  spectrum.

None of the Crab-like emission tails are significantly detected in the phase-averaged analysis. For the two Crab-like pulsars with a pulsation detection we obtained a TS of 0 in the case of PSR~J1838$-$0537, and a TS of 1.7 in the case PSR~J1833$-$1034. The low TS values are due to source-confusion effects for these two sources surrounded by multiple overlapping brighter components, the modelling of which is imperfect in the catalogues. This illustrates the challenges in searching for faint sources in complex regions of the Galactic plane.
The results on the pulsation search and phase-averaged emission are summarised in table~\ref{tab:pulsars_analysis}.}

\begin{table*}
\centering
\begin{tabular}{lcccc}
\hline
Name & \multicolumn{2}{c}{Periodicity search} & \multicolumn{2}{c}{Phase-averaged analysis}   \\
 & H-test & significance & TS &  spectral model\\
\hline
\multicolumn{5}{c}{\textit{Crab-like pulsars}}	\\

PSR J1833$-$1034	& 37  &5$\sigma$ 	& 1.7 & PWL 	\\
PSR J1838$-$0537	& 70.3  &7.2$\sigma$ 	& 0 & PWL 	\\

\multicolumn{5}{c}{\textit{Vela pulsar}} \\
PSR J0835$-$4510 	& 984.9  &$\gg$8$\sigma$  &  301 & PWL+PLEC		\\
\multicolumn{5}{c}{\textit{Hypothetical Vela-like pulsars}} \\
PSR J1413$-$6205	& 855.2  &$\gg$8$\sigma$ 	& 863  & PLEC		\\
PSR J1709$-$4429	& 135.6  &$\gg$8$\sigma$ 	& 110  & PLEC 	        \\
PSR J1732$-$3131	& 258.7  &$\gg$8$\sigma$ 	& 60   & PLEC 	\\
PSR J1813$-$1246	& 74.8  &7.4$\sigma$ 		& 249  & PLEC         \\
PSR J1952+3252		& 578.4  &$\gg$8$\sigma$ 	& 109  & PLEC 	\\
PSR J2021+3651		& 435.8  &$\gg$8$\sigma$ 	& 139  & PLEC 	\\
PSR J2021+4026		& 55.6  &6.3$\sigma$     	& 406  & PLEC  	\\

\hline
\end{tabular}
\caption{Pulsars with evidence of emission from simulated GPS data. The significance levels predicted are subject to caution given possible contamination from the PWN emission in the true sky model.}
\label{tab:pulsars_analysis}
\end{table*}

{\changetwo The detection of pulsations from two Crab-like pulsars illustrates the potential of the 
GPS data to be used as a path-finder to probe and constrain the possible extension of some pulsars' GeV component into the VHE range,
and thereby to help optimising the follow-up observations with CTA. 
On the other hand, the clear detection of the TeV component in the Vela \citep{HessVela2020} and hypothetical Vela-like pulsars (with diverse assumptions on their emission spectra) shows
that if such a population existed, the GPS would be able to discover it. Thanks to
the high sensitivity of CTA GPS in the multi-TeV range, this could even be the case for pulsars which remain still
undetected in the GeV range.}

\subsection{PeVatrons}

We assess the potential of the GPS to find PeVatrons (or candidates) following the approach presented in \citet{PeVatrons_Consortium_paper}. For this analysis we have considered catalogue B from which we extracted a list of sources with TS > 25 in the 1-200 TeV energy range and with a number of predicted events from the fitted spectral model at energies higher than 50 TeV greater than 1. In this way we obtained a sample of 231 sources showing spectra extending up to $\sim$100~TeV energies. 
These sources have been subsequently re-analysed in the full energy range (0.07 to 200 TeV). The analysis setup is the same as for the catalogue generation, with morphological parameters of the source of interest and all parameters of nearby sources and the background models fixed to the catalogue values. The only free parameters of the fit are therefore the spectral parameters of the source of interest. 

To identify a source as a PeVatron we used the PeVatron Test Statistic \citep[$PTS$,][]{PeVatrons_Consortium_paper} which is a likelihood ratio test with the null hypothesis corresponding to a spectral model with proton energy cutoff equal to 1 PeV.
Following the methodology in \citealt{PeVatrons_Consortium_paper}, a source is identified as a PeVatron if one can exclude the null hypothesis with a statistical significance of $5\sigma$ which corresponds to a threshold value of the PeVatron Test Statistic (PTS) of 35.6, taking into account the {\changetwo trials} factor in the analysis of 231 sources \citep{wasserman2010statistics,PeVatrons_Consortium_paper}. 
If a source is not identified as a PeVatron (PTS < 35.6), it is considered as a PeVatron candidate when it presents a 95\% CL lower limit on the proton energy cutoff ($LL_p$) higher than 1 PeV.

We found 32 PeVatron candidates 
and we ranked them according to $LL_p$ (see table \ref{tab:pevatrons}). Most of these candidates are known sources that have been simulated with a power-law model (see  appendix \ref{sec:cutoffs}), with the exception of HESS J1813-178 which has been simulated with an exponential cutoff power-law model with a gamma-ray energy cutoff at 127 TeV. We identified {\changetwo three} PeVatron sources ($PTS$> 35.6, reported in bold in table \ref{tab:pevatrons}) which are among the brightest  sources in the list of the PeVatron candidates. 

\changeref{We note that the two known sources Westerlund~1 and HESS J1641$-$463 were simulated with a hard power-law spectrum but were not identified as PeVatron candidates in our analysis. Both sources are located in complex regions, therefore the results are affected by source confusion. Furthermore, Westerlund~1 was modelled in the catalogue analysis as multiple objects that are not associated to the complex template simulated according to the strict association criteria that we adopted. We conclude that a more sophisticated analysis method is needed to identify PeVatron candidates with complex morphologies and affected by source confusion. This is beyond the scope of this paper and left for future work.}

We note that our test to define a PeVatron candidate is based on the derived 95\% CL lower limit on the underlying proton cutoff energy. However, the test does not differentiate between hadronic and leptonic emission mechanisms. We notice, indeed, the presence of seven synthetic PWNe appearing as PeVatron candidates. The presence of leptonic sources as PeVatron candidates is not surprising since leptonic accelerators can have spectra extending to very high energies, as suggested also by recent LHAASO measurements \citep{2021Natur.594...33C,WilhelmideOna:2022zmp}. Therefore, it is clear that the PeVatron identification methodology discussed in \citet{PeVatrons_Consortium_paper} and used in our work needs to be expanded in order to properly distinguish between leptonic and hadronic PeVatrons. However, this is beyond the scope of this paper and left for future work.

Our results show that bright PeVatrons can be \changereftwo{detected} in the GPS dataset itself, while the survey of the entire Galactic disk provides an ideal pathfinder to pinpoint candidates to follow-up with deeper observations using CTA.

\begin{table*}
 \setlength{\tabcolsep}{1.2pt}
\centering
\begin{tabular}{ccccc}
 \hline
 Source Name& {\changetwo Differential} flux at 1 TeV  & Spectral Index  & $LL_p$ [TeV]& PTS \\
 \hline
   
  \textbf{HESS J1841$-$055}     	&	$1.364\pm 0.008 \times 10^{-11}$ 	&	$ 2.426 \pm 0.004 $	&	22931	&	119	\\
   HESS J1800-240C     	&	$6.20\pm 0.16 \times 10^{-13}$	&	$ 2.608 \pm 0.018$	&	9639	&	18	\\
   \textbf{2HWC J1837-065}     	&	$2.40 \pm 0.01 \times 10^{-11}$	&	$ 2.905 \pm 0.002$	&	7397	&	42	\\
   HESS J1708-410     	&	$5.83 \pm 0.05 \times 10^{-12}$	&	$2.605 \pm 0.005$	&	7079	&	31	\\
   2HWC J1902+048*     	&	$4.17 \pm 0.04 \times 10^{-12}$	&	$ 3.290 \pm 0.004 $	&	5592	&	15	\\
   HESS J1834-087     	&	$2.72 \pm 0.03 \times 10^{-12}$	&	$2.62 \pm 0.008 $	&	5423	&	23	\\
   \textbf{HESS J1614-518}    	&	$8.47 \pm 0.04 \times 10^{-12}$	&	$2.425 \pm 0.004$	&	3660	&	44	\\
pwn1772	&	$2.59 \pm 0.16 \times 10^{-13}$	&	$2.50 \pm 0.05 $	&	3123	&	2	\\
   SNR G323.7-1.0     	&	$2.81 \pm 0.04 \times 10^{-12}$	&	$2.491 \pm 0.011 $	&	3101	&	12	\\
pwn438	&	$1.39 \pm 0.11 \times 10^{-13}$	&	$1.93 \pm 0.07 $	&	3006	&	1	\\
   HESS J1018-589 A     	&	$4.19 \pm 0.19 \times 10^{-13}$	&	$ 2.13 \pm 0.03 $	&	2438	&	8	\\
pwn521	&	$5.92 \pm 0.15 \times 10^{-13}$	&	$ 2.733 \pm 0.019 $	&	2337	&	6	\\
   pwn813     	&	$1.78\pm 0.02 \times 10^{-12}$	&	$ 2.307 \pm 0.009$	&	2217	&	14	\\
   Westerlund 2     	&	$2.90\pm 0.06 \times 10^{-12}$	&	$2.631 \pm 0.015$	&	2064	&	5	\\
   2HWC J1914+117*     	&	$2.13 \pm 0.03 \times 10^{-12}$	&	$ 2.853 \pm 0.008 $	&	2007	&	7	\\
   pwn2733     	&	$1.97\pm 0.15 \times 10^{-13}$	&	$ 2.26 \pm 0.05 $	&	1994	&	5	\\
   HESS J1503-582     	&	$2.37\pm 0.06 \times 10^{-12}$	&	$ 2.399 \pm 0.019$	&	1974	&	6	\\
pwn2252	&	$1.09\pm 0.14 \times 10^{-13}$	&	$ 1.85 \pm 0.07$	&	1959	&	1	\\
HESS J1119-614	&	$1.60\pm 0.04 \times 10^{-12}$	&	$2.637 \pm 0.017 $	&	1930	&	5	\\
W 51C	&	$1.02\pm 0.02 \times 10^{-12}$	&	$ 2.606\pm 0.018$	&	1913	&	5	\\
   HESS J1844-030     	&	$4.49\pm 0.15 \times 10^{-13}$	&	$ 2.57 \pm 0.02 $	&	1900	&	5	\\
   2HWC J1819-150*     	&	$1.596 \pm 0.008 \times 10^{-11}$	&	$2.89 \pm 0.07$	&	1780	&	12	\\
   pwn2913     	&	$8.3 \pm 1.1 \times 10^{-13}$	&	$1.953 \pm 0.005$	&	1616	&	4	\\
   isnr99     	&	$9.92 \pm 0.06 \times 10^{-12}$	&	$2.412 \pm 0.005$	&	1494	&	13	\\
   2HWC J1852+013*     	&	$5.21\pm 0.05 \times 10^{-12}$	&	$2.906 \pm 0.005$	&	1364	&	5	\\
   HESS J1832-093     	&	$5.92 \pm 0.17 \times 10^{-13}$	&	$2.63 \pm 0.02$	&	1313	&	4	\\
   pwn2934     	&	$1.37\pm 0.13 \times 10^{-13}$	&	$2.11\pm 0.05$	&	1232	&	3	\\
   HESS J1804-216     	&	$6.09\pm 0.07 \times 10^{-12}$	&	$2.721 \pm 0.007$	&	1230	&	4	\\
   CTB 37B     	&	$6.42\pm 0.19 \times 10^{-13}$	&	$2.675 \pm 0.019$	&	1218	&	3	\\
   2HWC J1921+131     	&	$1.766\pm 0.015 \times 10^{-12}$	&	$2.743 \pm 0.009$	&	1186	&	4	\\
   HESS J1813-178     	&	$2.81\pm 0.02 \times 10^{-12}$	&	$2.133 \pm 0.008$	&	1049	&	4	\\
   HESS J1846-029     	&	$6.89\pm 0.15 \times 10^{-13}$	&	$2.356 \pm 0.017$	&	1018	&	3	\\

 \hline
\end{tabular}
\caption{ List of the best PeVatron candidates ($LL_p$ > 1 PeV). The {\changetwo differential} flux values are given in cm$^{-2}$ s$^{-1}$ TeV$^{-1}$. The spectral index is obtained from the gamma-ray \BO{PLEC} fit and is generally in very good agreement with the simulated one.
\changetwo{$LL_p$ is the lower limit on the cut-off energy of the proton spectrum. The PTS is the test statistic used to identify PeVatron candidates as explained in the text.}
{\changethree Bold text highlights the three identified PeVatron sources (with $PTS$> 35.6).
  }
}
\label{tab:pevatrons}
\end{table*}

\section{Diffuse emission}
\label{sec:diffuse}

The study of Galactic diffuse gamma-ray emission with IACTs is challenging as, over large scales, the signal is largely dominated by the CR background. Moreover, diffuse emission is comprised of a contribution from unresolved source populations and from interstellar emission from CR interactions in the Galaxy. In this section we will show that: 1) we can extract from the source catalogue an estimate of the contribution from unresolved sources based on minimal modelling assumptions, and 2) CTA will make it possible to detect TeV interstellar emission and to statistically distinguish between different scenarios.

\subsection{Unresolved sources}\label{sec:unreso}

Unresolved sources, i.e. sources that individually fall below the detection threshold of the measurement, make a cumulative contribution to the measurable diffuse emission signal.
For the determination of the amount of this contribution, a model of the entire gamma-ray population must first be developed based on the catalogue of detected sources.
In a second step, source populations can be divided into detectable and unresolved sources by applying the detection threshold of the measurement.
For the development of a source model, two different approaches can be followed. In the first approach, a single source class or several source classes are modelled based on the underlying physics and the model is verified by comparison with the detected sources (as was done in the construction of our sky model in section~\ref{sec:skymodel}). In the second, more data-driven approach, fitting based on minimal modelling assumptions of the detected sources results in a model for a generic gamma-ray source population. Here, we adopted the latter approach to describe the gamma-ray source population based on the catalogue B described in section~\ref{sec:catalogs}. This allows us to assess our capability to characterise emission from unresolved sources independently from our prior knowledge of the true sky model {\changetwo (described in section~\ref{sec:skymodel})}. Specifically, we adopt the simple source population model from \cite{SteppaEgberts2020}.

{\changetwo The detection threshold in flux for extended sources as function of sky position, $F_\mathrm{min} (l,b,\sigma_\mathrm{source})$, is derived by scaling the detection threshold in flux for point sources (figure~\ref{fig:sensmap}),  $F_{\mathrm{min},0}$, such as :}
\begin{equation}
F_\mathrm{min} (l,b,\sigma_\mathrm{source}) = F_{\mathrm{min},0} (l,b) \, \sqrt{\frac{\sigma_\mathrm{source}^2 + \sigma_\mathrm{min}^2}{\sigma_\mathrm{min}^2}} 
\end{equation}
for source sizes \LT{$\sigma_\mathrm{source} \leq 2^\circ$ in radius}. 
Above an angular extension of $2^\circ$ the source is assumed to be undetectable given the applied background subtraction technique.
The quantity $\sigma_\mathrm{min}$ represents the minimum resolvable source size due to the instrument resolution and characteristics of the observations. For this exercise it is fixed to a value of 0.052\degr, which is the smallest angular extension of any source found in catalogue B.

The free parameters of the source population were then determined by a likelihood fitting procedure (based on Poisson statistics) to the sources detected in catalogue B. For all sources in the true sky model it is assumed that the source distance is known and therefore the flux and angular extent can be converted into the source luminosity and radius, respectively.

Simulation of  gamma-ray observations of the source population so characterised makes it possible to apply the detection threshold so as to divide {\changethree the population}
into detectable sources and unresolved sources for {\changetwo the two energy ranges below and above 1 TeV considered in catalogue B (section~\ref{sec:catalogs})}. 
To estimate the flux of unresolved sources and its dispersion, we simulate $1000$ realisations of the source {\changethree population}. \changereftwo{The final maps of unresolved sources in the two energy ranges are obtained by taking a bin-wise average of the multiple realisation. An example map is shown in appendix~\ref{app:unresolved}.}
The fluxes and distributions of the detectable sources agree well with the sources from the catalogue. We show, as an example, the \changereftwo{longitude} {\changetwo distribution of the integral} flux $> 1$~TeV in figure~\ref{fig:unresolved}.
\begin{figure}
    \centering
    \includegraphics[width=0.7\columnwidth]{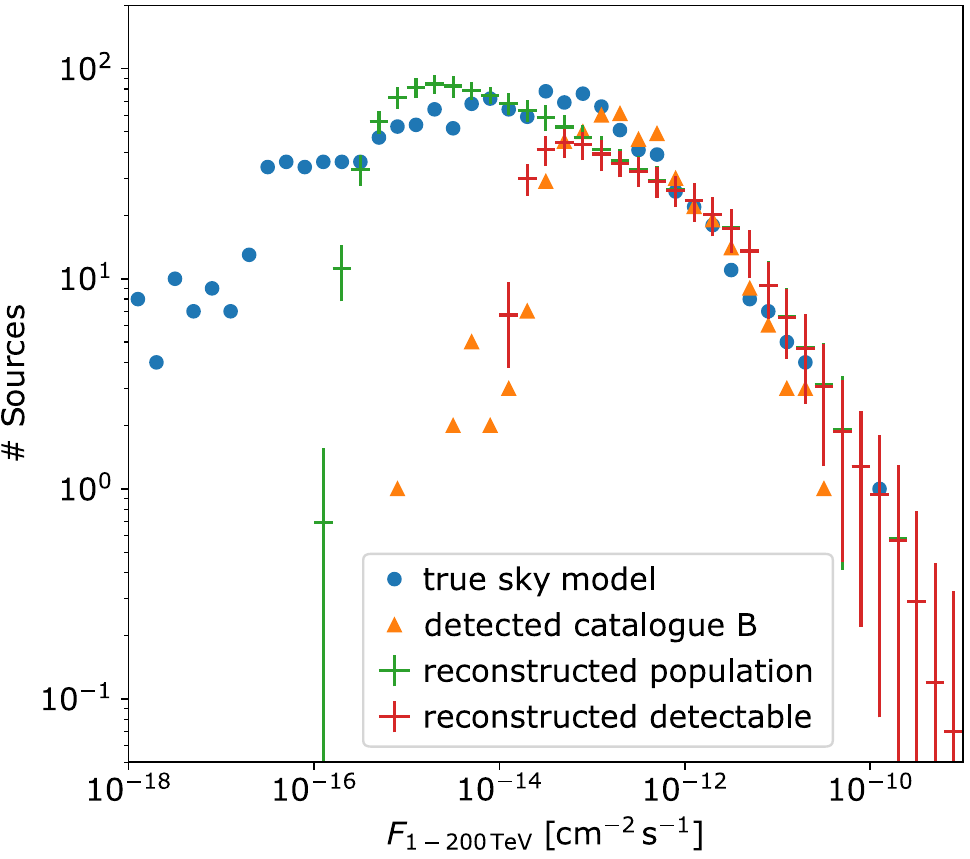}
    \caption{\LT{Number of sources as a function of {\changetwo integral} source flux in the 1-200 TeV energy range. We show the distributions for {\change the true sky model}, for detected sources (catalogue B), for the {\change population reconstructed} blindly from catalogue B, and the corresponding subsample of sources that should have been detected.} {\change The values shown for the reconstructed population consist of the mean from 1000 realisations with error bars spanning the 25\% to 75\% quantiles of the distribution.}}
    \label{fig:unresolved}
\end{figure}

In spite of the relatively simple assumptions, which are different from those used in the generation of the true sky model, one can see that the reconstructed population agrees well with the catalogue results, and that we obtain from the data a reasonable description of sub-threshold sources in agreement with the {\change true sky model}. We cannot reproduce all details of the complex true sky model, in particular in the range of very low fluxes ($<10^{-15}$~cm$^{-2}$~s$^{-1}$) and small source numbers, but this has a negligible impact on the properties of the collective emission from unresolved sources. {\change Conversely, we see that, for the realisation considered, catalogue B does not perfectly capture the flux distribution of the brightest sources in the true sky model due to a combination of a few bright sources lying at the edge of the survey region, notably the Crab Nebula, and because of fragmentation effects for sources with complex morphology discussed earlier. Nevertheless, the methodology compensates, at least partially, for these effects, and the reconstructed population matches the flux distribution of the true sky model at high fluxes.}

{\change The estimated contribution of unresolved sources is shown in figure \ref{fig:flux_IEMs}, together with the contribution of the true-source model,  the three alternative interstellar models, and the CR background. We can see that, for the models considered, the {\changetwo integral} flux of the unresolved sources is smaller by an order of magnitude than interstellar emission. 
Overall the level of the predicted flux matches that of the true sources below the catalogue threshold (TS<25). The  differences between {\change the true sky model} below the detection threshold and the unresolved source template extracted from the catalogue are expected as the mean template derived from multiple realisations cannot match exactly a single observed sky. These deviations are larger than statistical fluctuations on the number of detected photons. They will need to be taken into account when interpreting real observations, along with systematic uncertainties not dealt with in the present work.}

\begin{figure*}
\centering
    \includegraphics[width=\textwidth]{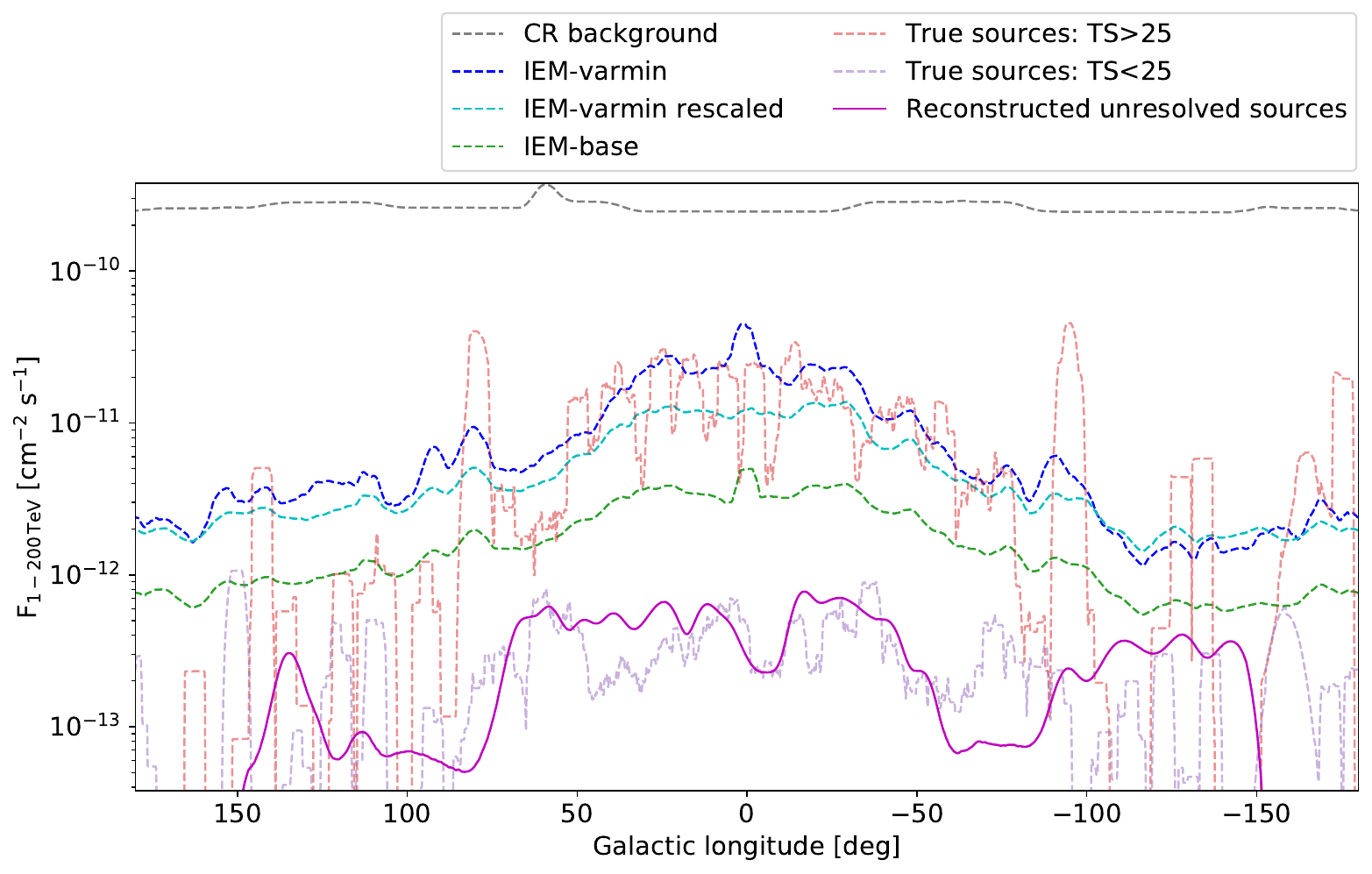}
    \caption{{\changetwo Integral} flux profile {\changetwo in the 1-200 TeV energy range as a function of} Galactic longitude for the CR background, for different variations of the interstellar emission model, for the true source models above and below the detection threshold of the catalogue, and for the unresolved sources template {\change reconstructed from} catalogue B (see Sect. \ref{sec:unreso}). Fluxes are integrated over latitudes of $\pm 6\degr$ and over a 6$^\circ$ sliding window in longitude.}
    \label{fig:flux_IEMs}
\end{figure*}

\subsection{Interstellar emission}

In this section we summarise our results on
interstellar emission.
{\changethree For the production of the catalogue the sources' parameters were fitted  to the data together with several adjustable parameters for {\change the true} CR background and IEM {\changetwo models}. Namely, the Galactic plane was split in overlapping regions, each spanning 10\degr in longitude and 12\degr in latitude, with their centres shifted in steps of $\pm5$\degr.} In each of these regions the CR background and IEM were each adjusted via a free normalisation and a power-law correction to the spectrum with free spectral index.
}
The comparison of the different emission components in the fitted and true models for catalogue B is shown in figure \ref{fig:flux_all}. We note that the relative {\change deviation} from the true flux is larger for the interstellar emission than for the sources, because most sources are detectable on smaller scales and are therefore easier to spatially distinguish from the CR background.

In order to illustrate how the measurement of the interstellar emission is affected by the confusion with the CR background, we show in figure \ref{fig:TS_IEMs} the Pearson correlation coefficient between the predicted counts of the CR background and the IEM {\change (for two different models and energy bands)}, the TS of the IEM, and the fitted IEM normalisations as a function of the Galactic longitude\footnote{\change We note that all models are fit to data simulated based on the IEM-base model.}. In the outer Galaxy ($|l|>90^\circ$), where the interstellar emission is fainter and extends over a larger scale height, the TS is lower and the correlation coefficient is larger (as there are less counts and they are more sparsely distributed). In this case the normalisation of the IEM is poorly constrained.
If we include energies below $1$~TeV we find an increasing bias of the IEM normalisation with longitude associated with a steeper decrease of the value of the TS with longitude and a global increase of the correlation coefficient. This can be explained by the degradation of the instrument performance at lower energies. Indeed the higher CR background lowers the TS, while the broader PSF and energy dispersion increase the confusion.
In the $|l|<60^\circ$ longitude range and $>1$~TeV energy range, the interstellar emission can be significantly detected and the normalisation of the model can be correctly determined from the data for the IEM-base model used in the simulation.

The normalisation of the IEM-varmin model, fitted to the data simulated based on IEM-base model, is compatible with the flux ratio of the two models in the region where interstellar emission is significantly detected.
Moreover, the difference in flux between IEM-base and the two IEM-varmin variants is larger than the statistical error on the fitted model and its deviation from the true model (figure \ref{fig:flux_all} and \ref{fig:TS_IEMs}), which suggests that CTA should be able to distinguish between such scenarios.

However, the systematic uncertainties on the \bkg modelling may further {\change complicate} the exercise. Even if the error on the \bkg does not exceed a few percent, it can lead to several tens of percent error on the IEM. We performed a simple test by flipping the \bkg template in latitude, which leads to a mean relative error on the background of 2\%, with a standard deviation of 40\%. In this case the longitude range where the fitted normalisation of the interstellar model remains compatible with {\change the true values} is reduced, but the results for the innermost region ($|l|<60^\circ$) remain similar.

Several questions on the effect of systematic uncertainties on interstellar emission studies are still open. {\change Notably}, the CR background normalisation usually has to be re-fitted for each observation outside of the region of interest, which is not well-defined for interstellar emission as it spans over the whole field-of-view. Alternatively, the \bkg of each observation could be jointly fitted together with the interstellar emission and source models but this would be computationally intensive.
Moreover one could investigate {\changethree the effect of potential systematic errors} in the IRFs and \bkg  modelling. We defer further investigations on these questions to future studies.  {\changetwo In this domain we expect important synergies between CTA and neutrino observatories \cite{icecubeGP2023,2023ApJ...949...16S,antaresGR2023,2015ApJ...815L..25G}.} {\changethree Synergies with HAWC, LHAASO and \textit{Fermi}-LAT will also be crucial as these instruments have a much better handle on the \bkg modelling.}

\begin{figure*}
    \includegraphics[width=\textwidth]{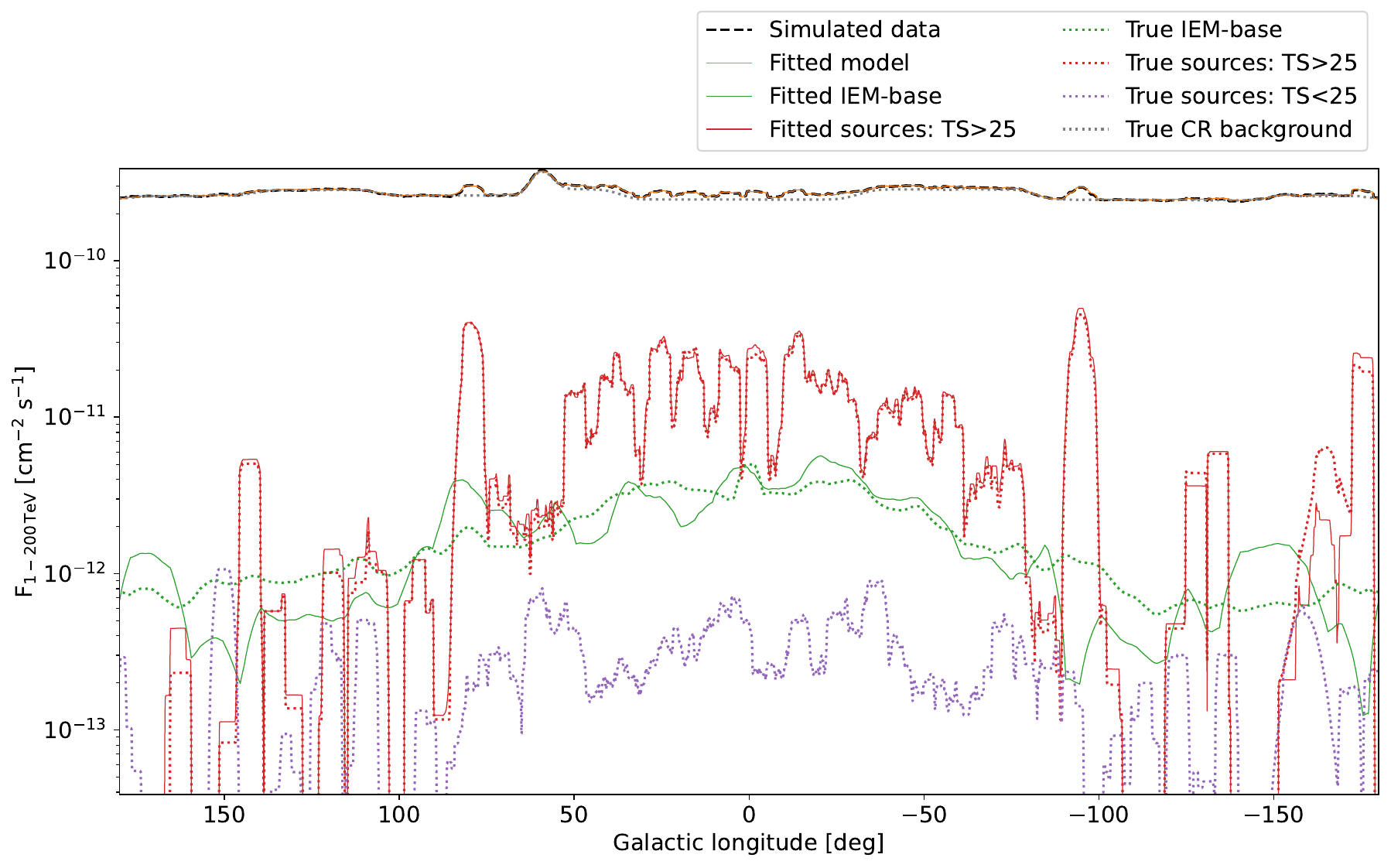}
    \caption{\LT{{\change Flux distribution in Galactic longitude} from different source and background components. Fluxes} are integrated over latitudes of $\pm 6\degr$ and over a 6$^\circ$ sliding window in longitude for the 1-200 TeV energy range. The fitted models are displayed as solid lines and the simulated models as dotted lines.}
    \label{fig:flux_all}
\end{figure*}

\begin{figure*}
    \includegraphics[width=\textwidth]{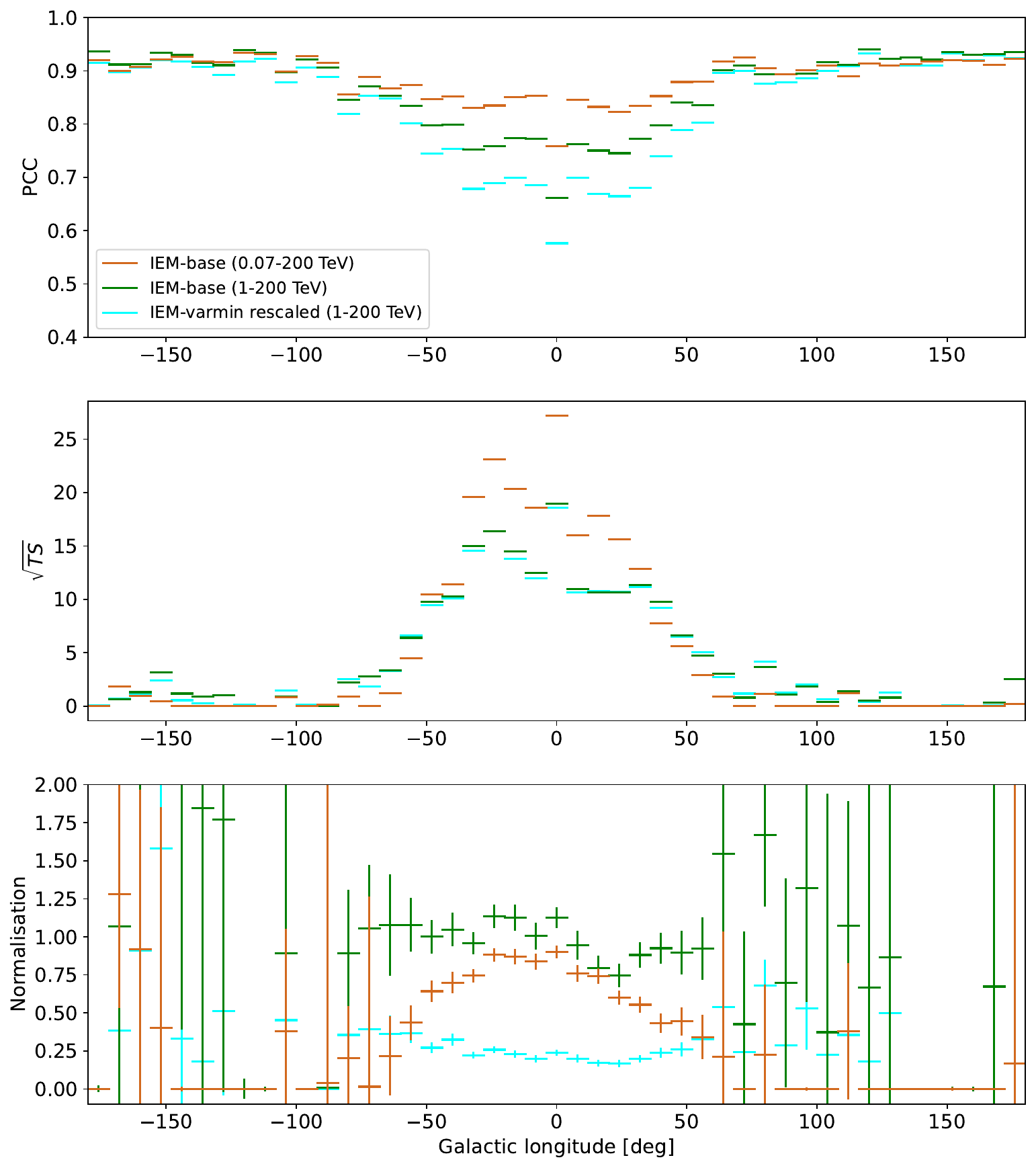}
    \caption{\LT{Longitude profiles showing from top to bottom: 1) Pearson correlation coefficient (PCC) of the predicted counts from interstellar emission and CR background; 2) $\sqrt{TS}$ of the interstellar emission models; 3) normalisation of the interstellar emission models {\change fitted to data simulated using the IEM-base model}. All these quantities are computed in 8$^\circ$ by 12$^\circ$ windows in Galactic longitude and latitude, respectively.}}
    \label{fig:TS_IEMs}
\end{figure*}

\section{Summary and perspectives}\label{sec:summary}

A survey of the entire Galactic plane has been proposed as a Key Science Project for CTAO. In this article we presented a {\change snapshot of the current status of the project preparation and predictions of the expected results}. With respect to previous assessments of the CTA GPS we employed an improved sky model {\change (made publicly available along with this paper)} that combines data from recent observations of known gamma-ray emitters with physically-driven models of synthetic populations of the three main classes of established Galactic VHE sources (PWNe, young and interacting SNRs, and compact binary systems), as well as of interstellar emission from CR interactions in the Milky Way. The article also illustrates the ongoing optimisation of the observation strategy (pointing pattern and scheduling), and the development and testing of the methods and software tools that will be later used to build source catalogues from data, to characterise specific source classes (binaries, pulsars, PeVatrons), and to study diffuse emission.

The approach used has several limitations. First of all, for most of the work we considered only one single realisation {\change of a given synthetic source population model and of the simulated data}. The effects of the variance of the synthetic populations {\change and the gamma-ray event lists are not assessed in the paper. Furthermore,} uncertainties in the population modelling may lead to variations in the number of sources detected and the properties of the sources. We also assumed that the IRFs were perfectly known and used the true background model in the analysis with simple scaling factors, while with real data the systematic uncertainties on both are certainly going to complicate the analysis.  {Note also that in this exercise, the association of a detected object with its counterpart is done by cross-matching  the catalogue results with the true sky model.
For real data, assigning a detected object to a source class will require the use of multiwavelength data and will be noticeably more difficult. A fraction of the detected sources may therefore remain unidentified.}

Moreover, as discussed in section~\ref{sec:observations}, we presented results for the CTA baseline/Omega configuration, while so far only the reduced Alpha configuration was approved for construction. For this reason, and in order to take into account other changes in the instrument configuration and performance yet to come, as well as advances in the field made possible by other instruments, the details of the programme proposal are likely to evolve in the coming years with respect the current snapshot presented in this paper.

Nevertheless, the results presented provide a plausible ballpark estimate of what can be expected from the CTA GPS {\change for the anticipated observation time of 1620 hours. 
\begin{itemize}
\item \LT{We show that, under our model assumptions and for the realisation considered, the GPS has the potential to increase by a factor five the number of known Galactic VHE emitters. }
\item In particular, we expect to be able to detect over two hundred PWNe and several tens of SNRs. PWNe should be detected across the entire Milky Way, at average {\changetwo integral} fluxes $> 1$~TeV one order of magnitude lower than in the existing sample, and with good coverage for the parent pulsar intrinsic spindown luminosity $>10^{36}$ erg s$^{-1}$ (50\% completeness for the model considered). Similar increases in the number of objects and a reduction in the typical {\changetwo integral} flux observable are obtained for SNRs. For about half of the newly-detected SNRs a significant {\changetwo angular} extension could be measured, which is valuable for source identification and for constraining physical models of the emission. 
\item The GPS also has the potential to provide several new VHE detections of gamma-ray binary systems {\changetwo and pulsars}, to confirm the existence of a hypothetical population of gamma-ray pulsars with an additional TeV emission component, and to \changereftwo{detect bright} PeVatrons {\change with {\changetwo differential} fluxes at 1 TeV $\gtrsim 10^{-11}$~cm$^{-2}$~s$^{-1}$~TeV$^{-1}$}. Furthermore, the GPS data will provide a pathfinder for deeper follow-up observations of these classes of sources.
\item {\change The GPS data will make it possible to detect interstellar emission for all models considered in the $|l|<60^\circ$ longitude range and $>1$~TeV energy range, and to statistically distinguish different scenarios. We can also extract from the GPS source catalogue an estimate of the contribution to diffuse emission from unresolved sources based on minimal modelling assumptions.} 
\end{itemize}

In addition to improvements in the source detection and characterisation methods, a major development still needed for the {\change scientific exploration of the} GPS is to gather multiwavelength/multimessenger data and to develop methods for the statistical association of sources. Indeed, the large number of sources detected will make it impractical to associate objects detected by CTA to emitters known from other wavelengths {\change via visual inspection of each source}. Advanced methods already exist in other energy bands \citep[e.g.,][]{1fgl,salvato2018,barkus2022}, but they must be adapted to the specific attributes of the VHE domain, notably {\changetwo spatial}  extension and confusion, and exploit as much as possible our understanding of the physical mechanisms underlying the multiwavelength emission.

The GPS will be accompanied by deep observations of individual Galactic sources, either from the {\change population} already known or discovered by the GPS itself. Such deep observations will provide fine morphological, spectral, and temporal characterisation of the sources, possibly complemented by a richer set of information on the objects extracted from multiwavelength/multimessenger data and/or detailed modelling \citep[e.g.,][]{ctarxj1713} and will be highly complementary to the population studies made possible by the GPS. Other proposed CTAO Key Science Projects will also address Galactic sources and interstellar emission through observations of the Galactic centre region, transient Galactic sources, the Large Magellanic Cloud, and star-forming systems \citep{2019CTAscience}. The combination of these programs will enable a transformational advance in our understanding of  gamma-ray source populations and of the physics of particle acceleration and transport in galaxies.

\acknowledgments

We gratefully acknowledge financial support from the following agencies and organisations:

\bigskip

State Committee of Science of Armenia, Armenia;
The Australian Research Council, Astronomy Australia Ltd, The University of Adelaide, Australian National University, Monash University, The University of New South Wales, The University of Sydney, Western Sydney University, Australia; Federal Ministry of Education, Science and Research, and Innsbruck University, Austria;
Conselho Nacional de Desenvolvimento Cient\'{\i}fico e Tecnol\'{o}gico (CNPq), Funda\c{c}\~{a}o de Amparo \`{a} Pesquisa do Estado do Rio de Janeiro (FAPERJ), Funda\c{c}\~{a}o de Amparo \`{a} Pesquisa do Estado de S\~{a}o Paulo (FAPESP), Funda\c{c}\~{a}o de Apoio \`{a} Ci\^encia, Tecnologia e Inova\c{c}\~{a}o do Paran\'a - Funda\c{c}\~{a}o Arauc\'aria, Ministry of Science, Technology, Innovations and Communications (MCTIC), Brasil;
Ministry of Education and Science, National RI Roadmap Project DO1-153/28.08.2018, Bulgaria; The Natural Sciences and Engineering Research Council of Canada and the Canadian Space Agency, Canada; CONICYT-Chile grants CATA AFB 170002, ANID PIA/APOYO AFB 180002, ACT 1406, FONDECYT-Chile grants, 1161463, 1170171, 1190886, 1171421, 1170345, 1201582, Gemini-ANID 32180007, Chile, W.M. gratefully acknowledges support by the ANID BASAL projects ACE210002 and FB210003, and FONDECYT 11190853; Croatian Science Foundation, Rudjer Boskovic Institute, University of Osijek, University of Rijeka, University of Split, Faculty of Electrical Engineering, Mechanical Engineering and Naval Architecture, University of Zagreb, Faculty of Electrical Engineering and Computing, Croatia;
Ministry of Education, Youth and Sports, MEYS LM2015046, LM2018105, LTT17006, EU/MEYS CZ.02.1.01/0.0/0.0/16\_013/0001403,\\ CZ.02.1.01/0.0/0.0/18\_046/0016007 and CZ.02.1.01/0.0/0.0/16\_019/0000754, Czech Republic; Academy of Finland (grant nr.317636 and 320045), Finland;
Ministry of Higher Education and Research, CNRS-INSU and CNRS-IN2P3, CEA-Irfu, ANR, Regional Council Ile de France, Labex ENIGMASS, OCEVU, OSUG2020 and P2IO, France; The German Ministry for Education and Research (BMBF), the Max Planck Society, the German Research Foundation (DFG, with Collaborative Research Centres 876 \& 1491), and the Helmholtz Association, Germany; Department of Atomic Energy, Department of Science and Technology, India; Istituto Nazionale di Astrofisica (INAF), Istituto Nazionale di Fisica Nucleare (INFN), MIUR, Istituto Nazionale di Astrofisica (INAF-OABRERA) Grant Fondazione Cariplo/Regione Lombardia ID 2014-1980/RST\_ERC, Italy; ICRR, University of Tokyo, JSPS, MEXT, Japan; Netherlands Research School for Astronomy (NOVA), Netherlands Organization for Scientific Research (NWO), Netherlands; University of Oslo, Norway; Ministry of Science and Higher Education, DIR/WK/2017/12, the National Centre for Research and Development and the National Science Centre, UMO-2016/22/M/ST9/00583, Poland; Slovenian Research Agency, grants P1-0031, P1-0385, I0-0033, J1-9146, J1-1700, N1-0111, and the Young Researcher program, Slovenia; South African Department of Science and Technology and National Research Foundation through the South African Gamma-Ray Astronomy Programme, South Africa; The Spanish groups acknowledge the Spanish Ministry of Science and Innovation and the Spanish Research State Agency (AEI) through the government budget lines PGE2021/28.06.000X.411.01, PGE2022/28.06.000X.411.01 and PGE2022/28.06.000X.711.04, and grants PID2022-139117NB-C44, PID2019-104114RB-C31,  PID2019-107847RB-C44, PID2019-104114RB-C32, PID2019-105510GB-C31, PID2019-104114RB-C33, PID2019-107847RB-C41, PID2019-107847RB-C43, PID2019-107847RB-C42, PID2019-107988GB-C22, PID2021-124581OB-I00, PID2021-125331NB-I00; the ``Centro de Excelencia Severo Ochoa'' program through grants no. CEX2019-000920-S, CEX2020-001007-S, CEX2021-001131-S; the ``Unidad de Excelencia Mar\'ia de Maeztu'' program through grants no. CEX2019-000918-M, CEX2020-001058-M; the ``Ram\'on y Cajal'' program through grants RYC2021-032552-I, RYC2021-032991-I, RYC2020-028639-I and RYC-2017-22665; the ``Juan de la Cierva-Incorporaci\'on'' program through grants no. IJC2018-037195-I, IJC2019-040315-I. They also acknowledge the ``Atracci\'on de Talento'' program of Comunidad de Madrid through grant no. 2019-T2/TIC-12900; the project ``Tecnologi\'as avanzadas para la exploracio\'n del universo y sus componentes'' (PR47/21 TAU), funded by Comunidad de Madrid, by the Recovery, Transformation and Resilience Plan from the Spanish State, and by NextGenerationEU from the European Union through the Recovery and Resilience Facility; the La Caixa Banking Foundation, grant no. LCF/BQ/PI21/11830030; the ``Programa Operativo'' FEDER 2014-2020, Consejer\'ia de Econom\'ia y Conocimiento de la Junta de Andaluc\'ia (Ref. 1257737), PAIDI 2020 (Ref. P18-FR-1580) and Universidad de Ja\'en; ``Programa Operativo de Crecimiento Inteligente'' FEDER 2014-2020 (Ref. ESFRI-2017-IAC-12), Ministerio de Ciencia e Innovaci\'on, 15\% co-financed by Consejer\'ia de Econom\'ia, Industria, Comercio y Conocimiento del Gobierno de Canarias; the ``CERCA'' program and the grant 2021SGR00426, both funded by the Generalitat de Catalunya; and the European Union's Horizon 2020 GA:824064 and NextGenerationEU (PRTR-C17.I1); Swedish Research Council, Royal Physiographic Society of Lund, Royal Swedish Academy of Sciences, The Swedish National Infrastructure for Computing (SNIC) at Lunarc (Lund), Sweden; State Secretariat for Education, Research and Innovation (SERI) and Swiss National Science Foundation (SNSF), Switzerland; Durham University, Leverhulme Trust, Liverpool University, University of Leicester, University of Oxford, Royal Society, Science and Technology Facilities Council, UK;
U.S. National Science Foundation, U.S. Department of Energy, Argonne National Laboratory, Barnard College, University of California, University of Chicago, Columbia University, Georgia Institute of Technology, Institute for Nuclear and Particle Astrophysics (INPAC-MRPI program), Iowa State University, the Smithsonian Institution, V.V.D. is funded by NSF grant AST-1911061, Washington University McDonnell Center for the Space Sciences, The University of Wisconsin and the Wisconsin Alumni Research Foundation, USA.

\bigskip

The research leading to these results has received funding from the European Union's Seventh Framework Programme (FP7/2007-2013) under grant agreements No~262053 and No~317446.
This project is receiving funding from the European Union's Horizon 2020 research and innovation programs under agreement No~676134.

\bigskip

This research made use of ctools, a community-developed gamma-ray astronomy science analysis software. ctools is based on GammaLib, a community-developed toolbox for the scientific analysis of astronomical gamma-ray data. This research made use of gammapy,\footnote{\url{https://www.gammapy.org}} a community-developed core Python package for TeV gamma-ray astronomy.




\bibliographystyle{JHEP}
\bibliography{references.bib} 



\appendix 

\section{Details on the construction of the sky model}\label{app:skymodel}

\subsection{Pulsar selection and models}\label{app:skymodelpsr}

We preselected from among more than 250 gamma-ray pulsars detected with the \textit{Fermi}-LAT \citep{2013ApJS..208...17A,2017ApJS..232...18A} those  whose {\changetwo integral} energy flux exceeds $1\%$ of that of the brightest pulsar, Vela (PSR J0835-4510), in the energy range {from} 1$-$10 GeV. This amounted to 38 pulsars, including the four brightest GeV gamma-ray pulsars detected with IACTs: Vela, Geminga (PSR J0633+1746), Crab (PSR J0534+2200) and PSR J1709$-$4429.

Two classes of gamma-ray emitting pulsars were constructed using the information on the VHE spectra of the Crab and Vela pulsars known to date. The first class (``Crab-like'' sources, 25 objects; see table \ref{tab:pulsars_sky model}) comprises the sources with a spectrum consistent with a power law (PWL) in the high-energy (HE) range from one to a few hundred GeV (called hereafter the ``GeV component'' for the sake of simplicity). Their VHE spectrum above 100 GeV was modelled simply extrapolating the ``GeV component'' without spectral cutoff, as observed in the Crab pulsar (see e.g. \citealt{2016AA...585A.133A}). This approach is similar to the one adopted in \citealt{2017MNRAS.471..431B}. The spectral parameters of Crab-like pulsars were taken from 3FHL whereas, for the Crab pulsar, we adopted the power-law spectrum of the peak emission (P1+P2) calculated for the joint \textit{Fermi}-LAT and MAGIC spectral data as reported in \citet{2016AA...585A.133A}.

The second class (``Vela-like'' sources, 13 objects; see table \ref{tab:pulsars_sky model}) comprises the sources with a \textit{Fermi}-LAT HE spectrum consistent with a power-law with an exponential cutoff (PLEC) model. Their VHE spectra were modelled {by} summing the ``GeV component'' {with} an additional multi-TeV pulsed component (called ``TeV component'' hereafter), as observed in the Vela pulsar \citep{HessVela2020}. The parameters of the GeV component were taken from \citep{2013ApJS..208...17A} (2PC), except for the Vela and Geminga pulsars for which we used the phase-averaged spectra from   \citet{2018AA...620A..66H} and \citet{2016AA...591A.138A}, respectively. Although no other pulsar with multi-TeV emission is known to date besides Vela, a TeV component was included in the spectral models for the 12 remaining objects in the same class. For these additional components the spectral index was fixed according to the measurements below 1 GeV with the \textit{Fermi}-LAT, and the high-energy cutoff was randomly set following a normal distribution centred at 7~TeV with a $\sigma$ of 3~TeV. The flux of the TeV component was randomly generated assuming a TeV to GeV flux ratio $\eta_{\rm TeV/GeV}$ ranging from $5\times 10^{-5}$ to $10^{-2}$, while complying at the same time with upper-limits available in the literature \citep{HessPSRULs2007,Veritas13psrs2019}.

The pulse shapes of both classes of pulsars were generated using templates at energies above 10 GeV made available by the \textit{Fermi}-LAT team \citep{3PC}. No evolution with energy of the phasograms was considered. We modelled the frequency as a function of time as a third-order polynomial. The coordinates and timing properties (reference epoch, frequency and its derivatives) for most of the pulsars were taken from \citet{2015ApJ...814..128K} \footnote{see also \url{https://confluence.slac.stanford.edu/display/GLAMCOG/LAT+Gamma-ray+Pulsar+Timing+Models}}. For 4 pulsars that are missing in \citet{2015ApJ...814..128K} (PSRs J1119$-$6127, J1648$-$4611, J1838$-$0537 and J2215+5135) we used the position and timing parameters reported {\changethree in the Fermi 2PC \citep{2013ApJS..208...17A}}.
   
\renewcommand{\arraystretch}{1.}
\setlength{\tabcolsep}{0.14cm}
\begin{table}
\centering
\begin{tabular}{lll}
\hline
Pulsar Name & GeV model & TeV model \\
\hline
\multicolumn{3}{c}{\textit{Crab-like pulsars}}	\\
PSR J0534+2200 (Crab)	& PWL	& not present	\\
PSR J0633+0632	& PWL (3FHL)	& not present	\\
PSR J1016$-$5857	& PWL (3FHL)	& not present	\\
PSR J1028$-$5819	& PWL (3FHL)	& not present	\\
PSR J1048$-$5832	& PWL (3FHL)	& not present	\\
PSR J1119$-$6127	& PWL (3FHL)	& not present	\\
PSR J1459$-$6053	& PWL (3FHL)	& not present	\\
PSR J1509$-$5850	& PWL (3FHL)	& not present	\\
PSR J1514$-$4946	& PWL (3FHL)	& not present	\\
PSR J1620$-$4927	& PWL (3FHL)	& not present	\\
PSR J1648$-$4611	& PWL (3FHL)	& not present	\\
PSR J1747$-$2958	& PWL (3FHL)	& not present	\\
PSR J1803$-$2149	& PWL (3FHL)	& not present	\\
PSR J1809$-$2332	& PWL (3FHL)	& not present	\\
PSR J1826$-$1256	& PWL (3FHL)	& not present	\\
PSR J1833$-$1034	& PWL (3FHL)	& not present	\\
PSR J1838$-$0537	& PWL (3FHL)	& not present	\\
PSR J1907+0602	& PWL (3FHL)	& not present	\\
PSR J1954+2836	& PWL (3FHL)	& not present	\\
PSR J1958+2846	& PWL (3FHL)	& not present	\\
PSR J2032+4127	& PWL (3FHL)	& not present	\\
PSR J2111+4606	& PWL (3FHL)	& not present	\\
PSR J2215+5135	& PWL (3FHL)	& not present	\\
PSR J2229+6114	& PWL (3FHL)	& not present	\\
PSR J2238+5903	& PWL (3FHL)	& not present	\\

\multicolumn{3}{c}{\textit{Vela-like pulsars}}	\\

PSR J0007+7303	&  PLEC (2PC) & PLEC	\\
PSR J0633+1746 (Geminga)	& super PLEC	& PLEC	\\
PSR J0835$-$4510 (Vela)	& super PLEC	& PLEC	\\
PSR J1057$-$5226	&  PLEC (2PC) & PLEC	\\
PSR J1413$-$6205	&  PLEC (2PC) & PLEC	\\
PSR J1418$-$6058	&  PLEC (2PC) & PLEC	\\
PSR J1709$-$4429	&  super PLEC (2PC) & PLEC	\\
PSR J1732$-$3131	&  PLEC (2PC) & PLEC	\\
PSR J1813$-$1246	&  PLEC (2PC) & PLEC	\\
PSR J1846+0919	&  PLEC (2PC) & PLEC	\\
PSR J1952+3252	&  super PLEC (2PC) & PLEC	\\
PSR J2021+3651	&  super PLEC (2PC) & PLEC	\\
PSR J2021+4026	&  super PLEC (2PC) & PLEC	\\

\hline
\end{tabular}
\caption{Pulsars included in the CTA GPS simulations. The spectral properties of the GeV and TeV band are chosen to be either a power-law (PWL), power-law with exponential cutoff (PLEC) or a PWL with super exponential cutoff
(super PLEC).}
\label{tab:pulsars_sky model}
\end{table}

\subsection{Other dedicated models}\label{sec:app-othermodels}

For gamma-ray binaries, periods and relevant references are listed in table~\ref{tab:grlb}. When the sky model was built the orbital period of HESS~J1832$-$093 was unknown and in our sky model we assumed it to be equal to 1 year similar to HESS~J0632$+$057. Now a periodicity of 86 days has been reported for its X-ray counterpart \citep{hessj1832_2020}. The binary nature of one more gamma-ray source,  4FGL~J1405.1$-$6119, was discovered when our work was already in an advanced state and thus it is not included in the sky model. 

For some well studied extended sources, a spatial template providing a more complex morphology is used in the simulation. Templates are also used to model some diffuse features not captured by the large-scale interstellar emission models.
The templates are based either on multi-wavelength observations or models as described in table~\ref{tab:templates}.  Some of the {templates} present an energy-dependent morphology.

\begin{table} 
            \centering   
    \begin{tabular}{lcc}
    \hline
        Name & Period& References \\
        \hline
       PSR~B1259$-$63 & 3.4 yr  &  \cite{psrbHESS2020}\\ 
        PSR~J2032$+$4127& 50 yr &  \cite{psrj2032+4127_2018}  \\
       LSI~$+$61$\degr$303 & 26.5 d& \cite{LSI61303:Gregory2002} \\
        LS~5039 & 3.9 d & \citet{LS5039:2006} \\
       1FGL~J1018.6$-$5856&16.6 d& \cite{1FGLJ1018.6:2015} \\
        HESS~J0632$+$057 &315 d& \cite{HESSJ0632:2014}\\
        HESS~J1832$-$093 & 1 yr& \cite{abramowski15}\\
     \hline   
    \end{tabular}
    \caption{Known and candidate gamma-ray binaries within the GPS footprint.} 
\label{tab:grlb}
\end{table}

\begin{table*} 
            \centering   
    \begin{tabular}{lccc}
    \hline
    Name & Source class & Template type & References \\
    \hline
    {IC~443} & iSNR & gamma-ray map &   \cite{2015ICRC...34..875H}\\
    {SNR~G78.2$+$2.1} & SNR & gamma-ray map & \cite{2017ApJ...843..139A,2018ApJ...861..134A} \\  
    {Vela~X} & PWN & radio and X-ray maps &  \cite{2012AA...548A..38A, 1999AJ....117.1578B, 2007MNRAS.382..382M, 1999AA...349..389V} \\
    {W~28} & iSNR & gamma-ray map & \cite{2008AA...481..401A} \\  
    {Pup A} & SNR & X-ray map & \cite{1993Sci...260.1769T,2015AA...575A..81H} \\
    {Vela Junior} & SNR & X-ray map & \cite{1993Sci...260.1769T} \\
    HESS J1800-240 & iSNR &gamma-ray map & \cite{2008AA...481..401A} \\
    {RX~J1713.7-3946} & SNR & X-ray map & \cite{2009AA...505..157A} \\ 
    {HESS~J1825} & PWN & interpolation of gamma-ray profiles & \cite{2019AA...621A.116H} \\
    {SS 433} lobes & microquasar lobes  & two Gaussians & \cite{2018Natur.562...82A} \\
    {Geminga halo} & pulsar halo & electron diffusion & \cite{2018APh...102....1L} \\
    {PSR~B0656}$+$14 halo & pulsar halo & electron diffusion & \cite{2018APh...102....1L} \\
    Cygnus cocoon & SFR & $1/r$ nuclei profile + gas map & \cite{2011Sci...334.1103A,2019NatAs...3..561A} \\
    {Westerlund 1} & SFR & $1/r$ nuclei profile + gas map & \cite{2013MNRAS.434.2289O,2019NatAs...3..561A} \\
    Galactic centre ridge & diffuse feature & gamma-ray map & \cite{hessgcridge2006} \\
    \textit{Fermi} bubbles & diffuse feature & gamma-ray map & \cite{herold2009} \\
    \hline
        \end{tabular}
    \caption{Sources modelled by dedicated spatial templates and references.}
\label{tab:templates}
\end{table*}

\subsection{Spectral cutoffs}\label{sec:cutoffs}

For a number of sources the spectrum is not well constrained above 30 TeV with current generation telescopes. We impose an exponential cutoff for known sources described in catalogues (gamma-cat, 2HWC, 3FHL) using a power-law spectrum with hard {\changetwo spectral} index ($< 2.4$), with the exception of  two PeVatron candidates, Westerlund~1 \citep{2019NatAs...3..561A} and HESS~J1641$-$463 \citep{2015ICRC...34..834O}. 
This cutoff is added to prevent generating too many artificial PeVatrons due to a power-law extrapolation.
The cutoff energy is computed based on physically motivated principles {\change described in the following text} for sources associated with SNRs, PWNe, and AGNs. {\changetwo For each catalogue considered, unidentified sources are treated as if they belonged to the dominant source class (PWNe for gamma-cat and 2HWC, AGNs for 3FHL)}.

SNR properties are extracted from SNRcat \citep{SNRcat}. Thermal composite SNRs and SNRs interacting with molecular clouds are considered to be {produced} by type II supernovae and their gamma-ray emission to be dominated by hadronic processes. Other SNRs are randomly assigned to be type I with 20\% probability or type II with 80\% probability, and their emission is assumed to be dominated by leptonic processes. When an age estimate is not available the age is inferred from the size of the SNR based on {the} evolutionary model described in \citet{cardillo2015}. If the age  estimated is unrealistically large ($>10$~kyr), we assume an age drawn randomly between 0.5~kyr and 1~kyr. The maximum energy of accelerated particles is then estimated based on equation~9 from \citet{cardillo2015}. For leptonic emitters the age-limited maximum energy for electrons is corrected to account for synchrotron energy losses{. To do so the} strength of the amplified magnetic field downstream {is} computed by assuming that the non-resonant instability reaches saturation upstream of the shock. The gamma-ray cutoff energy is then estimated {to be} 10\% of the maximum proton energy or electron energy (in the deep Klein-Nishina regime).

For PWNe we assume that the associated pulsar is the one at the smallest angular distance with spin-down power $> 10^{34}$ erg/s found in the ATNF catalogue \citep{ATNFcat}. Based on the pulsar properties, the maximum electron energy is computed as the minimum between the maximum potential drop available in the pulsar magnetosphere and the synchrotron-loss-limited maximum energy based on a model for the termination shock magnetic field evolution \citep{bucciantini2004}. As for leptonic SNRs, the gamma-ray cutoff energy is then estimated as 10\% of the maximum electron energy.

For AGNs the gamma-ray cutoff energy is set as a function of redshift using an analytical approximation of the maximum energy shown in figure~17 of \citet{2017ApJS..232...18A}, consistent with expectations for absorption by the extragalactic background light. For AGNs of unknown redshift or unassociated 3FHL sources, a redshift value is randomly drawn following the distribution of 3FHL AGNs with known redshift.

\changeref{The cutoff energies are shown in figure~\ref{fig:cutoffs}. They mostly lie below 1 TeV for 3FHL sources and below 100 TeV for gamma-cat and 2HWC sources.}
\begin{figure}
\centering
  \includegraphics[width=.7\textwidth]{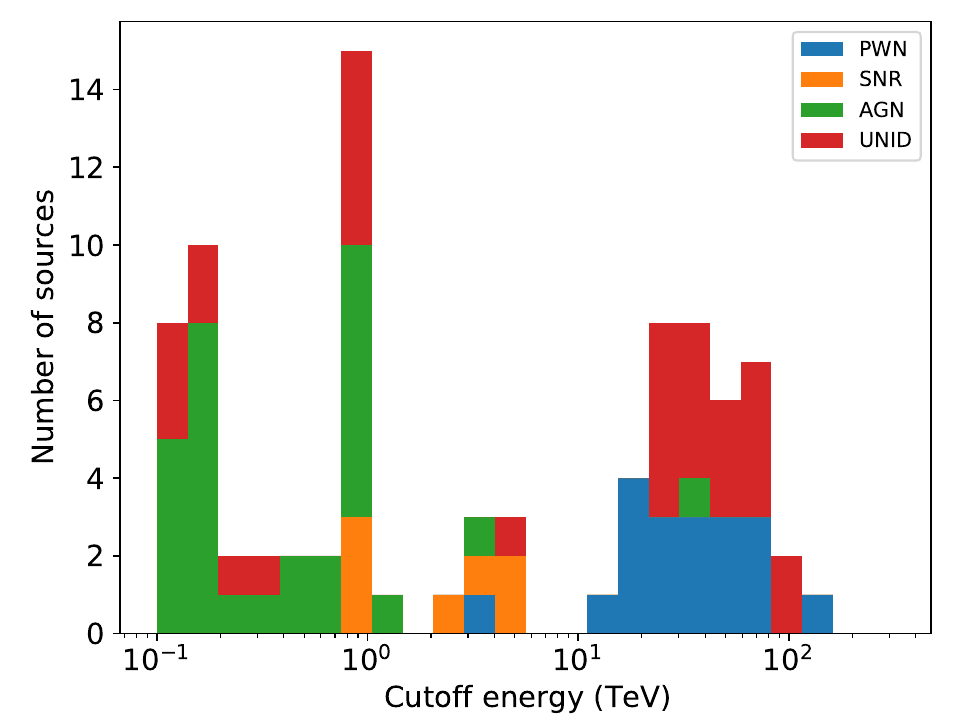}
  \caption{Cutoff energies estimated for sources described in catalogues (gamma-cat, 2HWC, 3FHL) using a power-law spectrum with hard spectral index ($< 2.4$). The histogram is colour-coded according to the source class: pulsar wind nebula (PWN), supernova remnants (SNR), active galactic nucleus (AGN) or unidentified (UNID). See text for details on how the cutoff energies are calculated \changereftwo{(appendix \ref{sec:cutoffs}).}
   \label{fig:cutoffs}}
\end{figure}

\subsection{Removal of bright synthetic sources}\label{sec:syn_source_del}

To remove bright synthetic sources we consider the following source properties:
\begin{itemize}
\item Galactic longitude, $l$ (of the centroid for extended sources);
\item Galactic latitude, $b$ (of the centroid for extended sources);
\item {\changetwo angular} extension radius, $r$ ({\change maximum radius or 99\% containment radius for models with non-zero intensities up to arbitrarily large distances from the centre such as Gaussians});
\item {\changetwo integral} photon flux $> 1$~TeV, $f$.
\end{itemize}

Each detected object is compared to synthetic sources belonging to the same class: SNRs, iSNRs, PWNe, composites, binaries. We note that, for this purpose, we identify a {\change subsample} of young SNRs and PWNe that are deemed to be observed as composite SNR/PWN systems. We do not use the information that SNR and PWN originate from the same progenitor but, as for real objects, we consider as composite systems the pairs of SNR and PWN that show (partial) overlap on the sky. The longitude/latitude of the composite system is assumed {to be} the flux-weighted average of the longitudes/latitudes of the individual objects. The extension radius is the radius of the smallest circle that encloses both objects, and the flux is the sum of the fluxes.

 For each detected object we compute a figure of merit to compare it to synthetic objects of the same class. For the $\imath^\mathrm{th}$ synthetic source the figure of merit is defined as: 
 \begin{equation}
 \mathcal{F}_\imath = \sqrt{\left(\frac{l_\imath - l}{u}\right)^2 + \left(\frac{b_\imath - b}{v}\right)^2 + \left(\frac{1}{w}\frac{r_\imath-r}{r_\imath+r}\right)^2+  \frac{1}{z^2}\log_{10}^2\left(\frac{f_\imath}{f}\right)}
 \end{equation}
with $u=180\degr$, $v=10\degr$, $w=2$, $z=0.6$ if $f_\imath < f$ or $z=0.3$ if $f_\imath > f$. The parameter values were determined empirically so that there was reasonable consistency between the spatial, {\changetwo angular}  extension, and {\changetwo integral} flux distribution for the synthetic source populations and the merged population. We note that:
\begin{itemize}
\item longitude and latitude are considered independently to preserve the population distributions for both quantities;
\item values of $r_\imath$ or $r$ lower than $0.05\degr$ are set to $0.05\degr$ for the calculation of $\mathcal{F}_\imath$ to account for the limited resolution of existing observations, based on the minimum resolved size in \citet{2018A&A...612A...1H};
\item the asymmetric value of $z$ around $f_\imath = f$ ensures that brighter synthetic sources, which could have been detected more easily, are more likely to get removed. 
\end{itemize}
The synthetic source with the smallest $ \mathcal{F}_\imath$ is considered as the most similar to the detected source and therefore removed.

\section{Details on observation scheduling}\label{app:scheduling}

We use the total observing times in several longitude ranges as provided in table~6.3 of \citet{2019CTAscience}. In order to distribute the STP and the LTP observations over 10 years, we somewhat arbitrarily chose the fractions illustrated in table~\ref{tab:obsfrac}. 

\begin{table*}
\centering
\begin{tabular}{lcc|cccccccc}
\hline
Programme & \multicolumn{2}{c}{STP} & \multicolumn{8}{c}{LTP} \\
\hline                                  
Year & 1 & 2 & 3 & 4 & 5 & 6 & 7 & 8 & 9 & 10 \\
\hline
Fraction & 0.6 & 0.4 & 0.15 & 0.2 & 0.2 & 0.15 & 0.1 & 0.1 & 0.05 & 0.05 \\
\hline                                             
\end{tabular}
\caption{Fractions of the total observing times as a function of year separated for STP and LTP.} 
\label{tab:obsfrac}
\end{table*}

\begin{table*}
\centering
\begin{tabular}{lcccccccccc}
\hline               
Year & 1 & 2 & 3 & 4 & 5 & 6 & 7 & 8 & 9 & 10 \\
\hline
South & 1.5 & 3 & 3 & 1.5 & 2.5 & 5 & 7 & 10 & 12 & 14 \\
North & 4 & 5 & 5 & 4 & 4 & 5 & 6 & 8 & 12 & 13 \\ 
\hline                                             
\end{tabular}
\caption{Length of the gap in days between nights over which GPS observations are performed.} 
\label{tab:leapday}
\end{table*}

The pointing positions were then determined independently for the Northern and Southern array for each year. For each array, the number of pointings to be scheduled was determined by dividing the total duration multiplied by the fraction of time used in a given year by the duration of one pointing (30 min). Given the resulting number of pointings, the longitude step size was then determined by dividing each longitude range by the number of pointings.

For each array, the first pointing was scheduled at a given start time. The start time was computed by adding 362.25 days for each year to the start time of the first year, which was taken to {be} January 1 2021, midnight. An additional offset of 29.5 days was added for the first year, and offsets of $(10 - \mathrm{year}) \times 59 - 365.25\, \mathrm{days}$ were added from the third year on, where we count the number of years starting from 0. In this way, the scheduling within a given year started with one- or two-month date shifts, assuring that similar pointing positions were not always observed during the same period in the year.

All pointings within a given year and for a given array were then scheduled independently. Before scheduling the first pointing, and after scheduling each pointing, the start time was incremented by adding 32 min, assuming 30 min exposure time and 2 min slew time. If, for the resulting time, the Sun had a zenith angle of less than $105\degr$ (i.e. the Sun started to rise), several days were added to advance in time. In {this} way, subsequent observation nights were separated by a few days, so that the GPS observations {covered} the entire year. The {number} of days that were added depended on the year of scheduling and the array site, and are specified in table~\ref{tab:leapday}.

After this addition, and as long as the Sun had a zenith angle of less than $105\degr$ and the Moon had a zenith angle of less than $90\degr$, time was advanced in steps of 4 minutes. In this way, pointings were only scheduled when the Sun was $15\degr$ below horizon, and when the Moon was close to or below the horizon. {\changetwo The suitability of observations under Moonlight for the Galactic plane survey was not investigated.} {\change Other factors that limit the time available and affect the observability such as weather and hardware failures were not taken into account. They will need to be addressed in future work.}

Once the time of the next pointing was determined, the pointing with the smallest zenith angle was determined in the list of pointings to be scheduled for that year and site. Toggling between pointings with negative and positive Galactic latitudes was enforced. Furthermore it was enforced that the next pointing differed by at least $2\degr$ and at maximum $10\degr$ in Galactic longitude, assuring some displacement while limiting the slew time between pointings. Pointings that had a zenith angle larger than $5\degr$ from the best achievable zenith angle within a given year were excluded. In {this} way, pointings were always scheduled near their smallest possible zenith angle. In {the} case that none of the unscheduled pointings satisfied all criteria, the constraints in Galactic longitude were removed and the pointing satisfying the remaining criteria was selected. If no pointing was found after relaxing the constraints, time was advanced according to the procedure described above, and a pointing was searched for the new time. This procedure led to the scheduling of all pointings in a given year.

This algorithm  made it possible to reach an exposure as a function of time difference between pointings within longitudes $\pm 5\degr$ from every position in the Galactic plane quite uniform and without any gaps in the interval between 5 days  and about 1 year. This covers the range of observed periods of known gamma-ray binaries and a large fraction of the binary population that was simulated for the GPS. 

\section{Details on  catalogue A  production}\label{app:catalog_A}

The catalogue production strategy from \cite{2018sf2a.conf..205C} {features an iterative process with two steps: detecting significant excesses (objects) and fitting a model for these excesses to the data}.

\subsection{Object detection}

{\changetwo The significance map used for the object detection was computed using a Poisson likelihood comparing the observed counts to predicted counts. The observed and predicted counts were binned into maps with pixel widths of $0.05^\circ$ and smoothed by a Gaussian with $\sigma=0.05^\circ$.} The predicted counts map initially includes only the CR and interstellar background models, while {subsequent} iterations also include {\change objects} that have been detected and fitted (see next section) in previous steps. 

Relevant structures in the significance map were identified using a technique based on the SExtractor algorithm \citep{1996A&AS..117..393B}. First, the map is filtered to keep only pixels above a threshold of 3$\sigma$, then groups of more than 3 pixels above threshold are isolated. {\changetwo Each group is then decomposed into multiple substructures through a deblending procedure that allows to better disentangle overlapping sources. The positions of the deblended objects are used as initial guess for the models fitted in the next step.}

\subsection{Model fitting}

The Galactic plane is divided {into} 180 regions of 3$^\circ$ by 12$^\circ$ in longitude and latitude ({with} 1$^\circ$ {of} overlap in longitude). Each analysis region is binned using a spatial pixel width of 0.02$^\circ$ and 20 logarithmically spaced energy bins from 0.07 TeV to 200 TeV. Each candidate object is fitted independently via a binned maximum-likelihood analysis in three dimensions. The spatial and spectral parameters of the candidate object, along with those of the background models, are fitted simultaneously. Spatial and spectral parameters of nearby sources that contribute to the analysis region are fixed at their previously fitted values. {Only sources that are at least marginally significant} ($\rm{TS} \geq 10$) are kept in subsequent iterations.
Initially, candidate objects from the detection step are represented as point sources with power-law spectra. The spatial models of the sources are sequentially {\change refined}, first {by comparing} a point-source model {with} a disk model. If the disk model is preferred  ($\rm{TS_{disk}} - \rm{TS_{point}} \geq 10$) then a Gaussian model is also tested, and chosen if its TS is larger than that of the disk model.
Note that the source {\changetwo angular} extension is restricted to a maximum value of 3$^\circ$ to minimise the odds of incorrectly detecting background emission as a source.

Once the best-fit spatial model has been determined for all sources, each source is then tested for spectral curvature. The default power-law model (PL) is {compared with} both log-parabola and exponentially cut-off spectral models. The curved spectral model with the largest TS value is chosen if $\rm{TS_{curved}} - T _{PL} \geq 10$.
Finally, all sources are refitted to find the best-fit global values. This is done to ensure {that} each source model accounts for the best fit spatial and spectral parameters of all other sources.

\section{Details on  catalogue B  production}\label{app:catalog_B}

\subsection{Datasets}

For each observation we select events within a 3$^\circ$ and 5$^\circ$ offset from the pointing position, below and above 1 TeV, respectively. We impose a smaller offset cut at low energy as the instrument effective field-of-view is smaller. 
All observations are stacked together and binned into three-dimensional maps. The energy axis contains 10 bins per decade in logarithmic scale from 0.07 TeV to 200 TeV. The spatial binning is 0.06$^\circ$ and 0.03$^\circ$, below and above 1 TeV, respectively. This energy-dependent spatial binning is motivated by the coarser angular-resolution at low energy, and allows to improve fitting performance while saving time and memory.

\subsection{Object detection}

The goal is to build, in a short amount of computational time, a list of
objects without prior case-specific morphological assumptions. This list provides source candidates, with robust guesses on their position and morphological parameters, to be tested subsequently by a conventional template-fitting analysis. 

The first step is to compute the significance of the excesses above a given background model (instrumental and interstellar). The significance maps are produced for different correlation radii, R$_{corr}$ = \BO{0.06$^\circ$, 0.1$^\circ$, 0.2$^\circ$} and for the two energy ranges considered $E=0.07-1$~TeV and $E=1-200$~TeV. This separation is useful as different populations of sources may have different optimal energy ranges for detection.
\change{The significance maps are filtered by hysteresis thresholding \citep{canny86} using the implementation provided by scikit-image \citep{scikit-image}. We use two thresholds: first, pixels above the higher threshold are selected, and then pixels between the two thresholds are preserved only if they are continuously connected to a pixel above the high threshold. The low and high thresholds are set to  2$\sigma$ and 4$\sigma$, respectively.}

Subsequent object detection combines three methods:
\begin{itemize}
\item Peak detection: we identify local maxima above 5$\sigma$ in the hysteresis-filtered significance maps. 
\item Circle detection: the contours of each group of pixels isolated by the hysteresis filtering are fitted as a circle. If less than \mbox{80 $\%$} of the pixels are included in the circle, the object is discarded.
\item Edge detection and Hough circle detection: for details see \cite{2020APh...12202462R} and references therein.
\end{itemize}
Peak detection is performed on significance maps with R$_{corr} = 0.06^\circ$, 0.1$^\circ$, 0.2$^\circ$  for each of the two energy ranges. The circle and Hough circle detection are performed only for R$_{corr}=0.1^\circ$ in the 1-200 TeV energy range in order to take advantage of the better angular-resolution.
\change{Removal of likely duplicate objects is performed when the results of the different detection methods are combined.
Groups of objects with an inter-center angular separation less than  $0.1^\circ$ and a difference in radius less than $0.25^\circ$ are replaced by a single object by averaging their position and radii.}

\subsection{Classification, spatial model selection, and candidate filtering}

\change{Extended source morphologies are usually more complex than the parametric models we use to describe them, so a single extended source can be detected as a cluster of objects. To address this, we start by classifying the objects depending on their degree of overlap.}
For every object of radius $R$ in the list we calculate the inter-center angular separation to any other object of radius $R_{other}$. The object is then classified as:
\begin{itemize}
\item "isolated", if $d>R+R_{other}$ for all other objects;
\item "outer sub-structure", if there exists another object such that $d \leqslant R+R_{other}$ and $R< R_{other}$;
\item "inner sub-structure", if there exists another object such that $d \leqslant R$ and $R < R_{other}$;
\item "parent", if there are any other objects such that $d \leqslant R+R_{other}$ and for all $R_{other} \leqslant R$;
\end{itemize}
In short: isolated objects are non-overlapping, while parents are large objects partially overlapping with smaller ones. These smaller objects are considered as sub-structures of their parent. 

{\change For each object, a baseline spatial model is determined using the} Pearson correlation coefficient ($PCC$) of a 5-point radial profile in flux. To this purpose, we integrated the flux map (with R$_{corr}$ = 0.1$^\circ$) in 5 rings of equal area between the centre of the object and its radius. \change{Then we computed the PCC  between the radius and flux values. The default spatial model is a generalised Gaussian (see next section). Alternatively, a shell is considered first if the candidate object has $PCC\, >1/3$, or if it has $|PCC|\, <1/3$ and overlaps with objects classified as inner sub-structures.}

For each object we compute a test statistic (TS) defined as the squared significance of the residual excess integrated within a correlation radius equivalent to the object radius. The candidate objects are filtered by requiring $ \rm TS$>10.
We then perform outlier detection using the isolation forest algorithm \citep{2018arXiv181102141H} implemented in scikit-learn \citep{scikit-learn}. The {\change object properties} considered when applying the outlier detection are: the mean distance of the 5 nearest neighbours, the distance of a sub-structure to its parent-object, and the PCC.
The distribution of objects associated with known sources {\change from existing catalogues}\footnote{The association procedure is described in \ref{sec:diag}. The only difference is that for candidate objects the surface used in the association criterion is computed as a disk using the radius estimated from the detection step.} in this parameter space informs us on the expected density of objects with a given morphology. {\change This information is used to train the outlier detection classifier and to set its selection score.
Sub-structures below a threshold in selection score are discarded in order to reduce spurious detections in complex sources (those that still remain after the removal of likely duplicate objects). The remaining objects are then ranked according to their selection score, which is used to determine their fitting order in the following (see next section).}

Furthermore, after the TS filtering and the outlier detection the selected candidates are divided into two lists:
\begin{itemize}
\item primary: objects associated with known sources, isolated objects, parent objects, and sub-structures more significant than their parent with a difference in TS larger than 25;
\item secondary: unassociated sub-structures less significant than their parent.
\end{itemize}

\subsection{Model fitting}\label{app:catalog_B5}

For model fitting, the Galactic plane is divided into 10$^\circ$ wide regions separated by 5$^\circ$ (half-overlapping). A 3$^\circ$ {\change border} is added to each sub-region to account {\changetwo for the spill-over into the analysis region of photons from the sources outside the region due to the instrument PSF}.
{\change Regions} containing less than five sources are merged with their neighbour in order to limit the number of regions fitted. The 56 sub-regions obtained are then fitted independently.
The objects {with centres outside of the \change fit region}, but within the {\change 3$^\circ$ border}, are merged into a unique background component.
So, for each energy range, we have three background {\change components: CR background}, IEM, and sources {\change centred outside the fitting region}.
Note that for the catalogue production, the true model for CR and IEM backgrounds are used and only a normalisation and spectral index correction are fitted. 

By default, candidate sources are fitted with a log-parabola\footnote{\url{https://docs.gammapy.org/1.0.1/user-guide/model-gallery/spectral/plot_logparabola.html}} as spectral model and a generalised-Gaussian\footnote{\url{https://docs.gammapy.org/1.0.1/user-guide/model-gallery/spatial/plot_gen_gauss.html}} as spatial model.
The generalised-Gaussian model has a shape parameter, $\eta$,  fitted for values in $[0.1, 1]$. This model is equivalent to a disk when $\eta$ tends to zero, a Gaussian for $\eta=0.5$ and a Laplace profile for $\eta=1$. 
The minimum size fitted is $0.06^\circ$ (slightly larger than the mean PSF radius at 1 TeV).
Alternatively, a shell is fitted as the spatial model based on the morphological estimate from the initial detection step (see previous section).

For each {\change region}, the candidates in the primary list are fitted first while those in the secondary list are added only if there is still a significant residual excess after the fitting of the primary candidates ($\rm TS_{resi}$>25).
Once the initial candidate lists are exhausted, more sources are added iteratively (up to 5 per region) at the position of the largest peak above $5\sigma$ in the residual significance map.
Finally, for each object, we compute the test statistic for the null hypothesis (no source) and keep only those with $\rm TS_{null}>25$ in at least one of the energy ranges.
Once the fitting of all regions is complete, we assemble the final global model. Models are taken from one {\change region} only if they are located within $2.5^\circ$ from the centre of the {\change region}  ({\change regions}  are overlapping).

\subsection{Models refinement}\label{app:catalog_B6}

{\change At this point, for each source, we compare generalised-Gaussian, shell, and point-like morphological models}.
The optimal parameters of a model are given by the likelihood maximisation, and the selection among alternative models is performed by minimisation of the Akaike information criterion \citep[AIC, ][]{1974ITAC...19..716A} in order to take into account the difference in the number of parameters between the models (which are not necessarily nested).

In order to simplify the global model we also search for groups of sources that could be replaced by an alternative model. In particular, we scan the source distribution to identify linear and circular patterns that could be regrouped into a single elliptical Gaussian or shell, respectively.
To do so, we first extract several groups of pixels by applying two thresholds to the flux map of the fitted model: the first (low) threshold (95\%  percentile of the flux distribution) isolates the different groups and the second (higher) threshold (10\% of the group maximum flux) separates the bright peaks from the more diffuse regions. Then, within each of these regions we search for clusters of sources whose positions best match a linear or circular pattern using the RANSAC algorithm \citep{Fischler1981RandomSC}, implemented in scikit-image \citep{scikit-image}.

\change{{\changetwo The clusters of sources forming a linear pattern} were replaced by an elliptical generalised Gaussian if the difference in AIC between multiple models and one elliptical model was greater than zero.} This test resulted in 13 new ellipticals. Similarly, {\changetwo the clusters of sources forming a circular pattern} were tested against a shell, but we also considered a shell plus a Gaussian \change{(that could best model a composite system made of a PWN and and SNRs from the same progenitor)}. This test added one composite object to the catalogue. Overall the clustering procedure removed about 30 objects. 
We also looked for sources surrounded by negative or positive residuals above $1\sigma$ with an ellipticity larger than 0.5, for which we tested an elliptical generalised Gaussian. This test resulted in one additional elliptical source.  

Finally, we reduced the catalogue to a list of complexes with similar spectral properties by merging the sub-structures with their parent object if they did not have a different association {\change to a known source} or a different spectral shape \change{(if the differences in index and curvature parameters to their mean values within the group were lower than their standard deviations within the group).}

\section{Maps of unresolved point sources}\label{app:unresolved}

\changereftwo{Figure~\ref{fig:unresolved_map} provides an example of a two-dimensional map of unresolved sources in the energy band $>1$~TeV reconstructed from the simulated CTA data as described in section~\ref{sec:unreso}.}

\begin{figure}
    \centering
    \includegraphics[width=1.1\columnwidth]{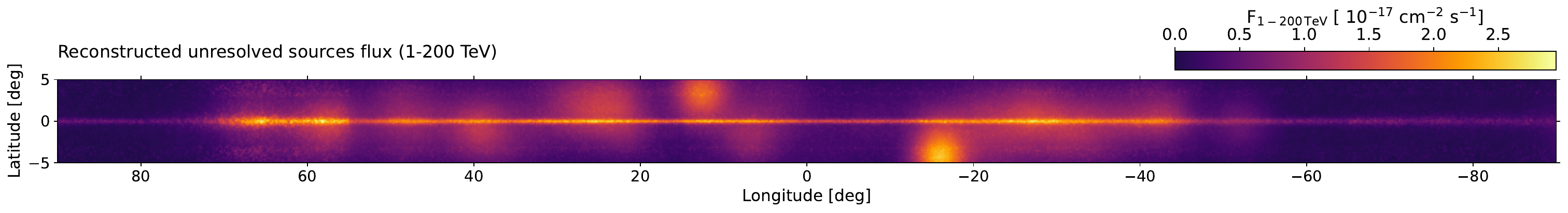}
    \caption{\changereftwo{Map of unresolved sources in the energy band $>1$~TeV reconstructed from the simulated CTA data as described in section~\ref{sec:unreso}. We show a zoom of the map in the longitude range $|l| < 90\degr$ where the emission is more prominent.}}
    \label{fig:unresolved_map}
\end{figure}


\end{document}